\documentclass[acmlarge,screen]{acmart}

\newif\ifcomments\commentstrue

\AtBeginDocument{%
  \providecommand\BibTeX{{%
    \normalfont B\kern-0.5em{\scshape i\kern-0.25em b}\kern-0.8em\TeX}}}

\setcopyright{cc}
\setcctype{by}
\acmJournal{IMWUT}
\acmYear{2025} \acmVolume{9} \acmNumber{4} \acmArticle{203}
\acmMonth{12} \acmPrice{}\acmDOI{10.1145/3770674}

\usepackage{boxedminipage}
\usepackage{xspace}

\usepackage{url,graphicx,changepage,subfiles,subfigure,listings}
\usepackage[mathscr]{euscript}
\usepackage{enumitem}

\newcommand{\modified}[1]{#1}

\usepackage{xcolor}
\usepackage{xspace}
\usepackage{cleveref}
\usepackage{multirow}
\usepackage{capt-of}

\newcommand{\capname}{{OrganicHAR}\xspace}
\newcommand{\name}{{OrganicHAR}\xspace}

\newcommand{\tabref}[1]{Table~\ref{#1}}
\newcommand{\figref}[1]{Figure~\ref{#1}}

\newcommand{\eqnref}[1]{Equation~\ref{#1}}
\newcommand{\appref}[1]{Appendix~\ref{#1}}

\newcommand{\etal}{\textit{et al.}}
\newcommand{\eg}{\textit{e.g.},~}
\newcommand{\ie}{\textit{i.e.},~}
\newcommand{\etc}{\textit{etc.}}

\newcommand{\fscoreAmbientIMUOAKDRelaxed}{56\%}

\newcommand{\overallfscoreavg}{68\%}

\newcommand{\accAmbientRelaxed}{70\%}

\newcommand{\accAmbientIMUOAKDConservative}{88\%}
\newcommand{\accAmbientIMUOAKDBalanced}{79\%}
\newcommand{\accAmbientIMUOAKDRelaxed}{73\%}

\newcommand{\overallAccAvg}{77\%}

\begin{document}

\title[OrganicHAR: Towards Activity Discovery in Organic Settings]{OrganicHAR: Towards Activity Discovery in Organic Settings for Privacy Preserving Sensors Using Efficient Video Analysis}

\author{Prasoon Patidar}
\orcid{0000-0003-0034-2767}
\affiliation{
  \institution{Carnegie Mellon University}
  \city{Pittsburgh}
  \country{United States}
}
\email{prasoonpatidar@cmu.edu}

\author{Riku Arakawa}
\orcid{0000-0001-7868-4754}
\affiliation{
  \institution{Carnegie Mellon University}
  \city{Pittsburgh}
  \country{United States}
}
\email{rarakawa@cs.cmu.edu}

\author{Ricardo Graça}
\orcid{0000-0003-3944-607X}
\affiliation{
  \institution{Fraunhofer Portugal AICOS}
  \city{Porto}
  \country{Portugal}
}
\email{ricardo.graca@aicos.fraunhofer.pt}

\author{Rúben Moutinho}
\orcid{0000-0002-5200-4260}
\affiliation{
  \institution{Fraunhofer Portugal AICOS}
  \city{Porto}
  \country{Portugal}
}
\email{ruben.moutinho@fraunhofer.pt}

\author{Adriano Soares}
\orcid{0009-0009-2205-7060}
\affiliation{
  \institution{Fraunhofer Portugal AICOS}
  \city{Porto}
  \country{Portugal}
}
\email{adriano.soares@aicos.fraunhofer.pt}

\author{Ana Vasconcelos}
\orcid{0000-0003-1842-172X}
\affiliation{
  \institution{Fraunhofer Portugal AICOS}
  \city{Porto}
  \country{Portugal}
}
\email{ana.vasconcelos@fraunhofer.pt}

\author{Filippo Talami}
\orcid{0000-0001-6577-1613}
\affiliation{
  \institution{Fraunhofer Portugal AICOS}
  \city{Porto}
  \country{Portugal}
}
\email{filippo.talami@fraunhofer.pt}

\author{Joana Couto da Silva}
\orcid{0000-0002-0411-4414}
\affiliation{
  \institution{Fraunhofer Portugal AICOS}
  \city{Porto}
  \country{Portugal}
}
\email{joana.couto@aicos.fraunhofer.pt}

\author{Inês Silva}
\orcid{0009-0008-4323-1382}
\affiliation{
  \institution{Fraunhofer Portugal AICOS}
  \city{Porto}
  \country{Portugal}
}
\email{ines.silva@aicos.fraunhofer.pt}

\author{Cristina Mendes Santos}
\orcid{0000-0003-1437-9157}
\affiliation{
  \institution{Fraunhofer Portugal AICOS}
  \city{Porto}
  \country{Portugal}
}
\email{cristina.santos@fraunhofer.pt}

\author{Mayank Goel}
\orcid{0000-0003-1237-7545}
\affiliation{
  \institution{Carnegie Mellon University}
  \city{Pittsburgh}
  \country{United States}
}
\email{mayankgoel@cmu.edu}

\author{Yuvraj Agarwal}
\orcid{0000-0001-9304-6080}
\affiliation{%
  \institution{Carnegie Mellon University}
  \city{Pittsburgh}
  \state{PA}
  \country{USA}
}
\email{yuvraj@cs.cmu.edu}

\renewcommand{\shortauthors}{Patidar et al.}

\begin{abstract}

Deploying human activity recognition (HAR) at home is still rare because sensor signals vary wildly across houses, people, and time, essentially requiring in-situ data collection and training.
Prior approaches use cameras to generate training labels for privacy-preserving sensors (LiDAR, RADAR, Thermal), but this forces sensors to detect predefined activities that cameras can see yet the sensors themselves cannot reliably distinguish.
In this work, we introduce \capname, an activity discovery framework that inverts this relationship by placing sensor capabilities at the center of activity discovery. Our approach identifies naturally occurring signal patterns using privacy-preserving sensors, leverages Vision Language Models (VLMs) only during these key moments for scene understanding, and discovers discrete activity labels at granularities that these sensors can reliably detect.
Our evaluation with 12 participants demonstrates \capname's effectiveness: it achieves \accAmbientIMUOAKDBalanced~ accuracy for coarse (4-5) activities using only basic ambient sensors (radar, lidar, thermal arrays), and \accAmbientIMUOAKDRelaxed~ accuracy for fine-grained (8-9) activities when a wearable IMU, depth, and pose sensor are added. \name maintains \overallAccAvg~ accuracy on average across configurations while discovering 4-8 categories per user (15 across all users) tailored to each environment and sensor capabilities. By triggering video processing only at key moments identified by local sensors, we reduce queries to VLM by 90\%, enabling practical and privacy-preserving activity recognition in natural settings.
\end{abstract}



\maketitle

\section{INTRODUCTION}
\label{sec:introduction}

Smart homes that can infer the activities of their occupants and offer them insights, automation, and assistance without privacy-invasive cameras or microphones have been a longstanding vision of the research community and companies \cite{applications_of_har,bouchabou_survey_2021, brush_home_2011}.
Yet despite significant advances in Human Activity Recognition (HAR) techniques, their deployment remains largely unrealized in most households \cite{vrigkas2015review, siirtola_incremental_2019}.
While these HAR systems achieve impressive performance using benchmarks and in controlled settings, they falter in real home settings.
This persistent disconnect stems not from technological limitations alone, but from a fundamental misalignment in how we have conceptualized the problem: we face the dual challenge of (i) deploying models in organic settings where users perform diverse, evolving activities in countless variations and (ii) simultaneously obtaining high-quality training labels to identify what activities are occurring in these complex, unstructured environments. 

Most existing HAR approaches start with a set of activities to detect and classify.
For instance, Patidar~\etal~\cite{patidar_vax_2023} trained privacy-preserving sensors (\eg radar, lidar) to detect 17 common activities of daily living at home.
While these approaches prove effective for short durations, the assumption that the same set of activities happens with similar consistencies across all environments breaks down in the long term, as users engage in everyday activities in their very own idiosyncratic ways \cite{hiremath_lifespan_2023,siirtola_incremental_2019}.
This reliance on pre-defined activity sets creates challenges in aligning with activities that matter to users and what activities these sensors can reliably detect in diverse environments.
Rather than imposing a fixed set of activities across all environments, we need HAR systems that can autonomously discover what activities are reliably detectable given the specific sensors deployed in each unique setting. This discovery capability is crucial for practical adoption—users must understand their sensors' actual detection capabilities before making informed decisions about how they want to use these systems.

\begin{figure}[ht]
    \centering
    \includegraphics[width=1.0\linewidth]{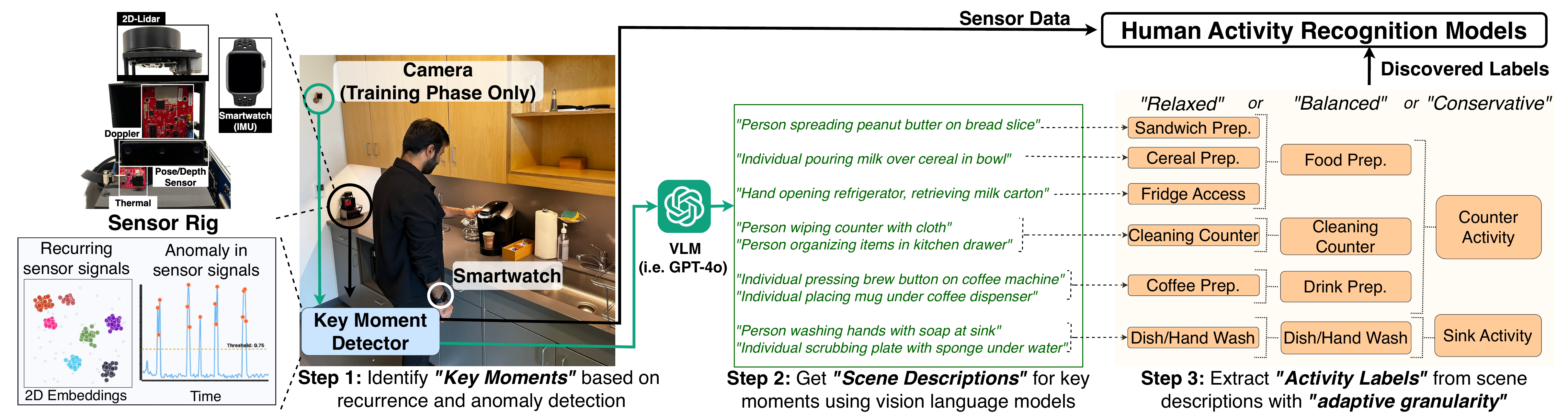}
    \caption{The \name framework discovers activity labels through a three-step sensor-first approach. \textbf{Step 1:} Privacy-preserving sensors identify ``key moments'' with recurring patterns (clusters) or anomalies (peaks). \textbf{Step 2:} A vision language model (VLM) generates natural language descriptions only for these key moments. \textbf{Step 3:} Descriptions are converted into activity labels with adaptive granularity tuned for downstream applications, following human-intuitive hierarchies where specific activities (\eg "Sandwich Prep") naturally group into broader categories ("Food Prep" → "Counter Activity") while maintaining semantic meaning. Unlike prior approaches that use predefined activities and continuous video monitoring (top right), \name discovers activities organically from sensor capabilities while reducing video processing by 90\%.}
    \Description[Three-step workflow showing sensor-first activity discovery with adaptive granularity]{
        Diagram showing OrganicHAR's three-step workflow for activity discovery in a kitchen environment. Left side shows a kitchen with person preparing food, monitored by privacy-preserving sensors (radar, lidar, thermal) and a camera used only during training phase. Center section illustrates Step 1 with two visual panels: top panel shows colored clusters representing recurring sensor signal patterns, bottom panel displays a time-series graph with orange dots marking temporal anomalies above a threshold line. These key moments feed into Step 2, where GPT-4o vision language model generates natural language descriptions like "Person spreading peanut butter on bread slice" and "Individual pouring milk over cereal in bowl." Step 3 shows hierarchical activity label organization at three granularity levels (Relaxed, Balanced, Conservative) with orange boxes representing discovered labels: specific activities like "Sandwich Prep" and "Cereal Prep" merge into "Food Prep," which further merges into "Counter Activity" at coarser levels. Similar hierarchies shown for Fridge Access, Cleaning Counter, Coffee Prep, and Dish/Hand Wash activities grouped by location (Food Prep, Cleaning, Drink Prep, Sink Activity). Top right corner contains comparison table contrasting Prior Approaches (pre-defined activity sets, fixed granularity, continuous video) versus OrganicHAR's sensor-first approach (discovered activities, adaptive granularity, key moments only).}
    \label{fig:organichar-teaser}
    \vspace{-1em}
\end{figure}

In this paper, we introduce \name \footnote{\label{footnote:organichargithubrepository} \url{https://github.com/synergylabs/OrganicHAR}} (Figure \ref{fig:organichar-teaser}), a novel framework that takes the first step towards this vision by discovering activities that naturally emerge from available sensing capabilities, rather than imposing predefined categories.
We start by identifying potentially meaningful patterns (\ie recurring spatial patterns and temporal fluctuations) within sensor data.
Next, we selectively leverage Vision Language Models~\cite{bordes_introduction_2024} (VLMs) to understand what activities these patterns represent, and convert descriptions (in natural language) into discrete labels to train HAR models.
Our approach offers two advantages critical for practical deployment: (1) it allows us to control the granularity of recognized activities for different environments and available sensing capabilities; and (2) it reduces computational overhead by processing only essential video segments during training.

Developing the \capname framework required solving two major technical challenges.
First, we developed novel techniques to identify meaningful interaction moments from multimodal time-series sensor data, without requiring prior activity models. 
Second, to address the inconsistent activity descriptions generated by VLMs, we implemented a clustering method that transforms these variable VLM descriptions into consistent, sensor-appropriate activity labels—effectively connecting rich semantic understanding of VLMs with the limited capabilities of privacy-preserving sensors. 

To build and evaluate the performance of \capname, we collected a comprehensive multimodal dataset (comprising a Doppler RADAR, a 2D LiDAR, a thermal array, a wearable IMU, pose and depth sensors) from 12 participants, who performed four different breakfast preparation tasks (preparing sandwiches, tea, coffee, and cereal).
Each participant performed these tasks naturally without instructions, allowing us to capture organic activity patterns as they would occur in real homes.
Our results show that \capname can reliably discover up to 4-8 labels per user (15 unique labels across 12 users) with 87\% accuracy.
Using these automatically discovered labels, we trained downstream activity recognition models for privacy-preserving sensors that achieved \accAmbientRelaxed-\accAmbientIMUOAKDConservative~ accuracy (F1-Score: \overallfscoreavg), with performance varying based on activity label granularity and sensor configuration. 
This performance is notable because it emerges from activity patterns discovered in an organic setting and is not limited to detecting manually defined activities. We further validated \name's effectiveness in the wild, through deployment in 5 homes for 7 days each, where the system adapted to natural daily routines and diverse kitchen environments, achieving 74.2\% accuracy while demonstrating effective learning over time.
Moreover, our approach is computation and cost-efficient since it only sends approximately 10\% of video data to VLMs in the cloud, which also reduces over time. 
These results highlight the potential for practical at-home HAR systems that require no human labor and minimize reliance on privacy-invasive sensors.
In summary, we make the following contributions:
\begin{itemize}[leftmargin=*]
    \item \capname, a framework that reorients HAR by discovering activities from sensor patterns rather than predefined categories, minimizing privacy concerns while enabling practical deployment in real homes without continuous video monitoring. The source code is publicly available.
    
    \item Evaluation of \capname, which demonstrated that it achieves \accAmbientRelaxed-\accAmbientIMUOAKDConservative~ accuracy for different hardware configurations while automatically discovering up to 4-8 activities per user (15 unique across 12 users) without manual annotation. Our real-world deployment in 5 homes further validates the practical effectiveness of \name, showing 74.2\% accuracy across natural daily routines.

    \item Multimodal dataset comprising controlled evaluation data from 12 participants performing breakfast preparation tasks across three distinct kitchen environments and real-world deployment data from 5 homes over 7 days each (11 hours), captured through six sensor modalities \cite{organichar-github}. We hope this dataset enables further research in the space of HAR in unconstrained settings.
\end{itemize}

\section{BACKGROUND AND MOTIVATION}
\label{sec:bg}

Ultimately, our goal is to develop context-aware systems that assist people in their homes—enabling aging in place, supporting complex daily activities, and promoting safety.
In deploying such systems, cameras and vision-based models are powerful, but continuous monitoring violates privacy. 
On the other hand, sensors, such as RADAR, Thermal sensor, and IMU, protect privacy but require extensive manual labeling of training data that users cannot realistically provide. 

We have worked toward this vision through a series of projects, each solving certain challenges while revealing others.
VAX \cite{patidar_vax_2023} demonstrated that privacy-preserving sensors (RADAR, LIDAR, Thermal sensor) could achieve 85\% accuracy after bootstrapping from camera data during the training phase, eliminating continuous camera-based monitoring. 
However, like several prior studies, VAX uses a rigid taxonomy of predefined activities and fails to capture real-world diversity, with unrecognized behaviors relegated to an ``Other'' category. 
Then, as one of the applications meaningful to users, PrISM \cite{arakawa_prism-q_2024,arakawa_prism-tracker_2023,arakawa_prism-observer_2024,DBLP:conf/uist/ArakawaPPLG25} was developed as a framework to support procedural tasks, such as cooking and self-care, with a context-aware, mixed-initiative assistant that combines HAR and dialogue interaction.
While demonstrating effectiveness in certain scenarios, the PrISM assistant often faces challenges in HAR; people improvise, combine tasks, and create their own routines—variations that predefined activity models cannot anticipate. 
These prior projects have motivated this work: we need to discover what activities (1) occur in an environment ; (2) can be sensed by the system; and (3) are beneficial to be tracked for the user. 

Hence, we built \name, which inverts the traditional paradigm by letting sensor capabilities drive activity discovery.
\name's ``sensor-first'' approach enables each deployment to develop its own activity taxonomy matched to both the available sensing resources and how individuals actually behave.
The convergence of these three technologies—VAX's privacy-preserving sensing, PrISM's procedural assistance, and OrganicHAR's organic discovery—now enables systems that learn individual household patterns, monitor for safety concerns, and provide assistance tailored to each person's unique routines.
We are currently working with people living with Dementia \cite{autonomousprojectwebsite} and post-operative skin cancer patients~\cite{Ha2024Self-narration,Vaccarello2024} to provide daily task assistance and safety monitoring. 
In our deployments, the system learns each person's unique patterns, monitors for safety concerns like forgotten appliances (\eg stove left on), and provides task assistance tailored to their specific routines and cognitive abilities.
By integrating bootstrapped sensing, adaptive interaction, and organic activity discovery, we move closer to systems that adapt to people's actual needs and routines instead of requiring people to adapt to predefined technological constraints.

\section{RELATED WORK}
\label{sec:relatedwork}

We begin with previous work using privacy-preserving sensors for recognizing human activities.
Then, we review machine-learning methods for addressing the challenges of training HAR models in new environments.
Finally, we focus on prior work for in-situ training to clarify the novelty of our approach.

\subsection{Human Activity Recognition (HAR) with Privacy-Preserving Sensors}
\label{sec:relatedwork-har}

Human Activity Recognition (HAR) has been extensively studied for its promising applications. Video-based HAR has traditionally dominated~\cite{pareek_aireview2021_har_survey}, benefiting from large-scale datasets~\cite{caba_cvpr2015_activitynet,carreira_2017cvpr_kinetics} and advanced understanding toolboxes~\cite{2020mmaction2,mmpose2020}. VLMs have recently revolutionized this field, enabling zero-shot activity recognition~\cite{lunia_can_2024,rocamonde_vision-language_2024}.
However, deploying video-based systems in real homes faces significant challenges: substantial computational resources (often gigabytes of GPU memory) and serious privacy concerns from continuous monitoring~\cite{abdi_2019soups_smartspeakerprivacy,zhang_review_2017}.
Researchers have investigated alternatives such as privacy-preserving ambient sensors—Doppler radars~\cite{ahuja_chi2021_vid2doppler,bhalla_ubicomp2021_imu2doppler}, lidars~\cite{laput_chi2019_surfacesight}, low-resolution thermal arrays, subsampled audio sensing~\cite{mollyn_ubicomp2022_samosa}, and environmental sensors~\cite{laput_CHI17_synthetic-sensors}—and wearable devices with IMUs~\cite{DBLP:journals/imwut/BhattacharyaAT22,DBLP:journals/imwut/XiaFAS22,DBLP:conf/uist/ZhouAAG25}.
Wang~\etal~provide a comprehensive review of multi-modal sensor fusion techniques~\cite{DBLP:journals/comsur/WangMLJYDH25}. Researchers have also explored multi-modal sensing~\cite{web-mites}, sensor fusion~\cite{aguileta_sensors2019_sensorfusion,munzner_acm2017_sensorfusion}, and generative training approaches~\cite{acm_wu_smartbuildingprivacy, Liu_2019_PAN} to enhance accuracy.

Most existing HAR approaches rely on supervised learning with predefined activity labels, requiring designers and developers to anticipate activities before data collection. This creates challenges in selecting the appropriate granularity: too fine-grained (\eg distinguishing between chopping different vegetables) reduces accuracy due to subtle signal differences and increased complexity; too coarse-grained (\eg simply detecting ``cooking'') diminishes utility by lacking meaningful and actionable task insights. These challenges highlight the need for approaches that \textit{discover} and recognize activities autonomously at appropriate granularity.
Notably, while clustering-based methods can identify patterns without predefined labels~\cite{cook_activity_2013,ariza_colpas_unsupervised_2020}, they require manual interpretation to assign meaningful labels~\cite{wu_automated_2020,hiremath_bootstrapping_2022}, limiting practical applications like monitoring specific home events. \name addresses this gap by streamlining the process through advances in vision and language foundational models.

\subsection{Transfer Learning and Domain Adaptation}
\label{sec:relatedwork-domainadaptation}

Transfer learning and domain adaptation techniques address HAR training challenges with privacy-preserving sensors by leveraging knowledge from data-rich but privacy-invasive modalities (video and raw audio) to generate synthetic training data for privacy-preserving sensors. Researchers have developed methods converting audio/video/images to synthetic IMU signals~\cite{kwon2020imutube,liang_iswc2022_audioimu,yoon_img2imu_2024}, video to doppler sensor readings~\cite{ahuja_chi2021_vid2doppler,deng_midas_2024}, or IMU signals to doppler sensor readings~\cite{bhalla_ubicomp2021_imu2doppler}. This paradigm has strengthened with advances in foundational models~\cite{leng_generating_2023,leng_imugpt_2024}, with researchers using VLMs as zero-shot or few-shot learners for HAR~\cite{arrotta_contextgpt_2024,rocamonde_vision-language_2024,leng_benefit_2023}.
These approaches require continuous video data access and substantial computational resources, making them impractical where privacy, cost, and latency matter. Self-supervised learning offers another direction to minimize reliance on labeled data~\cite{deldari2022beyond,jain_acm2022_collossl,presotto2022federated}, but still requires predefined activity sets and struggles to discover new activities autonomously.
Unlike approaches that transfer knowledge into predefined activity sets, our work explores a novel bootstrapping framework focusing on the signal representation of each sensor.
We use targeted video analysis of key moments to build robust models that subsequently operate using only privacy-preserving sensors, leveraging VLMs' understanding capabilities while maintaining privacy and practical deployability in real-world settings.

\subsection{Approaches for In-situ Training}
\label{sec:relatedwork-insitu}

In-situ training approaches aim to bootstrap HAR systems to new environments while minimizing data collection and annotation burden. Clustering-based methods group similar sensor patterns, and then request user labels for representative samples from each cluster~\cite{wu_automated_2020,hiremath_bootstrapping_2022}.
While reducing annotation burden, users must still manually provide labels, and systems remain limited to predefined activities.
For instance, some systems temporarily deploy cameras to capture ground truth labels for training models that operate on privacy-preserving sensors alone, but these typically struggle with complex or unknown activities~\cite{2020mmaction2,patidar_vax_2023}.
Other approaches explore interactive learning strategies that gradually refine activity models based on user feedback~\cite{karpekov_discover_2025}.
These often require significant user involvement or struggle to capture activity diversity.
Unlike prior approaches, \capname uses privacy-preserving sensors to identify key moments for targeted video analysis rather than requiring continuous video recording or extensive user annotation. By leveraging VLMs' scene understanding capabilities during these brief moments, we bootstrap location awareness and rich activity recognition without predefined labels. This enables autonomous activity discovery while minimizing privacy concerns and user burden.

\section{HARDWARE DESIGN}
\label{sec:hardware-design}

\begin{figure}[t]                   
    \centering
    \subfigure[Movement and position information using Doppler and Lidar sensors over time.]{%
        \includegraphics[width=\textwidth]{figures-final/OrganicHAR_Figure_1-doppler-lidar.pdf}}%
    
    \vspace{0.5cm}
    
    \subfigure[Thermal data (10x8) and Depth Array information at a single timestamp (snapshot).]{%
        \includegraphics[width=0.75\textwidth]{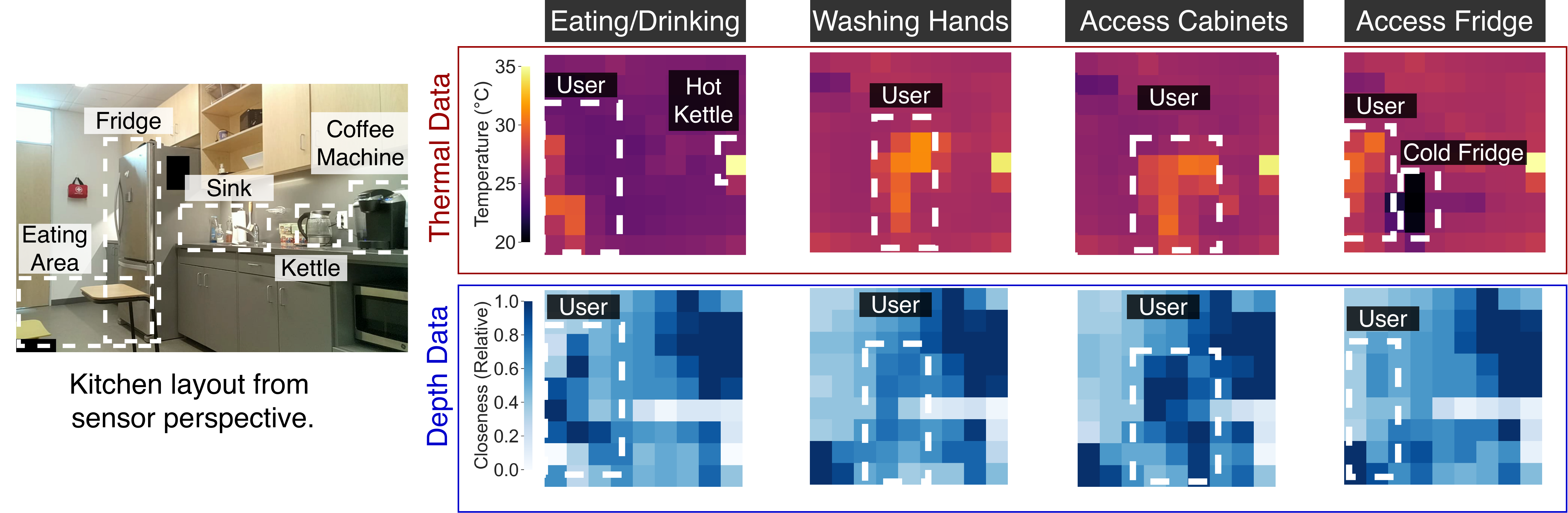}}%
    \hfill
    \subfigure[Hardware setup used for data collection.]{%
        \includegraphics[width=0.23\textwidth]{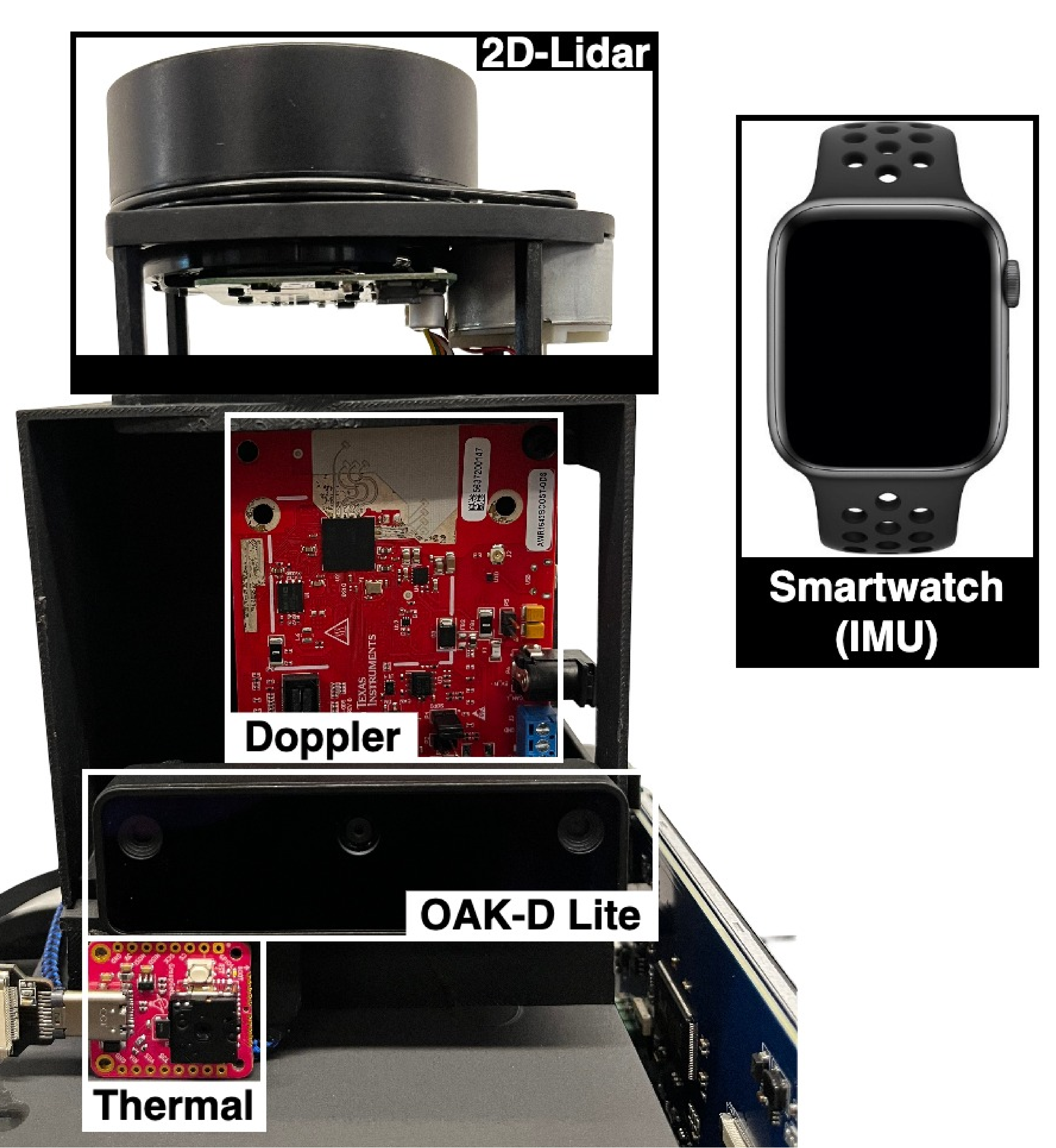}}%
    
    \caption{Visualizing information across various activities from various privacy-preserving sensors inspired by our prior work on VAX \cite{patidar_vax_2023}. (a) shows the movement range and radial velocity of moving objects using a Doppler sensor and object distance and angle using a Lidar sensor in the sensor plane across a sequence of activities happening in the living room. (b) shows snapshots of the Thermal and Depth Array at a given timestamp for various kitchen activities. For depth visualization, pixels grow darker when the user is closer to the sensor. (c) shows our hardware rig used to collect sensor data.}
    \Description[Multi-sensor privacy-preserving data visualization and hardware setup]{Three-part figure showing privacy-preserving sensor data and hardware. Part (a) displays two time-series panels spanning 50 seconds: top panel shows Doppler sensor movement data as colorful heatmap (blue, purple, yellow, green pixels) with movement range from 3.7 to negative 0.9 meters and radial velocity from 2 to 0 meters per second, annotated with activities including walking away (large variation, positive velocity), jumping jacks (small movements at fixed distance), moving to other side (angle change), forward lunges (small periodic radial movement), vacuuming (large periodic movements with high position-movement correlation), and walking towards sensor (negative velocity); bottom panel shows Lidar position data as line graph displaying object distance (3.6 to 0.0 meters) and angle (90 to negative 90 degrees) with annotations for consistent position and angle changes. Part (b) shows kitchen layout photo on left with labeled zones (Fridge, Coffee Machine, Sink, Kettle, Eating Area) and two rows of sensor data on right: top row displays thermal arrays (10x8, temperature range 20-35°C) showing user heat signatures and environmental objects (hot kettle, cold fridge) in purple and orange across four activities (Eating/Drinking, Washing Hands, Access Cabinets, Access Fridge); bottom row shows depth arrays with closeness scale 0.0 to 1.0, displaying user silhouettes as white dashed outlines against blue gradient backgrounds where darker blue indicates closer proximity. Part (c) photographs stacked hardware rig on black platform with labeled components from top to bottom: cylindrical 2D-Lidar with exposed rotating mechanism, red Doppler radar circuit board, black OAK-D Lite stereo camera with three apertures, pink thermal sensor board at base, and separate Apple Smartwatch with black sport band labeled as IMU sensor.}
    \label{fig:privacy-sensor-examples}
\end{figure}

Our system implements a modular sensing infrastructure that can be configured with different combinations of sensors based on application requirements and user privacy preferences. We categorize our sensors into three tiers: (1) a basic ambient sensing setup (\textit{Ambient (Basic) Only}) ambient sensors including position sensing with 2D lidar, movement sensing with a doppler radar, and infrared sensing using low-resolution (10x10) thermal arrays; (2) an intermediate sensing setup (\textit{Ambient (Basic) + Wearable (IMU)}) combining ambient sensors with wearable motion (IMU) data from smartwatches; and (3) an advanced setup (\textit{Ambient (Advanced) + Wearable (IMU)}) that includes on-device human pose and depth estimation.
Our current prototype integrates these sensing modalities (See Figure~\ref{fig:privacy-sensor-examples}c) through a small form factor PC (Intel NUC, 8-core, 16GB RAM) that handles all processing locally. For the initial training phase only, we use an iPhone for video capture to enable VLM-based scene understanding.

\textbf{Basic Ambient Sensing:} Our basic configuration focuses on non-optical sensors that capture coarse movement and presence information. This includes: 
(i) \textit{FMCW Doppler Radar} using a 77GHz mmWave radar (AWR1642BOOST-ODS) operating at 5Hz that captures movement through RF reflection, providing velocity and range information for detecting significant movements (Figure \ref{fig:privacy-sensor-examples}a), 
(ii) \textit{2D Lidar} that measures distance using Time of Flight (ToF) or laser beam parallax. We utilize a Slamtec RPLIDAR A1M8, which delivers 360-degree horizontal measurements at 6-8Hz with 1-degree angular resolution. This data identifies spatial occupancy patterns and environmental changes (Figure \ref{fig:privacy-sensor-examples}a),
and (iii) \textit{Low-Resolution Thermal Camera} that provides a 10x10 pixel thermal map at 8Hz. While the original sensor (FLIR Lepton 3.5) offers higher resolution, we reduce it to preserve privacy \cite{patidar_vax_2023} while still capturing thermal signatures associated with different activities, such as using appliances or engaging in tasks that have thermal signatures (\eg heating kettle, accessing a fridge; see Figure \ref{fig:privacy-sensor-examples}b).

\textbf{Wearable Sensing:} We also support configurations where a user-worn wearable device is utilized, enabling the capture of personalized motion data using \textit{Smartwatch IMU}, and a custom iOS app to capture detailed motion data from Apple Watch. The IMU provides acceleration, gyroscope, and magnetometer data at 50Hz, enabling recognition of hand movements and gestures that complement the ambient sensing data. This sensor captures personal movement data and does not record any environmental information.

\textbf{Advanced Ambient Sensing:} Our advanced configuration adds on-device processing capabilities through the OAK-D Lite platform \cite{Sharma_2022_OAK-D}. It includes: 
(i) \textit{On-Device Pose Sensing} that outputs normalized pose coordinates (640x480 frame) from device, and
(ii) \textit{Low-Resolution Depth Maps} using stereo vision capabilities on OAK-D Lite (Figure \ref{fig:privacy-sensor-examples}b). This low-resolution depth information provides valuable context about spatial relationships while making it extremely hard to reconstruct any part of the original images \cite{jalal_depth_2017}.

Each sensing tier presents distinct privacy-capability tradeoffs. The basic ambient configuration maximizes privacy as its sensors cannot capture personally identifiable information, requiring no user interaction after installation. The wearable configuration enables personalized motion tracking but requires the device to be worn consistently. The advanced configuration provides detailed user information through processed pose and depth information, though optical sensors may raise privacy concerns despite no raw image output, making it better suited for common areas. Notably, future hardware-limited sensors that constrain functionality at the hardware level~\cite{pete_2022arxiv_mlsensors,mollyn_imuposer_2023,devrio_smartposer_2023} could potentially address these concerns.
As we demonstrate later, OrganicHAR's multiple configuration options enable deployments that balance privacy preferences with recognition capabilities based on individual needs and spatial contexts.

\section{SYSTEM DESIGN}
\label{sec:systemdesign}

\begin{figure}
    \centering
    \includegraphics[width=0.95\linewidth]{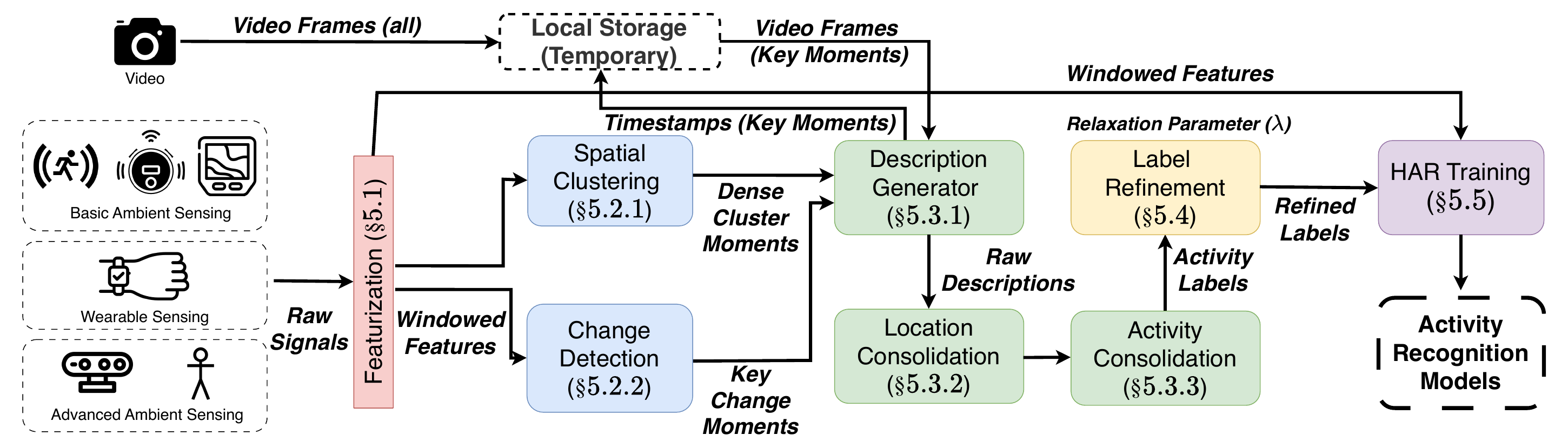}
    \caption{Overall architecture of \name. Raw sensor signals from different hardware configurations (\S \ref{sec:hardware-design}) are featurized for each sensor separately (\S \ref{sec:featurization}), and used to identify \textit{key moments} (\S \ref{sec:key-moments-generation}) for targeted video processing using VLMs. The raw descriptions are then converted to discrete activity labels using LLMs (\S \ref{sec:label-discovery}), and further refined using a relaxation parameter $\lambda$ (\S \ref{sec:semantic-granularity-control}). The final activity labels are used to train activity recognition models (\S \ref{sec:training-x}) for deployment.}
    \Description[System architecture flowchart showing sensor-to-model pipeline]{Pipeline flowchart with five main stages from left to right. Stage 1: Three sensor input tiers (Basic Ambient Sensing with radar/lidar/thermal icons, Wearable Sensing with smartwatch icon, Advanced Ambient Sensing with camera and pose icons) plus video camera marked "key moments only". Stage 2: Pink Featurization box processes raw signals into windowed features. Stage 3: Two parallel blue boxes for key moment identification - Spatial Clustering produces dense cluster moments, Change Detection produces key change moments. Stage 4: Two parallel processing streams - top green stream shows Description Generator converting moments to raw descriptions, then Label Refinement with relaxation parameter lambda producing activity labels; bottom green stream shows Working Zone Consolidation and Zone-Specific Actions. Stage 5: Purple HAR Training box receives refined labels and produces Activity Recognition Models for deployment. Arrows show data flow between components with labels indicating intermediate outputs.}
    \label{fig:system-overview}
\end{figure}

Figure~\ref{fig:system-overview} presents the overall architecture of our \name framework, highlighting its major components. \name operates through a sensor-first approach where privacy-preserving sensors drive activity discovery. Raw sensor signals from different hardware configurations (\S \ref{sec:hardware-design}) are first featurized for each modality separately (\S \ref{sec:featurization}). These features are then used to identify \textit{key moments} (\S \ref{sec:key-moments-generation}) that warrant targeted video processing using VLMs. The resulting raw descriptions from VLMs are converted to discrete activity labels using LLM-based clustering (\S \ref{sec:label-discovery}) and further refined using a relaxation parameter $\lambda$ (\S \ref{sec:semantic-granularity-control}) to control semantic granularity. Finally, these discovered activity labels are used to train activity recognition models (\S \ref{sec:training-x}) that operate solely on privacy-preserving sensors during deployment, eliminating the need for continuous video monitoring.

\subsection{Featurization}
\label{sec:featurization}

Each sensor modality produces unique signal patterns with inherent capabilities and limitations, requiring tailored featurization methods to maximize their strengths. 
\name employs a sliding window approach with fixed-length 5-second segments and 0.5-second stride length for both training and inference. For feature extraction at timestamp $t_s$, we compute features using sensor data from interval $[t_s-5, t_s]$ seconds. This window size is consistent with recent sensor fusion approaches \cite{tarekegn_enhancing_2023, banos_window_2014} to capture sufficient temporal context while maintaining computational efficiency. During final activity recognition, we aggregate consecutive predictions from overlapping windows to determine activity boundaries and durations, accommodating both brief interactions and extended tasks.
This multimodal approach achieves complementary fusion, which enables our system to identify scene-relevant moments and activity patterns across different environments and sensor configurations more robustly than any single modality. We provide sensor-specific featurization details in Appendix \ref{sec:appendix-featurization}.

\subsection{Key Moments Identification}
\label{sec:key-moments-generation}

Once we have featurized data, we start by identifying \textit{key moments} for each modality. This critical module determines which segments of sensor data warrant further analysis by the VLM. The module ingests featurized sensor streams and processes them through two complementary approaches: (1) a spatial clustering module that identifies recurring patterns representing consistent activities in the feature space, and (2) a temporal change detection module that identifies segments with notable signal shifts indicating activity transitions.

\subsubsection{Spatial Clustering}
\label{sec:spatial-clustering}
We employ clustering in the feature space of each sensor modality to identify recurring patterns representing distinct activity signatures. While we refer to this as ``spatial clustering,'' it operates not in the physical space but in the high-dimensional feature space specific to each sensor, where sensor signals from similar activities form natural clusters regardless of their physical location.
We leverage HDBSCAN (Hierarchical Density-Based Spatial Clustering of Applications with Noise)~\cite{mcinnes_hdbscan_2017}, selected for its ability to discover clusters of varying densities without requiring a predefined cluster count, crucial for unsupervised activity discovery in unfamiliar environments. After applying standard preprocessing (robust scaling and dimensionality reduction using principal component analysis), we optimize the clustering process through careful hyperparameter tuning.
The key hyperparameters governing our feature space clustering include:

\begin{itemize}
    \item $\text{min\_cluster\_size} \in \{m_1, m_2, \ldots, m_a\}$: Determines the minimum points required to form a cluster, directly influencing the granularity of detected activity patterns.
    \item $\text{min\_samples} \in \{s_1, s_2, \ldots, s_b\}$: Controls clustering conservativeness, with higher values producing more stringent cluster formation.
    \item $\text{n\_components} \in \{c_1, c_2, \ldots, c_c\}$: Defines the dimensionality of the feature space after reduction.
\end{itemize}

Central to our approach is a carefully designed scoring function that evaluates clustering quality across all sensor modalities for a given combination of hyperparameters:

\begin{equation}
\label{eq:clustering_score}
\text{score} = w_1 \cdot S_{count} + w_2 \cdot S_{noise} + w_3 \cdot S_{prob}
\end{equation}

This function balances three essential aspects of effective activity clustering for our use case:
\begin{itemize}
    \item $S_{count}$: Evaluates how well the number of discovered clusters aligns with activity diversity (between $C_{min}$ and $C_{max}$ clusters), penalizing overly granular and overly coarse solutions.
    \item $S_{noise}$: Assesses the proportion of data points assigned to meaningful clusters rather than noise, preferring solutions with noise ratios below $N_{max}$. It separates random sensor activations from meaningful patterns.
    \item $S_{prob}$: Captures the average cluster membership probability, reflecting the confidence in cluster assignments and favoring well-defined, distinct activity signatures.
\end{itemize}

This scoring framework consistently evaluates cluster quality across diverse sensing modalities while accommodating their varying signal characteristics. The weighted combination balances cluster count appropriateness ($w_1$) with noise handling and assignment confidence ($w_2$ and $w_3$). From each cluster, we select representative windows with high membership confidence as key moments for VLM analysis, significantly reducing the required video processing requests while capturing essential activity patterns. Our approach excels at activity discovery by identifying natural signal groupings that correspond to distinct behaviors — cooking activities might cluster in Doppler feature space despite different kitchen locations, while refrigerator interactions might form clusters in thermal feature space despite varying user movements. Through empirical evaluation across sensors, we selected hyperparameter ranges that provide optimal clustering: min\_cluster\_size $\in \{3-8\}$, min\_samples $\in \{2-5\}$, and n\_components $\in \{8-80\}$, with modality-specific adjustments (\eg thermal sensing uses cluster\_selection\_epsilon $\in \{0.01-0.03\}$ for tighter clusters). The scoring weights ($w_1 = 0.3-0.4, w_2 = 0.15-0.3, w_3 = 0.1-0.3$) and target parameters (desired clusters between 20-40, maximum noise ratio of 0.8) are consistent across modalities.

\subsubsection{Change Detection}
\label{sec:change-detection}

We also leverage the temporal changes in signals of each modality, expecting that these moments can capture useful actions.
We used an algorithm based on multimodal anomaly detection, specifically the one proposed by Yamanishi~\etal~\cite{DBLP:journals/datamine/YamanishiTWM04}.
This algorithm is online, unsupervised outlier detection using the Gaussian Mixture Model (GMM), which can capture changes in time-series signals without prior knowledge.
Whenever a new data sample is received, it computes an anomaly score based on its likelihood under the current GMM and simultaneously updates the model parameters.
Specifically, let $\mathbf{y}_t \in \mathbb{R}^M$ denote the input sample at time $t$, and let $\pi_{i,t}$, $\mu_{i,t}$, and $\Sigma_{i,t}$ represent the weight, mean vector, and covariance matrix of the $i$th component (for $i=1,\ldots,K$) of the GMM at time $t$. The anomaly score for $\mathbf{y}_t$ is then defined as
\begin{equation}
	\label{eq:anomaly_score}
	s_t = - \ln \left( \sum_{i=1}^{K} \pi_{i,t-1} \, \mathcal{N} \left( \mathbf{y}_t \mid \mu_{i,t-1}, \Sigma_{i,t-1} \right) \right).
\end{equation}

Following this, the parameters of the GMM are updated according to
\begin{align}
	\label{eq:gmm_update}
	\lambda_{i,t} &= \frac{\pi_{i,t-1} \, \mathcal{N} \left( \mathbf{y}_t \mid \mu_{i,t-1}, \Sigma_{i,t-1} \right)}{\sum_{j=1}^{K} \pi_{j,t-1} \, \mathcal{N} \left( \mathbf{y}_t \mid \mu_{j,t-1}, \Sigma_{j,t-1} \right)}, \nonumber \\
	\pi_{i,t} &= (1 - \alpha) \, \pi_{i,t-1} + \alpha \, \lambda_{i,t}, \nonumber \\
	\tilde{\mu}_{i,t} &= (1 - \alpha) \, \tilde{\mu}_{i,t-1} + \alpha \, \lambda_{i,t} \, \mathbf{y}_t, \\
	\mu_{i,t} &= \frac{\tilde{\mu}_{i,t}}{\pi_{i,t}}, \nonumber \\
	\tilde{\Sigma}_{i,t} &= (1 - \alpha) \, \tilde{\Sigma}_{i,t-1} + \alpha \, \lambda_{i,t} \, \mathbf{y}_t {\mathbf{y}_t}^T, \nonumber \\
	\Sigma_{i,t} &= \frac{\tilde{\Sigma}_{i,t}}{\pi_{i,t}} - \mu_{i,t} \mu_{i,t}^T, \nonumber
\end{align}
where $\alpha$ is a forgetting factor that controls the influence of past observations.
After computing the anomaly score for each \( t \) using \eqnref{eq:anomaly_score}, we selected the top \( N \) points as detected moments.
The parameters \( K \), \( M \) and \( \alpha \) are hyperparameters of this component.
We empirically set \( K=2 \), \( M=10 \), and \( \alpha=0.1 \), which works well on average across all modalities.

\subsection{Label Discovery}
\label{sec:label-discovery}

After identifying \textit{key moments} through spatial clustering and temporal change detection, our framework translates these sensor-identified moments into meaningful activity labels using a multi-stage approach described below:

\subsubsection{Semantic Description Generation}

For each \textit{key moment}, we prompt a VLM to generate semantic descriptions from video frames—the only stage where video data is processed, occurring exclusively during initial training. We iteratively tested multiple configurations to determine optimal video input parameters. Higher frame rates and resolutions frequently led to hallucinations and inconsistent descriptions. \modified{Our empirical testing revealed that low-resolution frames (640×480) sampled at 1 FPS for 5-second clips provided the ideal balance between detail and consistency for current (as of July 2025) VLM capabilities, though this balance may shift as vision-language models improve.}

A critical challenge in prompt design was achieving dual objectives: controlling VLM speculation while balancing specificity and consistency. VLMs tend to infer intentions beyond what is directly observable and struggle with appropriate granularity—either overgeneralizing activities (\eg labeling everything as ``kitchen activity'') or focusing on irrelevant details. We addressed these tensions by explicitly instructing the model to \textit{``focus only on what you can clearly see''} while providing structured categories guiding appropriate detail levels. We directed the VLM to analyze four aspects: (1) \textit{actions}, including movements between locations; (2) \textit{objects} the person interacts with; (3) \textit{location} where activity occurs; and (4) \textit{activity structure} (initial conditions, main actions, results). For each category, we provided examples at appropriate granularity levels to establish consistent reference points. By requiring JSON-structured output with self-assessed confidence scores, we enabled systematic filtering of observations ($\theta_{conf}=0.8$), rejecting both speculative and overly generic descriptions. While accurate, the resulting descriptions exhibited inconsistent terminology and granularity across similar activities—addressed in our label consolidation process. We provide our complete prompt in Appendix \ref{sec:appendix-prompts}.

\subsubsection{Location Consolidation}

We implemented a two-step label consolidation process: location consolidation followed by activity description consolidation.
The location-based approach works well with privacy-preserving ambient sensors, which often provide clean separation in signals for spatial differentiation.
Additionally, the VLM outputs were more consistent for location references than for action descriptions, providing a stable foundation for generating location-specific activity labels. 
We design an LLM prompt to consolidate various location references from VLM outputs (\eg ``at sink,'' ``by sink,'' ``near sink basin'') into consistent \textit{functional zones} like ``sink area,'' ``counter area,'' and ``coffee machine area.'' The prompt instructs the LLM to group locations based on supported activities, merge functionally similar spaces, and maintain separation between distinct activity zones. This spatial organization helps disambiguate semantically similar actions through their context (\eg distinguishing ``washing hands'' from ``washing dishes'' based on precise sink location) and creates a foundation for more accurate activity recognition that leverages the spatial detection capabilities of our sensors.

\subsubsection{Activity Consolidation}

With established functional zones, we transform unstructured action descriptions from the VLM into discrete activity labels.  For each functional zone, we use an LLM prompt to create high-level clusters based on action descriptions from this zone. The prompt differentiates between actions (physical movements like ``pouring'' or ``washing'') and purpose/context (the goal or situation, such as ``breakfast preparation'' or ``dishwashing''). This distinction helps maintain important separations -- for example, keeping ``pouring cereal'' and ``pouring coffee'' as different activities despite sharing the same physical action. In the sink area, ``washing dish with sponge'' and ``scrubbing plate with sponge'' merge into ``washing dishes with sponge,'' while remaining distinct from ``washing hands with soap.''

We then use LLM-assisted matching to assign each activity description to existing clusters based on interaction patterns. For each description, we extract action and object information, then use an LLM to calculate weighted similarity scores based on action alignment ($w_a=60$), object consistency ($w_o=25$), and location match ($w_l=15$). These empirically determined weights reflect the relative importance of each factor in determining activity similarity. The mechanism effectively consolidates variations like ``filling cup using coffee machine'' and ``using coffee machine with mug'' into ``preparing coffee drink'' while distinguishing them from functionally different activities occurring at the same location.

\subsection{Label Refinement and Semantic Granularity Control}
\label{sec:semantic-granularity-control}

While our two-step consolidation produces well-defined activity labels, we also need to provide users control over label merging at different granularities. From a human perspective, certain distinctions are more meaningful than others—``rinsing dishes'' and ``washing dishes with sponge'' should merge before ``washing hands with soap'' and ``washing dishes'' despite textual similarity. We analyze semantic relationships across six dimensions. We use an LLM to expand each activity label along: (1) action type (\eg ``cleaning,'' ``preparing''), (2) objects involved (\eg ``dishes,'' ``sponge,'' ``water''), (3) sub-location (\eg ``sink basin,'' ``counter edge''), (4) purpose/goal (\eg ``remove food residue,'' ``prepare beverage''), (5) access patterns (\eg ``reaching for scrubber,'' ``turning faucet''), and (6) related activities (\eg ``rinsing dishes'' relates to ``drying dishes''). This multi-dimensional characterization provides rich semantic representation.
Next, we compute a pairwise similarity matrix $S$ where the score $S_{ij}$ between activity labels $a_i$ and $a_j$ is calculated by combining weighted cosine similarities across these dimensions:
\begin{equation}
\begin{split}
S_{ij} = w_{\text{action}} \cdot \text{sim}_{\text{action}}(a_i, a_j) + 
w_{\text{object}} \cdot \text{sim}_{\text{object}}(a_i, a_j) + 
w_{\text{location}} \cdot \text{sim}_{\text{location}}(a_i, a_j) + \\
w_{\text{purpose}} \cdot \text{sim}_{\text{purpose}}(a_i, a_j) + 
w_{\text{access}} \cdot \text{sim}_{\text{access}}(a_i, a_j) + 
w_{\text{relation}} \cdot \text{sim}_{\text{relation}}(a_i, a_j)
\end{split}
\end{equation}

For each dimension, we generate textual embeddings using an embedding model and compute cosine similarities between them. The weights reflect each dimension's contribution to human-relevant distinction: action type ($w_{\text{action}}=0.20$), object involvement ($w_{\text{object}}=0.25$), sub-location ($w_{\text{location}}=0.15$), purpose/goal ($w_{\text{purpose}}=0.15$), interaction patterns ($w_{\text{access}}=0.15$), and explicit relationships ($w_{\text{relation}}=0.10$).
Finally, we apply hierarchical clustering with a relaxation parameter $\lambda$ to group activity labels into clusters $C$ such that $\forall a_i, a_j \in c_k, S_{ij} \geq 1 - \lambda$. The values of $\lambda$ are typically in the range from 0 (one cluster per activity) to 1 (merge all activities in a single cluster).
This gives users control over recognition granularity through a single parameter—smaller $\lambda$ values preserve fine distinctions between activity labels, while larger values merge semantically related activities.

\subsection{Training Privacy-preserving Human Activity Recognition Models}
\label{sec:training-x}

The final phase trains HAR models on privacy-preserving sensors without video after initial training. For each sensor modality, we evaluate multiple classifiers robust to imbalanced datasets, implementing leave-one-session-out cross-validation to assess generalizability. This addresses real-world sensor challenges including missing values and class imbalance across different environments.
After training individual classifiers, we employ grid search to identify suitable sensor-classifier combinations for each functional zone. We evaluate two ensemble methods: soft voting (combining probability distributions from multiple classifiers) and hard voting (weighted voting using each classifier's highest-confidence prediction)~\cite{awan-ur-rahman_understanding_2023}. Our framework computes balanced accuracy (macro-recall) for each configuration to determine effective deployment combinations.
Ultimately, after training, \name implements a hierarchical approach towards inferring activities: first identifying the user's functional zone, then applying appropriate zone-specific activity recognition models using the selected sensor ensemble. This context-aware structure recognizes that activities present different signatures in different zones while managing computational resources. By activating only necessary sensors and models, the system balances privacy preservation with recognition accuracy while operating on local sensors after training.

\subsection{Detailed Implementation}
\label{sec:endtoend-system}

\name is implemented in Python with approximately 8,000 lines of code and consists of three main components: (i) key moment identification from privacy-preserving sensors, (ii) semantic label discovery using VLMs, and (iii) training privacy-preserving sensor models with discovered activity labels. For sensor data processing, we utilize NumPy \cite{harris2020array} for array manipulation, Pandas \cite{reback2020pandas} for time-series analysis, and OpenCV \cite{opencv_library} for video processing. The key moment identification implements Gaussian Mixture Models from scikit-learn \cite{scikit-learn} for temporal change detection and HDBSCAN \cite{mcinnes_hdbscan_2017} for spatial clustering. During the training phase only, we leverage OpenAI's API \cite{OpenAI_API} with the \textit{GPT-4o} model to generate semantic descriptions for the key moments, which are then organized into hierarchical activity clusters using SciPy's \cite{2020SciPy-NMeth} squareform and linkage functions.
For model training, we employ multiple classifier implementations from scikit-learn and imbalanced-learn \cite{imbalanced-learn}, including RUSBoost~\cite{seiffert_rusboost_2010}, Balanced Random Forest~\cite{yagci_balanced_2016}, EasyEnsemble~\cite{liu_easyensemble_2009}, KNN~\cite{cannings2020classification}, and SVM~\cite{cannings2020classification}.
We open-sourced the \name implementation \cite{organichar-github}, which can be used with different combinations of privacy-preserving sensors with minimal implementation changes.

\section{DATA COLLECTION}
\label{sec:datacollection}

\begin{figure}[t]
\begin{center}
 \subfigure[Kitchen 1]{
    \centering
    \includegraphics[width=0.323\textwidth]{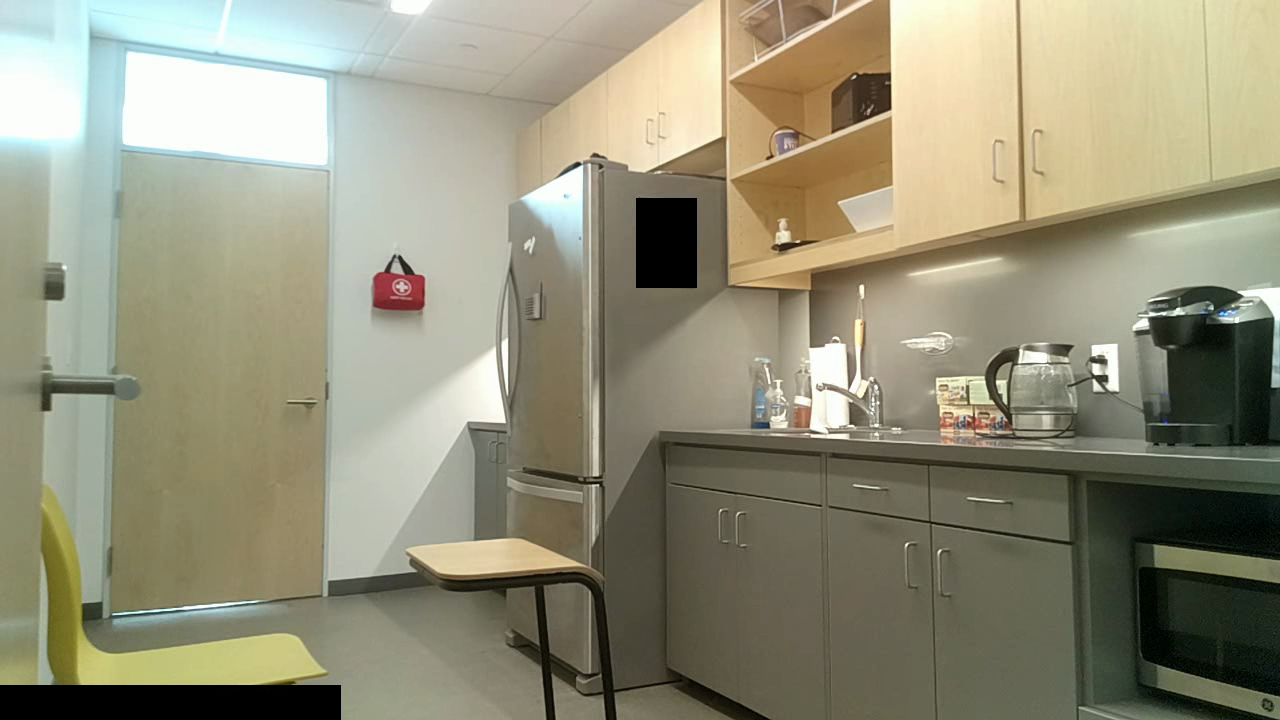}}
 \subfigure[Kitchen 2]{
     \centering
     \includegraphics[width=0.323\textwidth]{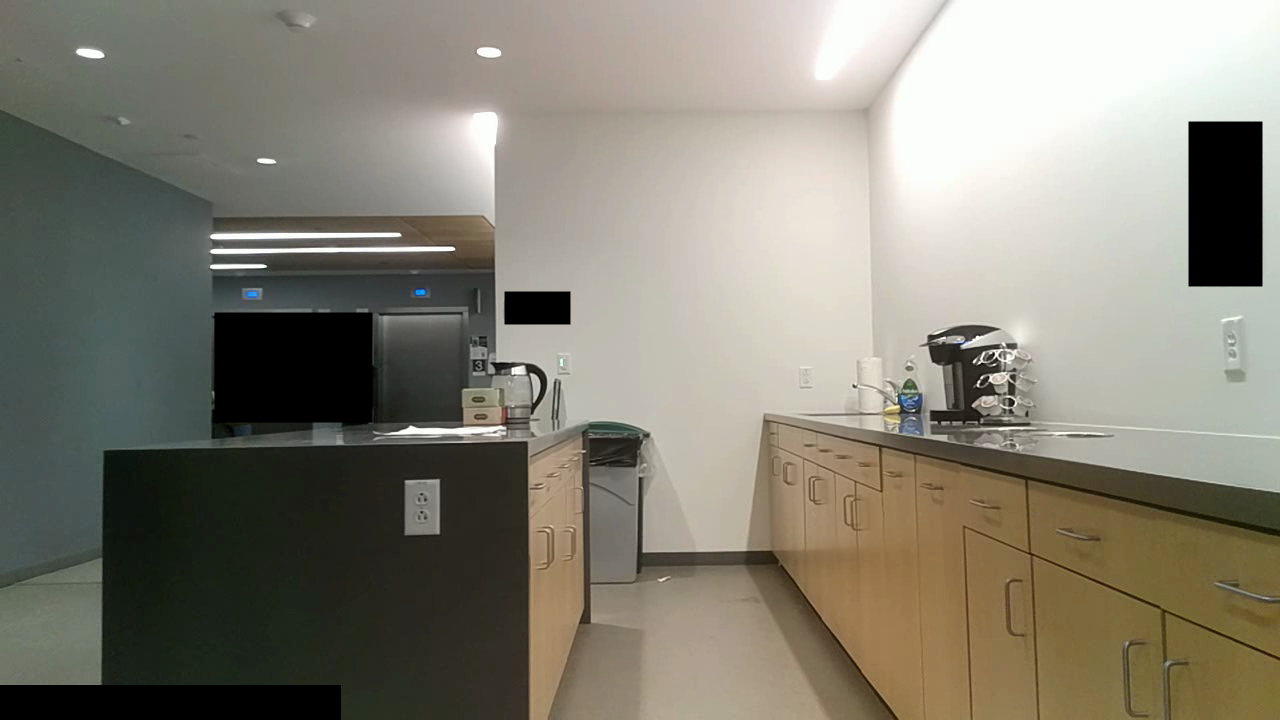}}
 \subfigure[Kitchen 3]{
     \centering
 \includegraphics[width=0.323\textwidth]{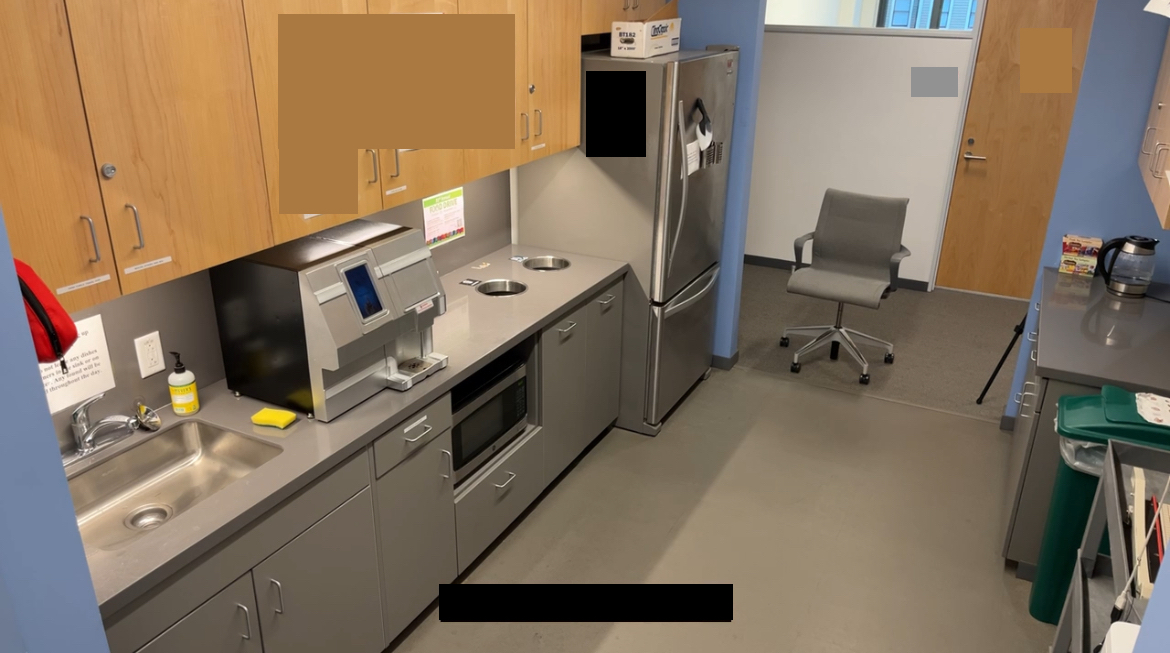}}
\end{center}
\caption{Kitchen environments used in our study: (left) Kitchen 1 with compact galley layout, (middle) Kitchen 2 with island counter, and (right) Kitchen 3 with integrated appliances.}
\Description[Three office kitchenette environments with different layouts]{Three photographs of office kitchen spaces used for data collection. Kitchen 1 shows compact galley-style layout with light wood cabinets, gray countertops along left wall featuring sink with soap dispenser, coffee machine, and stainless steel refrigerator, with gray office chair and eating area in background. Kitchen 2 displays refrigerator on left side, long continuous gray counter with sink in center area, coffee maker and pod organizer on right end, yellow chair in foreground, and light wood upper cabinets. Kitchen 3 features white island counter in foreground with two gray stools, and back wall with continuous gray countertop including sink and coffee maker, light wood lower cabinets, and integrated appliances configuration.}
\label{fig:camera-positioning}
\end{figure}

We collected data from 12 participants in three distinct kitchen environments. Participants performed four breakfast preparation tasks: \textit{preparing tea}, \textit{making coffee from a coffee machine}, \textit{preparing cereal}, and \textit{making sandwiches} with various spreads. Critically, participants received no explicit instructions on how to execute these tasks, and researchers did not intervene or provide prompts during the activities. Each participant approached the tasks in their own way, using natural movements and sequences that reflected genuine everyday behavior rather than scripted demonstrations. The data collection for each user lasted approximately 1.5 hours, with participants performing each task 2-3 times with natural variations. To ensure clear segmentation while maintaining ecological validity, participants washed their hands before each task and used a hand clap to mark segment boundaries.

We collected data across three kitchenettes (Figure \ref{fig:camera-positioning}): \textbf{Kitchen 1 (P1-P4)} featured a compact galley-style layout with a single countertop, refrigerator at one end, sink in the middle, and coffee machine at the opposite end; \textbf{Kitchen 2 (P5-P8)} had an open layout with an island counter separate from the main preparation area, featuring a sink and coffee machine; and \textbf{Kitchen 3 (P9-P12)} was a medium-sized kitchenette with microwave, coffee machine, and sink aligned along a single wall, with a refrigerator perpendicular to the main counter.

Our dataset comprises $\sim$4300 five-second snippets (approximately 6 hours) of breakfast preparation activities across six sensor modalities: doppler radar, 2D lidar, low-resolution thermal array, wearable IMU, on-device pose sensing, and depth processing. Unlike existing HAR datasets featuring scripted activities, our dataset captures natural variations in how people perform everyday tasks. Ground truth annotation was performed by an unbiased observer who labeled each snippet without prior knowledge of system-generated labels. Our dataset is open-sourced \cite{organichar-github}, including all sensor data, VLM annotations, and ground truth labels.

\section{EVALUATION}
\label{sec:evaluation}

Our evaluation of \name focuses on two primary research questions:
\begin{itemize}
    \item \textbf{RQ1:} How effective and efficient is our activity discovery pipeline in identifying meaningful activity labels compared to traditional approaches?
    \item \textbf{RQ2:} How do the HAR models trained on these discovered labels perform across different sensor configurations and semantic granularity settings?
    \item \textbf{RQ3:} How does \name perform in real-world deployment scenarios, and how does its performance evolve over time?
\end{itemize}
For each user and sensor configuration, we evaluated \name across three semantic granularity settings, \ie (\textit{Conservative}: $\lambda=0.4$, \textit{Balanced}: $\lambda=0.3$, and \textit{Relaxed}: $\lambda=0.2$)  to determine how permissively activities are merged based into single activity label based on their semantic similarity, as discussed in \S \ref{sec:semantic-granularity-control}.

\subsection{Performance of Label Discovery Pipeline}
\label{sec:evaluation-discovery}

\begin{figure}[t]
    \centering
    \begin{minipage}[c]{0.31\textwidth}
        \centering
        \includegraphics[width=\textwidth]{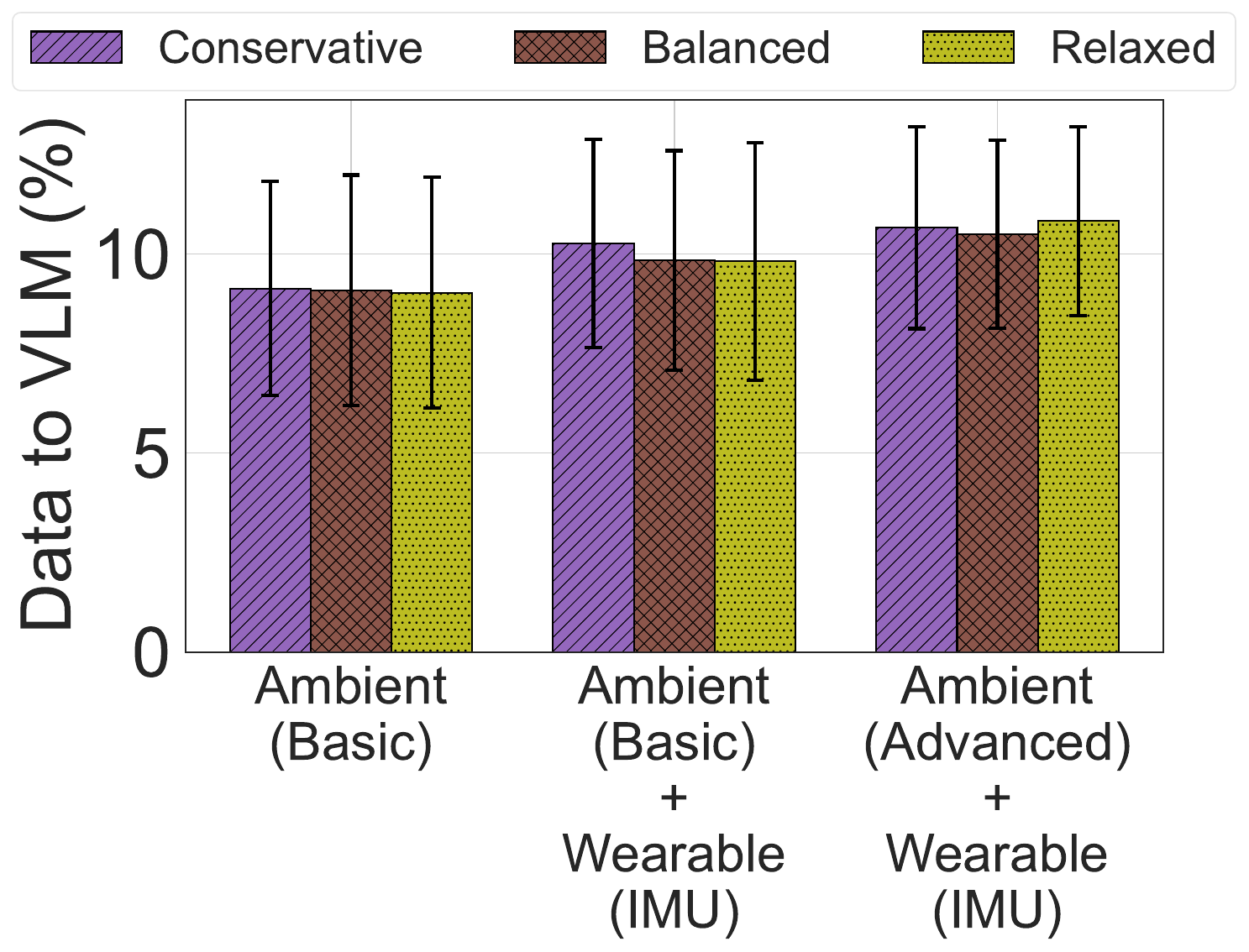}
        \vspace{0.2em}
        \caption{\small Percentage of video data requiring VLM analysis across configurations. Our approach processes only 9-11\% of total video data, demonstrating efficiency compared to continuous monitoring.}
        \Description[Bar chart showing video processing efficiency across sensor configurations]{Bar chart displaying percentage of video data sent to VLM, with y-axis from 0 to 12 percent labeled "Data to VLM (\%)". Three grouped bar clusters on x-axis representing sensor configurations: Ambient (Basic), Ambient (Basic) plus Wearable (IMU), and Ambient (Advanced) plus Wearable (IMU). Each cluster contains three bars with distinct patterns: purple diagonal stripes for Conservative, brown crosshatch for Balanced, and yellow dots for Relaxed granularity settings. All bars show values between 9 and 11 percent with error bars indicating standard deviation. Basic ambient configuration shows approximately 9 percent, intermediate configuration around 9.5 percent, and advanced configuration approximately 10.5 percent, with minimal variation across granularity settings within each configuration.}
        \label{fig:video-efficiency}
    \end{minipage}
    \hfill
    \begin{minipage}[c]{0.65\textwidth}
        \centering
        \renewcommand{\arraystretch}{1.0}
        \setlength{\tabcolsep}{3pt}
        \begin{tabular}{|c|c|c|c|c|}
        \hline
        \multirow{2}{*}{\textbf{Granularity}} & \multirow{2}{*}{\textbf{Metrics}} & \multicolumn{3}{c|}{\textbf{Sensor Config}} \\
        \cline{3-5}
        & & \textbf{\begin{tabular}[c]{@{}c@{}}Ambient\\(Basic)\\Only\end{tabular}} & \textbf{\begin{tabular}[c]{@{}c@{}}Ambient\\(Basic)+\\Wearable\\(IMU)\end{tabular}} & \textbf{\begin{tabular}[c]{@{}c@{}}Ambient\\(Advanced)+\\Wearable\\(IMU)\end{tabular}} \\
        \hline
        \multirow{2}{*}{\textbf{Conservative}} & \textbf{Accuracy} & 90.4\%±8.9\% & 91.1\%±6.7\% & 91.7\%±6.9\% \\
        \cline{2-5}
        & \textbf{F1 Score} & 89.4\%±9.8\% & 90.6\%±6.4\% & 90.3\%±7.4\% \\
        \hline
        \multirow{2}{*}{\textbf{Balanced}} & \textbf{Accuracy} & 89.9\%±6.3\% & 85.5\%±7.1\% & 87.1\%±7.9\% \\
        \cline{2-5}
        & \textbf{F1 Score} & 84.1\%±8.7\% & 78.0\%±9.9\% & 81.1\%±11.8\% \\
        \hline
        \multirow{2}{*}{\textbf{Relaxed}} & \textbf{Accuracy} & 86.7\%±4.9\% & 83.8\%±6.5\% & 85.9\%±6.7\% \\
        \cline{2-5}
        & \textbf{F1 Score} & 73.2\%±10.9\% & 72.4\%±8.7\% & 75.6\%±9.9\% \\
        \hline
        \end{tabular}
        \vspace{1em}
        \captionof{table}{Average accuracy and F1 scores (mean±std) of discovered activity labels compared to ground truth across three semantic granularity settings. Conservative ($\lambda=0.4$) represents coarse-grained activities, Balanced ($\lambda=0.3$) shows medium granularity, and Relaxed ($\lambda=0.2$) captures fine-grained activities. Higher performance is observed with more advanced sensor configurations, particularly for fine-grained recognition.}
        \Description{Average accuracy and F1 scores (mean±std) of discovered activity labels compared to ground truth across three semantic granularity settings. Conservative ($\lambda=0.4$) represents coarse-grained activities, Balanced ($\lambda=0.3$) shows medium granularity, and Relaxed ($\lambda=0.2$) captures fine-grained activities. Higher performance is observed with more advanced sensor configurations, particularly for fine-grained recognition.}
        \label{tab:discovery-accuracy}
    \end{minipage}
\end{figure}

\subsubsection{Efficiency of Key Moment Identification}
\label{sec:evaluation-efficiency}

One of the primary advantages of our sensor-first approach is minimizing the amount of video data that requires processing during the training phase. Figure \ref{fig:video-efficiency} shows the percentage of video data selected for VLM analysis across different sensor configurations and granularity settings. The error bars show variability (standard deviation) across different users.
Our key moment identification approach demonstrates remarkable efficiency, requiring analysis of only approximately 9-11\% of the total collected video data across all configurations. The \textit{Ambient(Basic) Only} configuration required the least video processing (around 9\%), while the advanced configuration with \textit{Ambient(Advanced)+Wearable(IMU)} utilized slightly more (around 10.5\%). This modest increase is expected as richer sensing capabilities can identify more nuanced activity transitions that warrant VLM analysis.

\begin{figure}[t]
    \centering    
    \begin{minipage}[b]{1.0\textwidth}
    \centering
    \includegraphics[width=\textwidth]{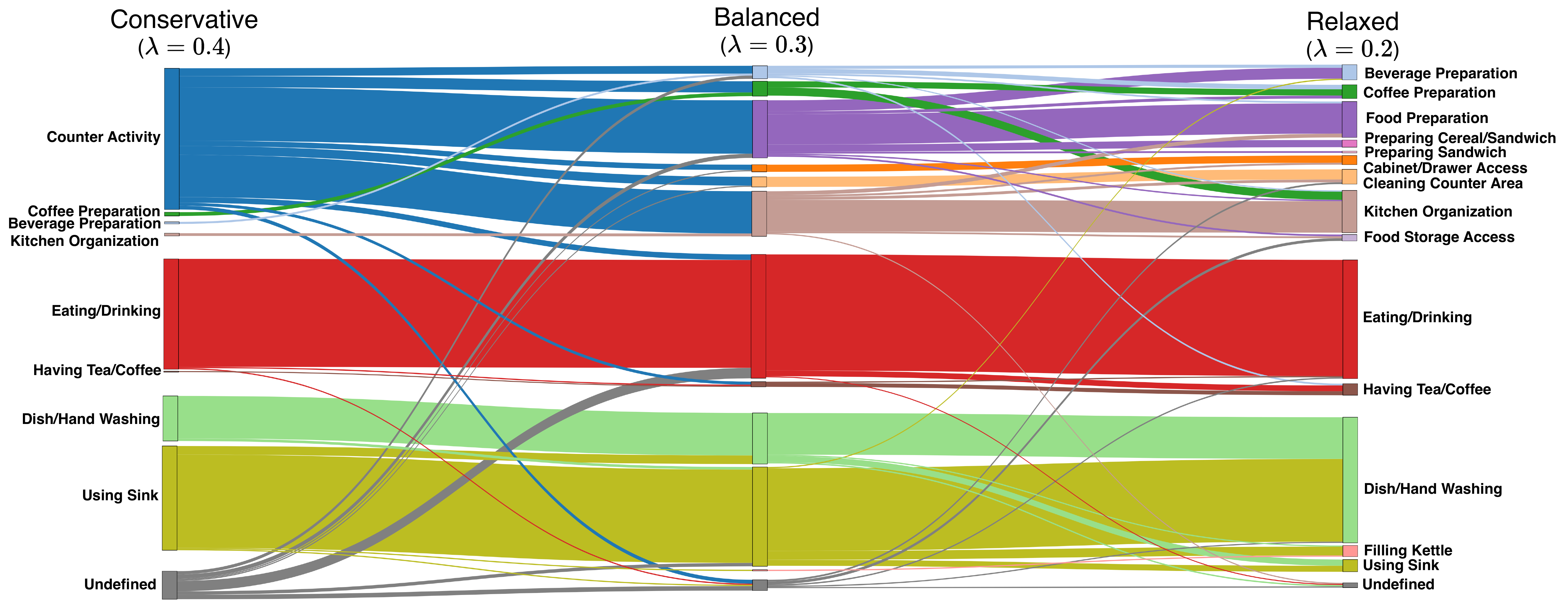}
    \caption{Discovered activity labels across three semantic granularity settings: Conservative ($\lambda=0.4$, left), Balanced ($\lambda=0.3$, middle), and Relaxed ($\lambda=0.2$, right). The flow width represents the proportion of data with each label, showing how broader categories branch into more fine-grained activities as granularity increases. For example, the ``Counter Activity'' broad category (left) differentiates into ``Food Preparation,'' ``Cleaning Counter Area,'' and other activities in the Balanced setting.}
    \Description[Sankey diagram showing hierarchical activity label discovery across three granularity levels]{Sankey flow diagram with three vertical columns representing granularity settings from coarse to fine. Left column (Conservative, lambda equals 0.4) shows six broad categories: Counter Activity (large blue bar at top), Coffee Preparation, Beverage Preparation, Kitchen Organization (thin bars in blue-brown tones), Eating/Drinking (wide red bar), Having Tea/Coffee (thin brown bar), Dish/Hand Washing (green bar), Using Sink (yellow-green bar), and Undefined (gray bar at bottom). Middle column (Balanced, lambda equals 0.3) displays branching flows where Counter Activity splits into multiple streams flowing to Food Preparation (purple), Cleaning Counter Area (orange), Cabinet/Drawer Access (orange), and other categories; Eating/Drinking remains largely intact as wide red flow; sink-related activities show moderate differentiation. Right column (Relaxed, lambda equals 0.2) shows finest granularity with eleven distinct categories: Beverage Preparation, Coffee Preparation, Food Preparation (purple), Preparing Cereal/Sandwich, Preparing Sandwich (orange tones), Cabinet/Drawer Access, Cleaning Counter Area (orange), Kitchen Organization (tan), Food Storage Access (tan), Eating/Drinking (red, still prominent), Having Tea/Coffee (brown), Dish/Hand Washing (green), Filling Kettle (pink), Using Sink (yellow), and Undefined (gray). Flow ribbons connecting columns vary in width proportional to data volume, with colored curves showing how coarse categories progressively differentiate into specific activities from left to right.}
    \label{fig:label-sankey}
    \end{minipage}
\end{figure}

\subsubsection{Discovered Activity with Different Granularity}
\label{sec:evaluation-discovery-granularity}

Figure \ref{fig:label-sankey} illustrates the discovered activity labels across granularity settings, revealing how broader categories in the Conservative setting (left) branch into fine-grained distinctions in the Balanced (middle) and Relaxed (right) settings.
The Conservative setting identified four primary activities: ``Counter Activity,'' ``Using Sink,'' ``Dish/Hand Washing,'' and ``Eating/Drinking,'' capturing kitchen zones and behaviors at a high level. Some users showed more nuanced counter activities, though with low discovery rates. In the Balanced setting, ``Counter Activity'' branches into specific categories like ``Food Preparation,'' ``Cleaning Counter Area,'' and ``Cabinet/Drawer Access,'' discovering activity hierarchies matching the task groupings. The Relaxed setting continues this differentiation, breaking ``Counter Activity'' into fine-grained parts, with ``Food Preparation'' splitting into ``Preparing Cereal/Sandwich'' and other specialized tasks.

We observe several patterns through this granularity analysis.
For instance, specific activities like ``Eating/Drinking'' remain stable across all settings, having distinctive sensor signatures consistently recognized regardless of granularity. ``Using Sink'' appears broadly in Conservative mode but becomes less prominent in Relaxed settings as the system refines it to ``Dish/Hand Washing'' specifically, demonstrating how our approach clarifies ambiguous activities as similarity thresholds change. Flow patterns follow logical parent-child relationships rather than arbitrary recombination—``Cabinet/Drawer Access'' emerges specifically from ``Counter Activity,'' not unrelated categories. Overall, our system discovered 15 distinct activity labels, with users typically having 4-8 relevant activities depending on settings and sensor setup. 
This result suggests that OrganicHAR can adapt to what each sensor configuration reliably detects while maintaining meaningful relationships matching human understanding of kitchen activities, offering clear advantages over predefined activity sets.

\subsubsection{Accuracy of the Discovered Activity}

We evaluated our discovered activity labels against ground truth annotations as shown in Table \ref{tab:discovery-accuracy}. Since discovered labels often use different terminology than human-annotated ground truth, we performed manual mapping to convert annotations into the discovered activity label space for comparison. The Conservative setting (coarse-grained activities) demonstrates high accuracy above 90\% and strong F1 scores across all sensor configurations. Performance moderately declines with increasing granularity, yet remains robust with accuracy above 83\% even in the most challenging scenarios. The \textit{Ambient(Basic) Only} configuration sometimes outperforms \textit{Ambient(Basic)+Wearable(IMU)} in finer-grained settings, suggesting wearable data occasionally introduces variability when distinguishing semantically similar activities. The \textit{Ambient(Advanced)+Wearable(IMU)} configuration delivers the strongest Relaxed setting performance (85.9\% accuracy), highlighting depth and pose information's value for fine-grained recognition. The more pronounced F1 score decline (72-75\%) in the Relaxed setting reflects inherent challenges in precisely identifying fine-grained activities with VLMs processing low-frame-rate data (1 FPS), camera angle limitations, and partial occlusions. Given these results, we recommend the Balanced setting ($\lambda=0.3$) as the optimal compromise for most deployments, offering meaningful activity differentiation while maintaining strong accuracy (85-89\%) and F1 scores (78-84\%). The high average performance demonstrates that \name can automatically identify meaningful activity labels aligning well with human observations.

Moreover, to validate robustness across VLM architectures, we tested three state-of-the-art models (GPT-4.1, Gemini 2.5 Flash, and Claude Sonnet 4). Our 1 FPS, 640×480 configuration remains optimal across all VLMs, though detection rates vary significantly (50.2-92\%). Consistent fine-grained recognition limitations across all models (F1 scores of 35.6-55.4\%) suggest these challenges reflect current VLM capabilities generally. Detailed analysis is presented in Appendix \ref{sec:vlm-analysis}.

\subsection{Performance of Human Activity Recognition Models}
\label{sec:evaluation-recognition}

In this section, we examine how HAR models trained on our discovered labels perform across different sensor configurations.
Regarding this, the label discovery evaluation in the above revealed low F1-scores (72-75\%) in the Relaxed setting, indicating imperfect alignment between some discovered fine-grained labels and ground truth. This misalignment would affect HAR models trained on these labels, where the system attempts to distinguish between highly similar activities. This informs us that the Conservative and Balanced settings ($\lambda=0.4$ and $\lambda=0.3$) represent more realistic deployment scenarios for \capname with current sensing capabilities.
In other words, the Relaxed setting results ($\lambda=0.2$) should be interpreted as exploratory findings that illustrate current technical boundaries rather than definitive performance benchmarks.

\begin{table}[t]
\centering
\renewcommand{\arraystretch}{1.1}
\caption{Average accuracy and F1 scores (mean±std) of the HAR models across sensor configurations and granularity settings. These models perform effectively despite being trained on organically discovered labels rather than predetermined categories. The Balanced setting ($\lambda=0.3$) offers an optimal compromise between recognition granularity and performance for most deployments, while Conservative settings excel in accuracy (84-88\%) when coarser activity recognition suffices. Relaxed setting results (shown in gray) represent exploratory findings illustrating current technical boundaries rather than recommended configurations.}
\begin{tabular}{|c|c|c|c|c|}
\hline
\multirow{2}{*}{\textbf{Granularity}} & \multirow{2}{*}{\textbf{Metrics}} & \multicolumn{3}{c|}{\textbf{Sensor Config}} \\
\cline{3-5}
 &  & \textbf{\begin{tabular}[c]{@{}c@{}}Ambient(Basic)\\Only\end{tabular}} & \textbf{\begin{tabular}[c]{@{}c@{}}Ambient(Basic)+\\Wearable(IMU)\end{tabular}} & \textbf{\begin{tabular}[c]{@{}c@{}}Ambient(Advanced)+\\Wearable(IMU)\end{tabular}} \\
\hline
\multirow{2}{*}{\textbf{Conservative}} & \textbf{Accuracy} & 83.9\%±10.1\% & 85.1\%±6.5\% & 87.8\%±8.4\% \\
\cline{2-5}
 & \textbf{F1 Score} & 82.9\%±10.6\% & 82.5\%±9.3\% & 86.8\%±9.3\% \\
\hline
\multirow{2}{*}{\textbf{Balanced}} & \textbf{Accuracy} & 75.2\%±9.2\% & 72.5\%±6.6\% & 78.8\%±8.9\% \\
\cline{2-5}
 & \textbf{F1 Score} & 65.5\%±13.3\% & 63.7\%±9.7\% & 70.6\%±11.1\% \\
\hline
\multirow{2}{*}{\textbf{Relaxed}} & \textbf{Accuracy} & \textcolor{gray}{70.4\%±11.5\%} & \textcolor{gray}{69.0\%±9.0\%} & \textcolor{gray}{73.2\%±10.4\%} \\
\cline{2-5}
 & \textbf{F1 Score} & \textcolor{gray}{51.4\%±10.3\%} & \textcolor{gray}{47.8\%±10.5\%} & \textcolor{gray}{55.6\%±14.8\%} \\
\hline
\end{tabular}
\vspace{0.5em}
\Description{Average accuracy and F1 scores (mean±std) of activity recognition models across sensor configurations and granularity settings. These models perform effectively despite being trained on organically discovered labels rather than predetermined categories. The Balanced setting ($\lambda=0.3$) offers an optimal compromise between recognition granularity and performance for most deployments, while Conservative settings excel in accuracy (84-88\%) when coarser activity recognition suffices. Relaxed setting results (shown in gray) represent exploratory findings illustrating current technical boundaries rather than recommended configurations.}
\label{tab:har-performance}
\end{table}

\begin{figure}[t]
\begin{center}
 \subfigure[Ambient(Basic) Only]{
    \centering
    \includegraphics[width=0.32\textwidth]{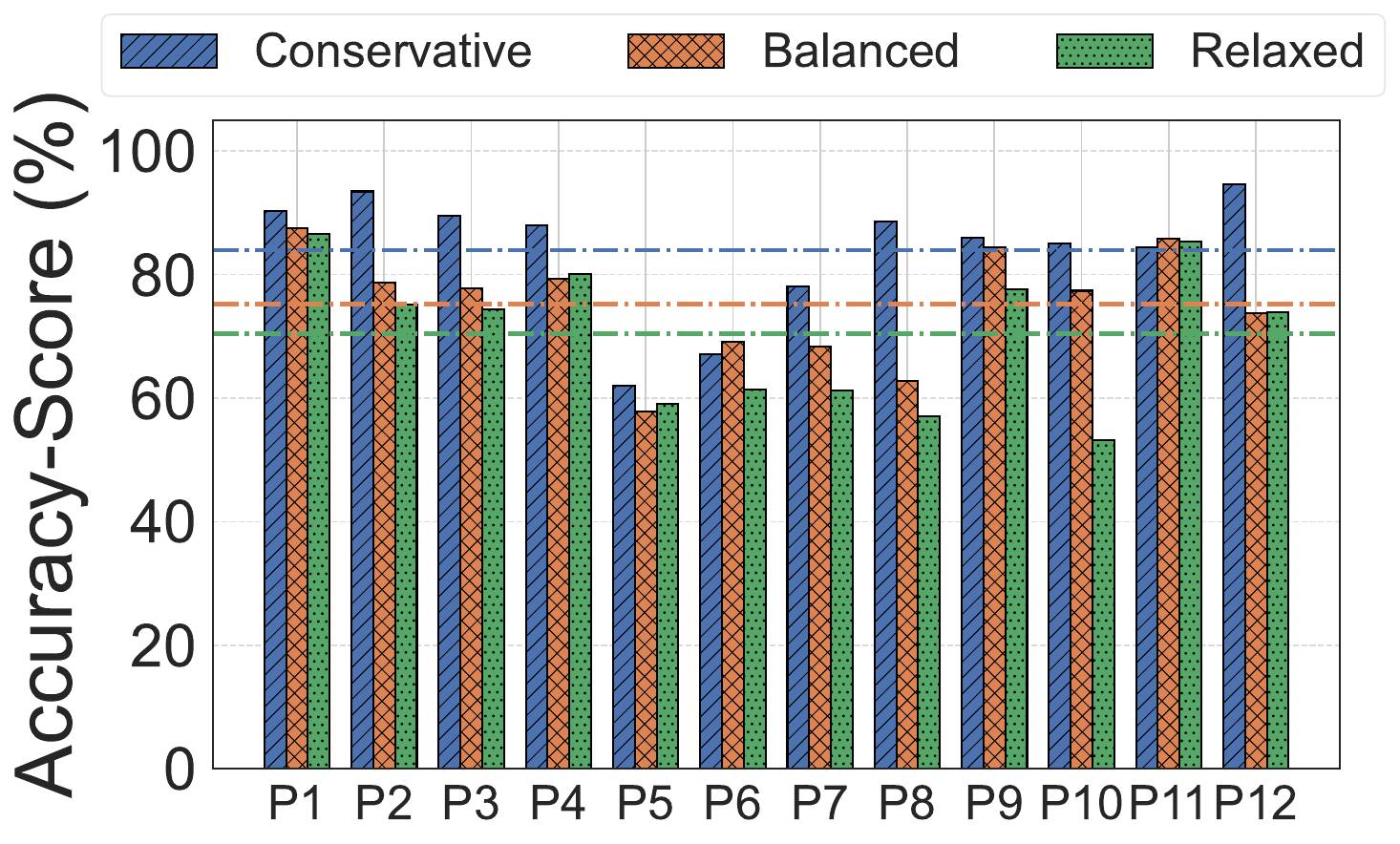}}
 \subfigure[Ambient(Basic)+Wearable(IMU)]{
     \centering
     \includegraphics[width=0.32\textwidth]{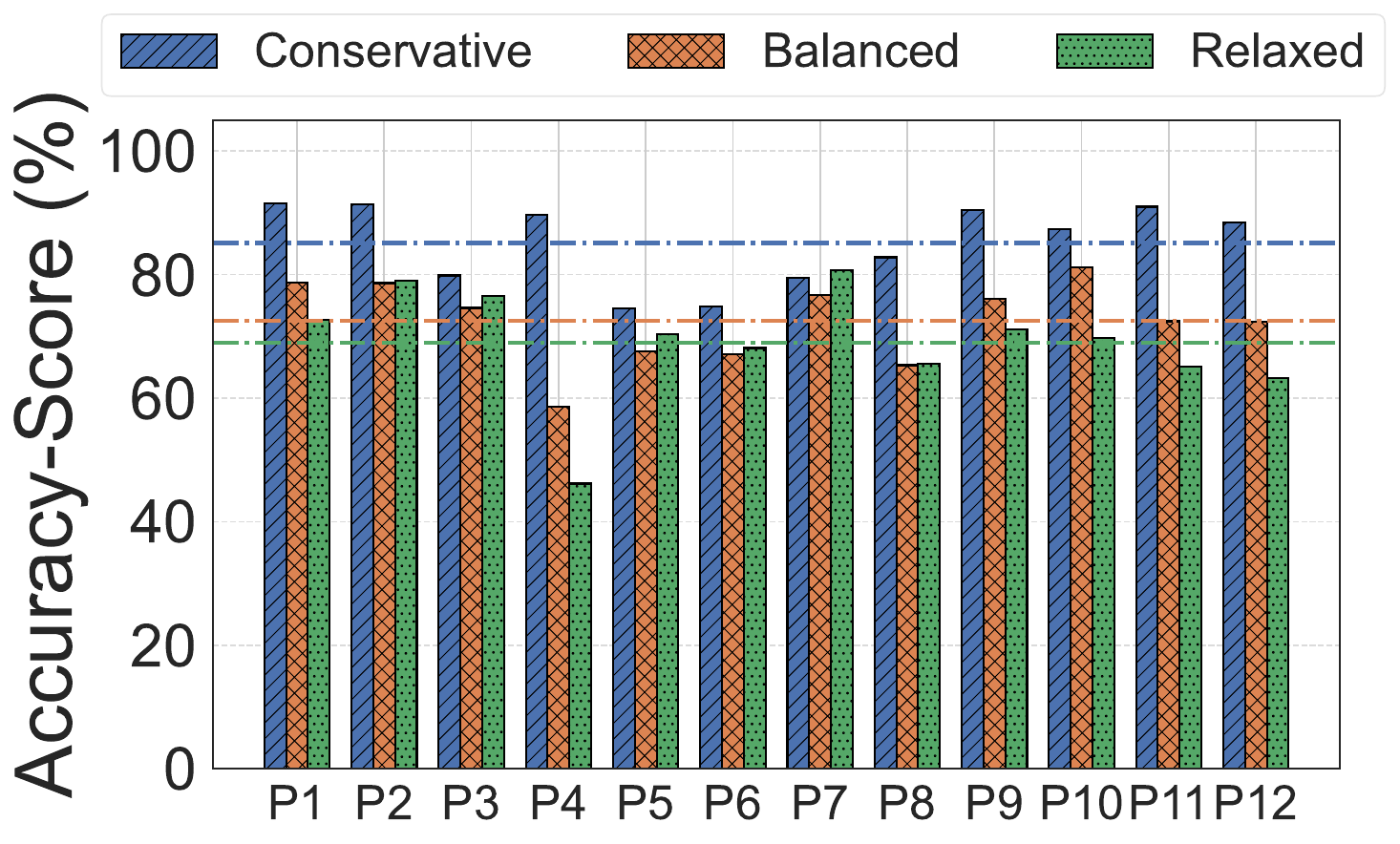}}
 \subfigure[Ambient(Advanced)+Wearable(IMU)]{
     \centering
     \includegraphics[width=0.32\textwidth]{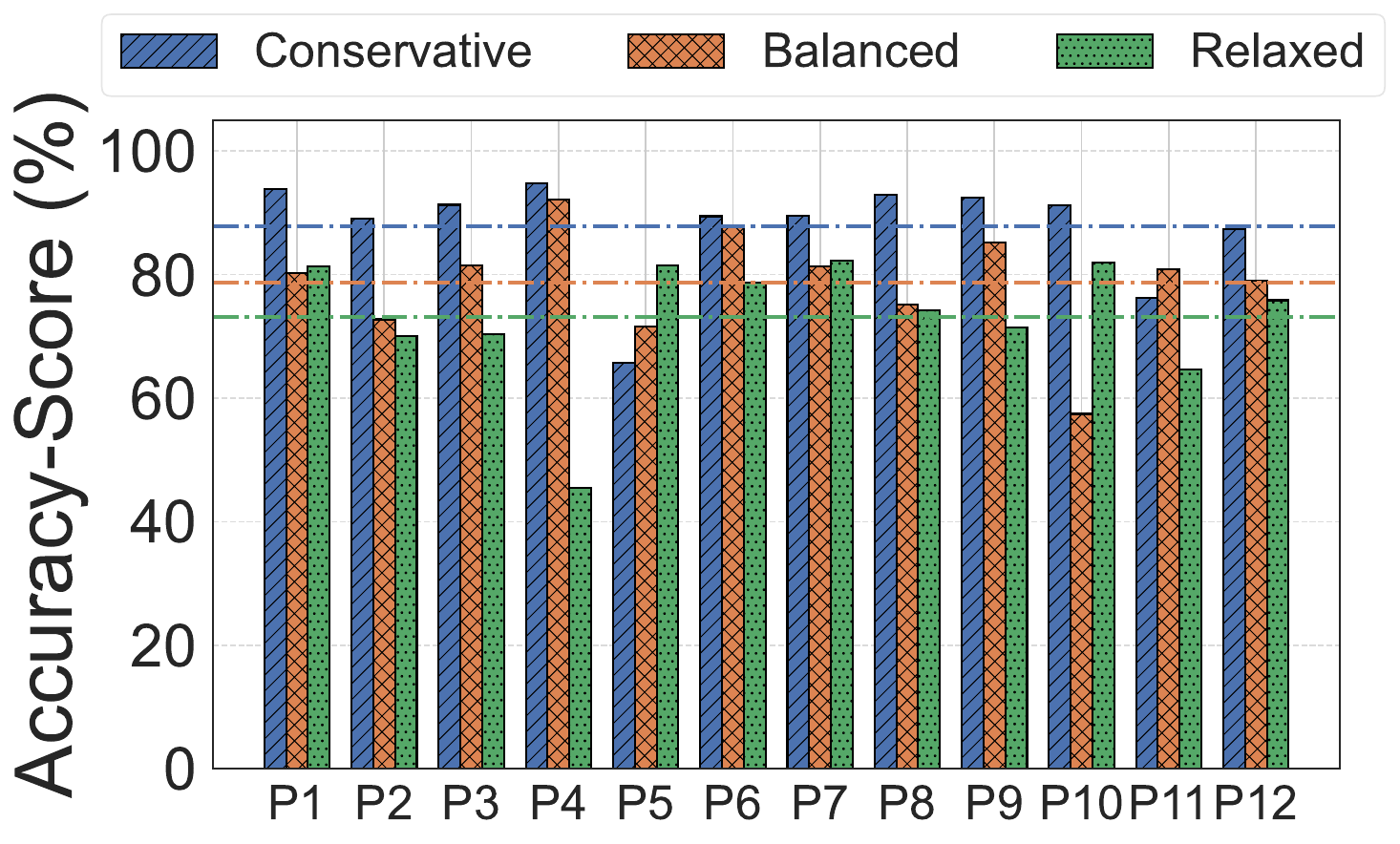}}
\end{center}

\caption{Per-participant accuracy across three sensing configurations. Kitchen 1 participants (P1-P4) show consistently high accuracy even with basic ambient sensing, Kitchen 2 participants (P5-P8) show greater degradation with finer granularity, and Kitchen 3 participants (P9-P12) benefit most from wearable IMU. Horizontal dashed lines indicate average performance per granularity setting.}
\Description[Per-participant accuracy bar charts across three sensor configurations and granularity levels]{Three grouped bar charts showing accuracy percentages ($0$-$100\%$) for $12$ participants (P1-P12) across three granularity settings. Each chart displays three bars per participant: blue diagonal-striped (Conservative), orange crosshatch (Balanced), and green dotted (Relaxed). Horizontal dashed reference lines indicate average performance per granularity. First chart (Ambient Basic Only) shows Conservative bars highest at $80$-$93\%$, Balanced at $66$-$79\%$, Relaxed at $46$-$79\%$, with Kitchen 1 participants (P1-P4) maintaining $80$-$93\%$ across all settings, Kitchen 2 participants (P5-P8) showing degradation to $58$-$74\%$ for Relaxed, and Kitchen 3 participants (P9-P12) showing $63$-$91\%$ range. Second chart (Ambient Basic plus Wearable IMU) displays similar patterns with Conservative $62$-$95\%$, Balanced $57$-$88\%$, Relaxed $53$-$87\%$, where P5 shows notable drop in Conservative to $62\%$. Third chart (Ambient Advanced plus Wearable IMU) shows highest overall performance with Conservative $65$-$95\%$, Balanced $57$-$92\%$, Relaxed $45$-$82\%$, demonstrating P4 reaching $95\%$ Conservative accuracy and most participants maintaining above $70\%$ across settings.}
\label{fig:accuracy-performance}
\end{figure}

\begin{figure}[t]
\begin{center}
 \subfigure[Ambient(Basic) Only]{
    \centering
    \includegraphics[width=0.32\textwidth]{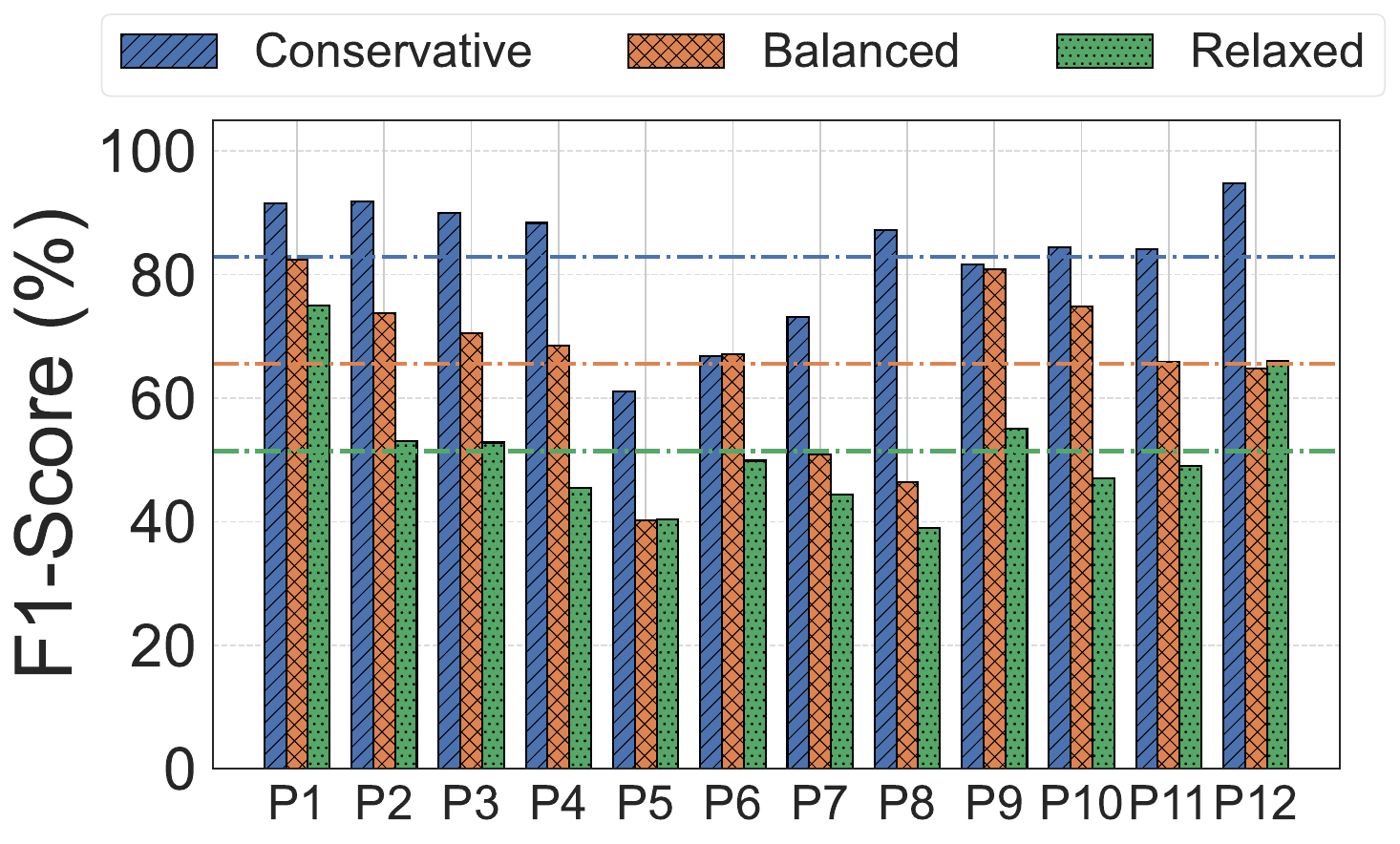}}
 \subfigure[Ambient(Basic)+Wearable(IMU)]{
     \centering
     \includegraphics[width=0.32\textwidth]{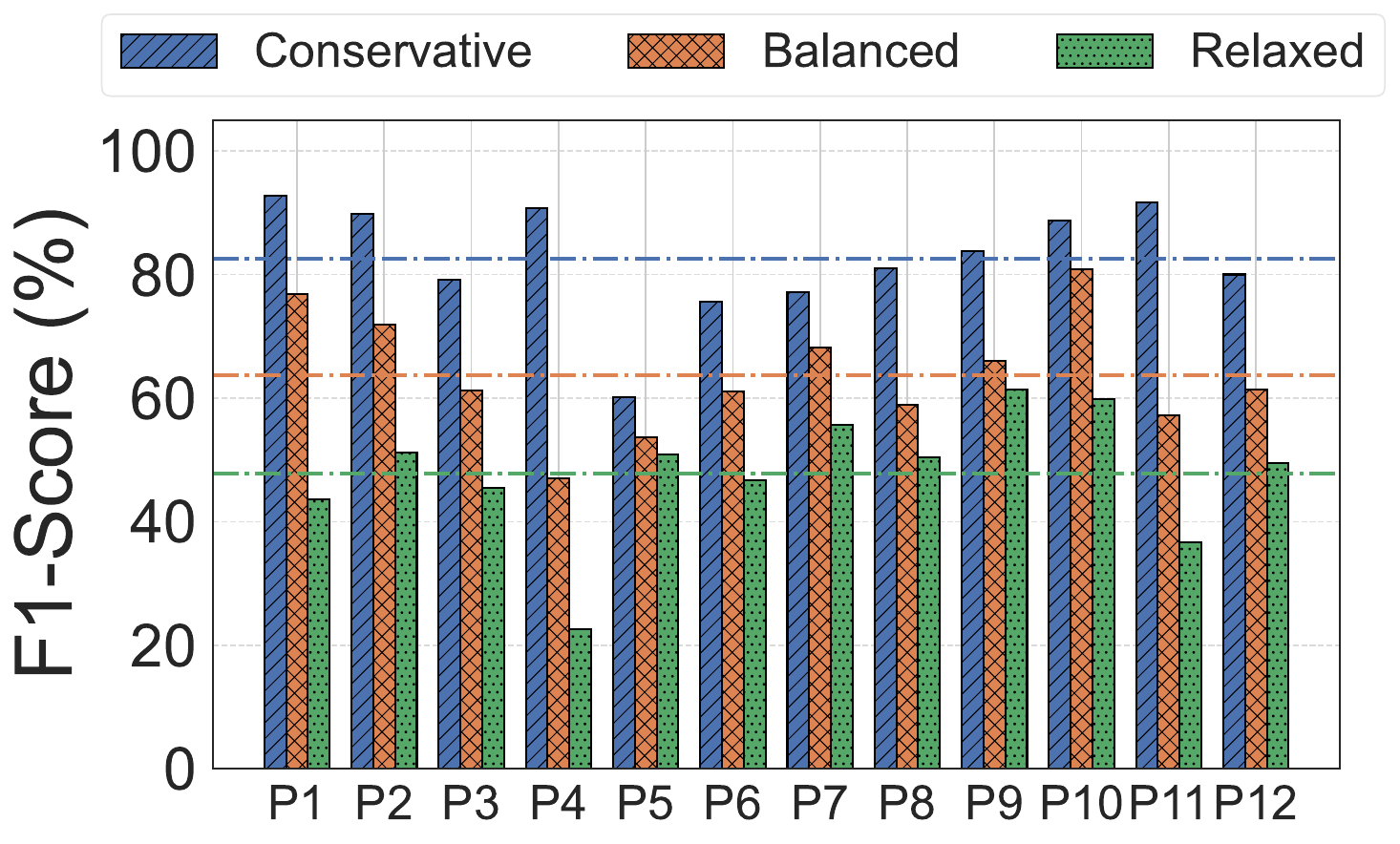}}
 \subfigure[Ambient(Advanced)+Wearable(IMU)]{
     \centering
 \includegraphics[width=0.32\textwidth]{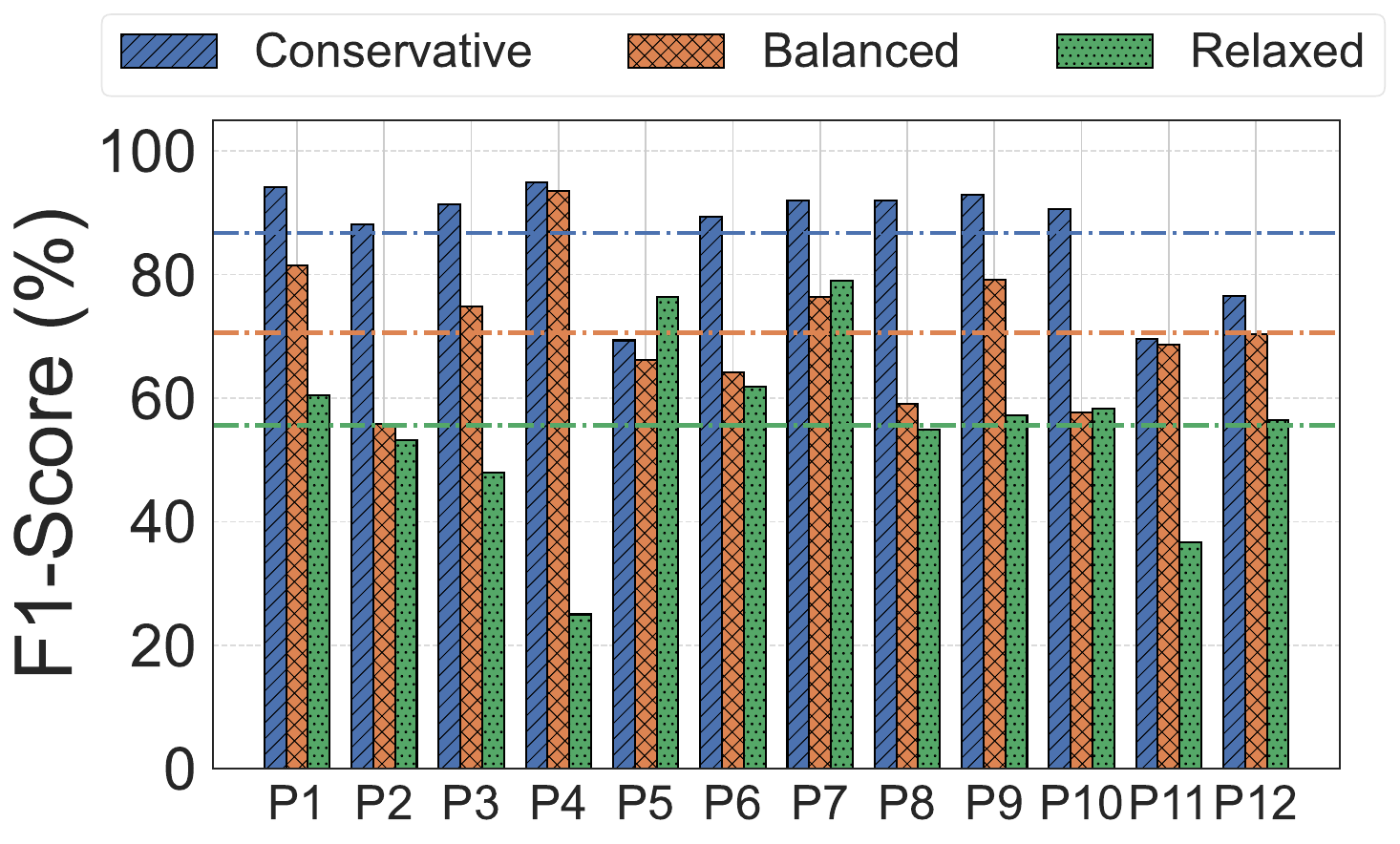}}
\end{center}
\caption{Per-participant F1 scores across sensing configurations, revealing sharper performance drops than accuracy metrics with increasing granularity. F1 scores for the Relaxed setting (green) show particularly large variance between participants and kitchens, illustrating the impact of environmental layout on the precision-recall balance in fine-grained activity recognition.}
\Description[Per-participant F1 score bar charts across three sensor configurations and granularity levels]{Three grouped bar charts showing F1 score percentages ($0$-$100\%$) for $12$ participants (P1-P12) across three granularity settings. Each chart displays three bars per participant: blue diagonal-striped (Conservative), orange crosshatch (Balanced), and green dotted (Relaxed). Horizontal dashed reference lines indicate average performance per granularity. First chart (Ambient Basic Only) shows Conservative bars ranging $60$-$93\%$, Balanced $47$-$81\%$, Relaxed $22$-$62\%$, with notably sharp drops from Conservative to Relaxed, particularly P4 dropping from $91\%$ to $22\%$ and P5 showing lowest Conservative at $60\%$. Second chart (Ambient Basic plus Wearable IMU) displays Conservative $61$-$95\%$, Balanced $40$-$82\%$, Relaxed $39$-$75\%$, where P12 achieves highest Conservative at $95\%$ and P5-P8 show consistent degradation with Kitchen 2 participants demonstrating larger variance in Balanced and Relaxed settings ($39$-$50\%$ range). Third chart (Ambient Advanced plus Wearable IMU) shows Conservative $69$-$95\%$, Balanced $55$-$93\%$, Relaxed $25$-$79\%$, with P4 reaching $95\%$ Conservative and $93\%$ Balanced but dropping to $25\%$ Relaxed, illustrating extreme sensitivity to granularity; P5-P7 show improved Relaxed performance ($62$-$79\%$) compared to other configurations, while overall variance remains high across all participants in Relaxed setting.}
\label{fig:f1-performance}
\end{figure}

\subsubsection{Overall Performance}
Table \ref{tab:har-performance} shows our trained HAR models' performance across sensor configurations. The HAR models demonstrate robust performance with 70-88\% accuracy and 48-87\% F1 scores, varying by sensor configuration and semantic granularity. Performance decreases with increasing granularity: Conservative ($\lambda=0.4$) achieves 84-88\% accuracy and 83-87\% F1 scores across all configurations, Balanced ($\lambda=0.3$) reaches 72-79\% accuracy and 64-71\% F1 scores, and Relaxed ($\lambda=0.2$) attains 69-73\% accuracy and 48-56\% F1 scores. The \textit{Ambient(Advanced)+Wearable(IMU)} configuration consistently outperforms others, particularly for fine-grained activities, achieving \accAmbientIMUOAKDRelaxed~ accuracy and \fscoreAmbientIMUOAKDRelaxed~ F1 score even in the challenging Relaxed setting. Additional spatial awareness from pose estimation and depth sensing helps resolve ambiguities between semantically similar activities. Interestingly, \textit{Ambient(Basic) Only} occasionally outperforms \textit{Ambient(Basic)+Wearable(IMU)}, indicating wearable data can introduce noise when movements aren't consistently captured.

\subsubsection{User-Level Performance Variability}
Figures \ref{fig:accuracy-performance} and \ref{fig:f1-performance} show accuracy and F1 scores across participants for three sensor configurations and granularity levels. Participants P1-P4 (Kitchen 1) maintain high performance across granularity settings with basic ambient sensing, while P5-P8 (Kitchen 2) show significant degradation with increased granularity. This suggests environment layout impacts recognition—Kitchen 1's constrained layout enables more reliable activity differentiation than Kitchen 2's open layout. For P9-P12 (Kitchen 3), wearable IMU data notably improves performance, providing valuable complementary information in certain environmental configurations.
F1 scores prove more sensitive to sensor configuration and granularity due to label imbalance, dropping 31-35 percentage points from Conservative to Relaxed settings versus 13-16 points for accuracy. This steeper decline highlights challenges in maintaining precision and recall when distinguishing semantically similar activities with limited sensing. The \textit{Ambient(Advanced)+Wearable(IMU)} configuration shows a smaller Conservative-to-Relaxed performance gap (15\% accuracy difference) than other configurations, indicating richer sensing enables fine-grained recognition without sacrificing coarse-grained performance. These results suggest that \name achieves robust performance across various hardware configurations by adapting recognition granularity to available sensing capabilities rather than forcing predetermined labels, 

\subsubsection{Activity Confusion Analysis}
\label{sec:activity-confusion}

\begin{figure}[t]
\begin{center}
 \subfigure[Conservative ($\lambda=0.4$)]{
    \centering
    \includegraphics[width=0.49\textwidth]{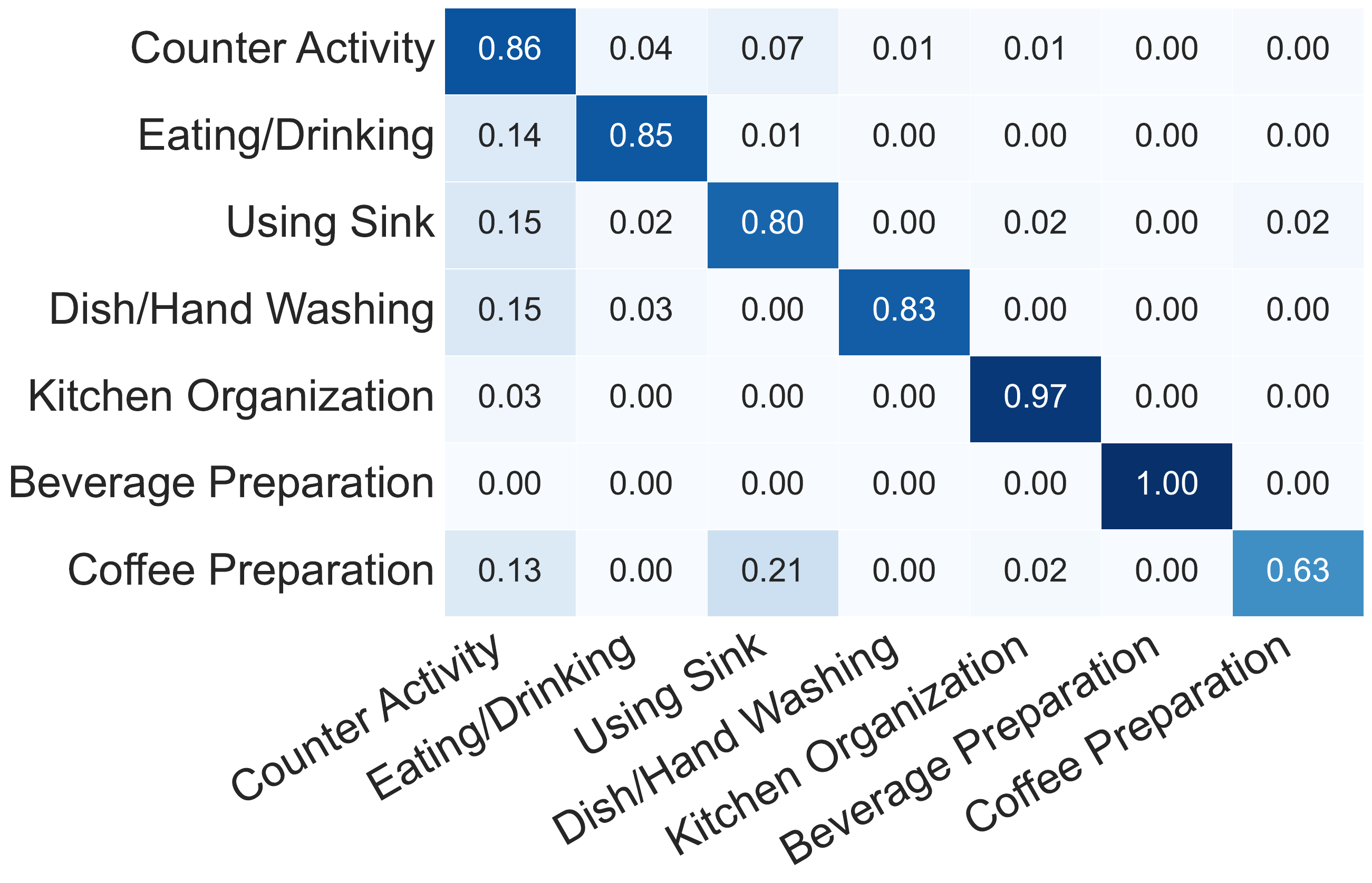}}
 \subfigure[Balanced ($\lambda=0.3$)]{
     \centering
     \includegraphics[width=0.49\textwidth]{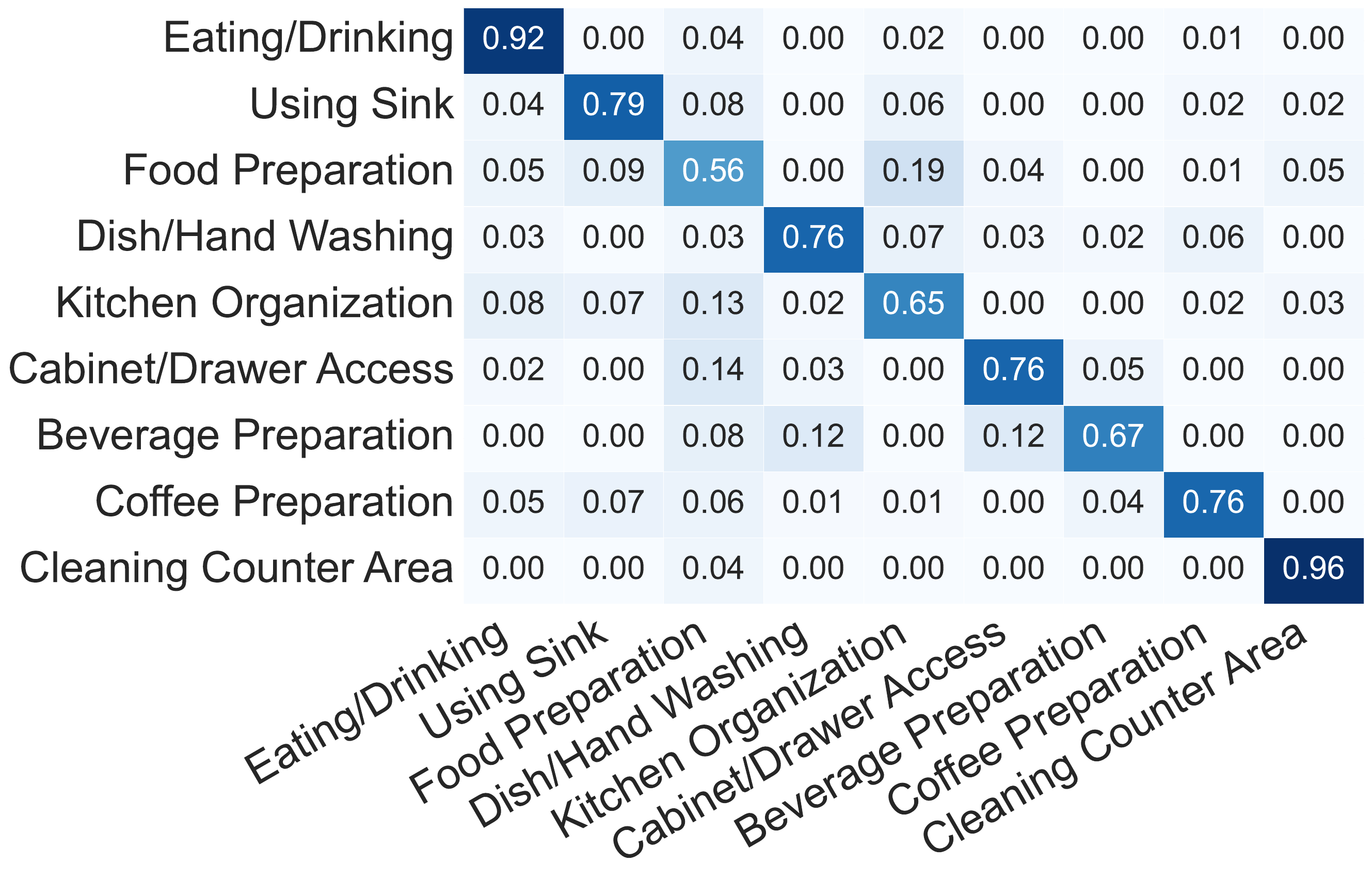}}
 \subfigure[Relaxed ($\lambda=0.2$)]{
     \centering
     \includegraphics[width=0.69\textwidth]{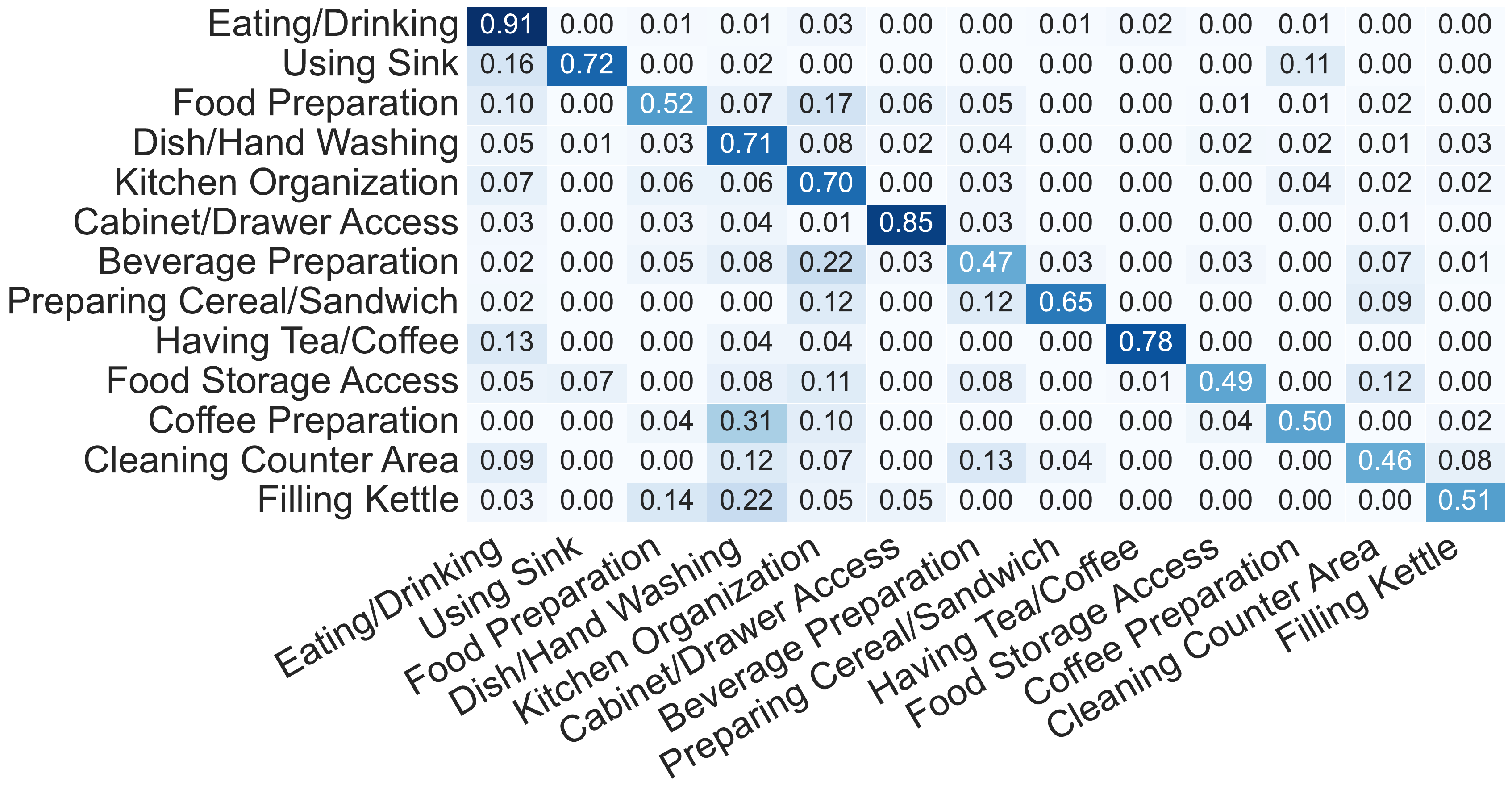}}
\end{center}
\caption{Confusion matrices showing the HAR performance using basic ambient sensors across three granularity settings. As granularity increases from Conservative (a) to Relaxed (c), more fine-grained activities emerge, revealing specific confusion patterns between semantically or spatially related activities.}
\Description[Three confusion matrices showing activity recognition performance across granularity levels]{Three confusion matrices with color-coded cells where darker blue indicates higher values and light blue indicates lower values. First matrix (Conservative, $\lambda=0.4$) shows $7 \times 7$ grid with activities: Counter Activity, Eating/Drinking, Using Sink, Dish/Hand Washing, Kitchen Organization, Beverage Preparation, and Coffee Preparation. Diagonal values (correct predictions) range from $0.63$ to $1.00$, with Beverage Preparation achieving perfect $1.00$ recognition and Kitchen Organization at $0.97$. Notable confusion patterns include Counter Activity confused with Using Sink ($0.15$), Dish/Hand Washing ($0.15$), and Coffee Preparation ($0.13$); Coffee Preparation shows lowest diagonal at $0.63$ with confusion with Using Sink ($0.21$). Second matrix (Balanced, $\lambda=0.3$) displays $13 \times 13$ grid with finer-grained activities including Food Preparation, Cabinet/Drawer Access, Preparing Cereal/Sandwich, Having Tea/Coffee, Food Storage Access, Cleaning Counter Area, and Filling Kettle. Diagonal values range from $0.46$ to $0.91$, with Eating/Drinking maintaining highest at $0.91$ and Cabinet/Drawer Access at $0.85$. Lower performing activities include Beverage Preparation ($0.47$, confused with Kitchen Organization $0.22$), Cleaning Counter Area ($0.46$), and Food Storage Access ($0.49$). Third matrix (Relaxed, $\lambda=0.2$) shows $9 \times 9$ grid with intermediate granularity. Diagonal values range from $0.56$ to $0.96$, with Cleaning Counter Area achieving $0.96$ and Eating/Drinking at $0.92$. Food Preparation shows lowest diagonal at $0.56$ with confusion spreading across Kitchen Organization ($0.19$), Cabinet/Drawer Access ($0.14$), and Using Sink ($0.09$). Beverage Preparation confused with Dish/Hand Washing ($0.12$) and Cabinet/Drawer Access ($0.12$). Overall pattern shows increasing confusion and decreasing diagonal values as granularity increases from Conservative to Balanced to Relaxed settings.}
\label{fig:confusion-matrices}
\end{figure}

Figure \ref{fig:confusion-matrices} shows confusion matrices for the \textit{Ambient(Basic) Only}configuration across three granularity settings. The Conservative setting (Figure \ref{fig:confusion-matrices}a) achieves high accuracy with minimal cross-category confusion. ``Counter Activity'' confuses with other activities as it encompasses multiple fine-grained activities; ``Coffee Preparation'' confuses with ``Using Sink'' due to spatial proximity. The Balanced setting (Figure \ref{fig:confusion-matrices}b) shows increased confusion with finer distinctions. ``Food Preparation'' has the lowest accuracy, confusing with ``Kitchen Organization'' and ``Using Sink'' due to similar ambient signatures. ``Beverage Preparation'' significantly confuses with cabinet access and washing activities since preparation incorporates these actions. In the Relaxed setting (Figure \ref{fig:confusion-matrices}c), ``Beverage Preparation'' degrades further, confusing primarily with ``Kitchen Organization'' as ambient sensors cannot distinguish fine manipulations in similar locations. Kettle-related and washing activities show substantial confusion, illustrating difficulty differentiating tasks with similar postures and locations using only ambient sensors.
Despite increasing confusion at finer granularities, ``Eating/Drinking'' maintains excellent performance across all settings with distinctive ambient signatures. ``Cabinet/Drawer Access'' similarly maintains high accuracy through distinctive movement patterns captured by basic ambient sensors.

These patterns validate our sensor-first approach—\name adapts activity granularity based on reliable detection capabilities rather than forcing potentially indistinguishable predetermined categories. This analysis identifies which sensor configurations suit specific recognition goals, enabling informed privacy-capability tradeoffs (Appendix \ref{sec:appendix-activity-confusion} covers other configurations).

\subsubsection{Incremental Training Analysis}
\label{sec:evaluation-incremental}

\begin{figure}[t]
  \centering
  \begin{minipage}[t]{0.4\textwidth}
    \vspace{5pt}
    \centering
    \includegraphics[width=\linewidth]{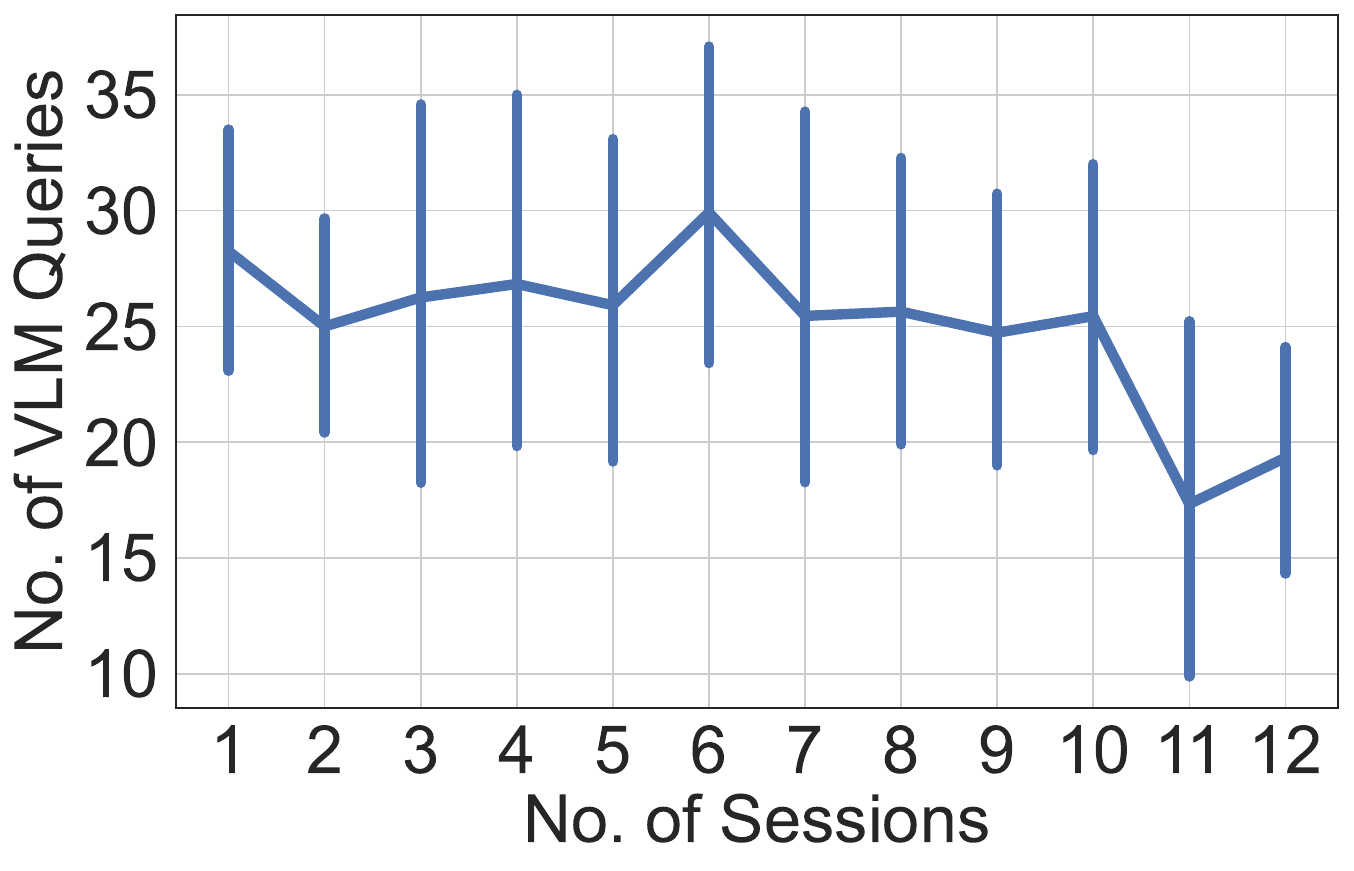}
    \vspace{-0.5em}
    \caption*{\small (a) VLM queries required across sessions}
  \end{minipage}
  \hspace{0.02\textwidth}  
  \begin{minipage}[t]{0.4\textwidth}
    \vspace{5pt}
    \centering
    \includegraphics[width=\linewidth]{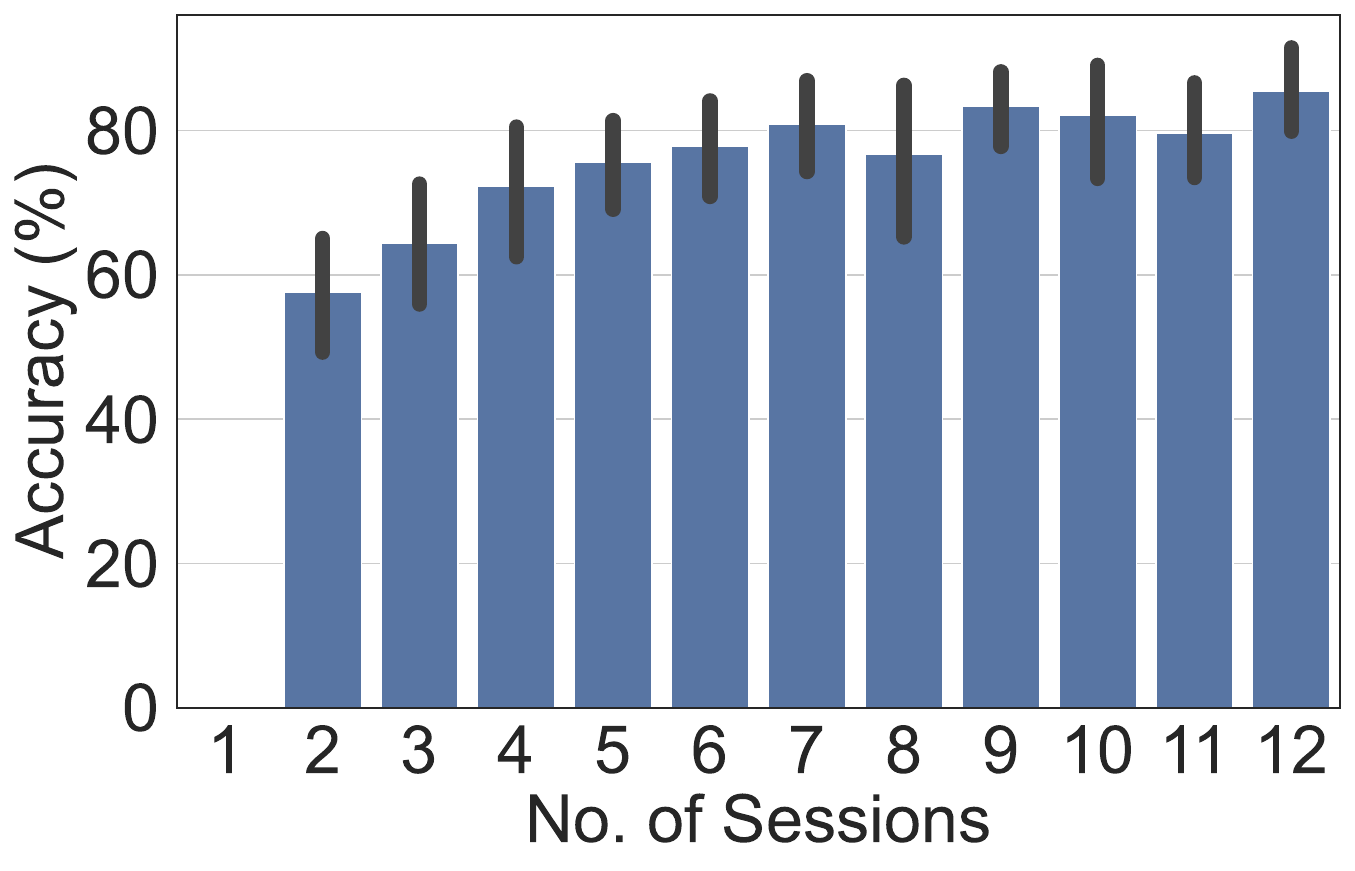}
    \vspace{-0.5em}
    \caption*{\small (b) Recognition accuracy across sessions}
  \end{minipage}

  \caption{Incremental training analysis of \name: (a) Count of VLM queries after incorporating new training session. (b) Recognition accuracy measured on all future sessions. Error bars represent 95\% confidence intervals across users.}
  \Description[Two-panel incremental training analysis showing VLM queries and accuracy progression]{Two-panel chart showing incremental training progression across $12$ sessions. Panel (a) displays line chart with vertical range bars showing number of VLM queries (y-axis $10$-$35$) per session. Each session shows vertical bar with upper and lower bounds connected by blue line. Session $1$ starts at approximately $28$ queries (range $23$-$34$), remains stable around $25$-$27$ queries through Session $10$, then declines notably in Session $11$ to $18$ (range $10$-$26$) and Session $12$ to $19$ (range $14$-$25$), suggesting potential efficiency gains with accumulated training data. Panel (b) displays bar chart showing recognition accuracy percentage (y-axis $0$-$90\%$) across sessions with blue bars and dark gray error bars representing $95\%$ confidence intervals. Clear upward trajectory from Session $1$ at $57\%$ accuracy (error bar $48$-$67\%$) through Session $12$ at $85\%$ ($78$-$90\%$). Steepest improvement occurs in first $5$ sessions, progressing from $57\%$ to $77\%$, with more gradual gains thereafter reaching $78$-$82\%$ by Sessions $6$-$11$, and final increase to $85\%$ in Session $12$. Combined panels demonstrate system learning effectiveness with improved accuracy while potentially reducing computational requirements over time.}
  \label{fig:incremental}
\end{figure}

To understand how cloud VLM query requirements and recognition accuracy evolve with increasing training data, we conducted an incremental training analysis simulating sequential session processing. We used \textit{Ambient(Advanced)+Wearable(IMU)} as it has the highest VLM usage among configurations. With each new session, \name identifies interesting segments across all data, applies the current model, invokes VLM analysis for low-confidence predictions or novel patterns, and retrains with the expanded dataset.
Figure \ref{fig:incremental}a shows total VLM queries required after incorporating each session, including queries for new data and previously-seen data that becomes significant due to pattern repetition across sessions. Query counts remain stable (25-30 per session) for the first 9 sessions despite the growing dataset, as the system continuously reassesses historical data alongside new observations, sometimes identifying patterns in earlier sessions not initially flagged when the dataset was smaller. The downward trend in sessions 10-11 (17-19 queries) is notable, though extended data collection is needed to confirm this pattern. Figure \ref{fig:incremental}b shows recognition accuracy (measured on sessions not included in training) improving from 57.6\% after the first session to 85.4\% after 11 sessions, with most gains in early sessions. 
These results demonstrate promising trends: steady accuracy improvement coupled with potential VLM query reduction in later sessions, suggesting \name could become both more effective and computationally efficient over time. 

\subsection{Real-world Deployment Evaluation}
\label{sec:evaluation-deployment}

\begin{figure}[t]
\begin{center}
 \subfigure[Home-1]{
    \centering
    \includegraphics[width=0.19\textwidth]{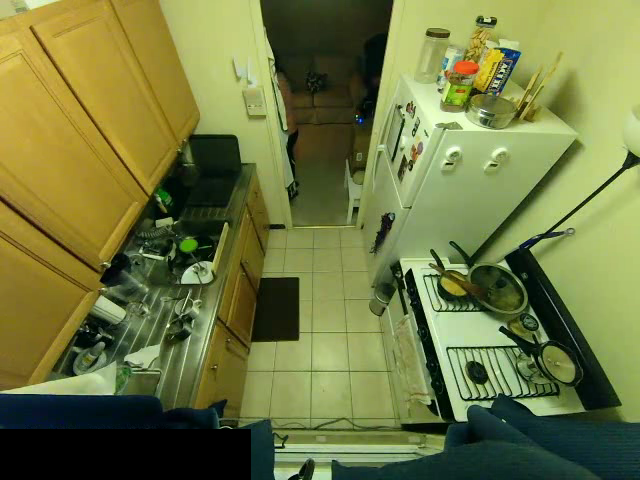}}
 \subfigure[Home-2]{
     \centering
     \includegraphics[width=0.19\textwidth]{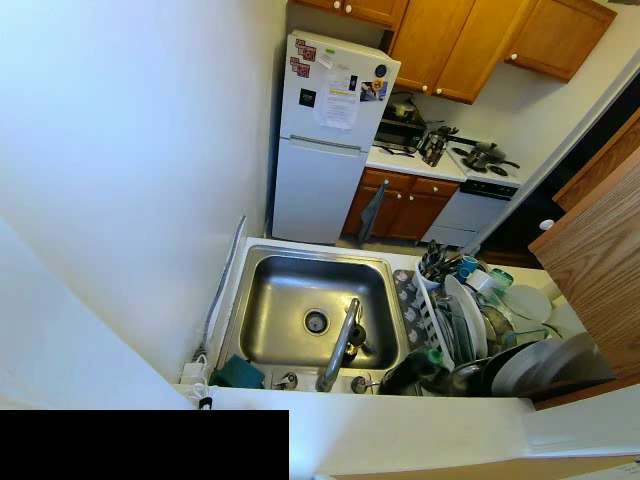}}
 \subfigure[Home-3]{
     \centering
     \includegraphics[width=0.19\textwidth]{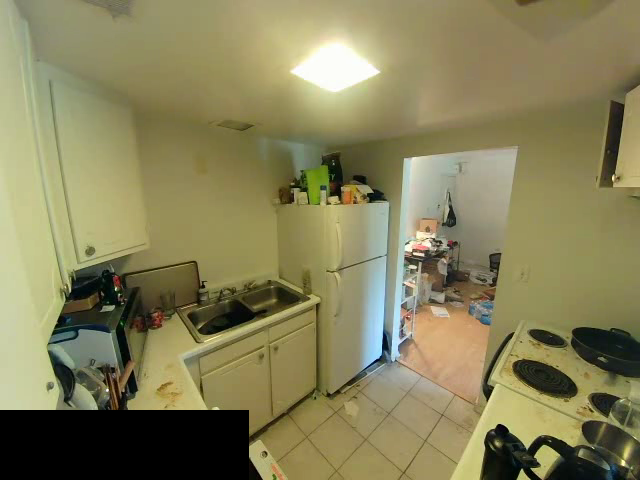}}
 \subfigure[Home-4]{
     \centering
     \includegraphics[width=0.19\textwidth]{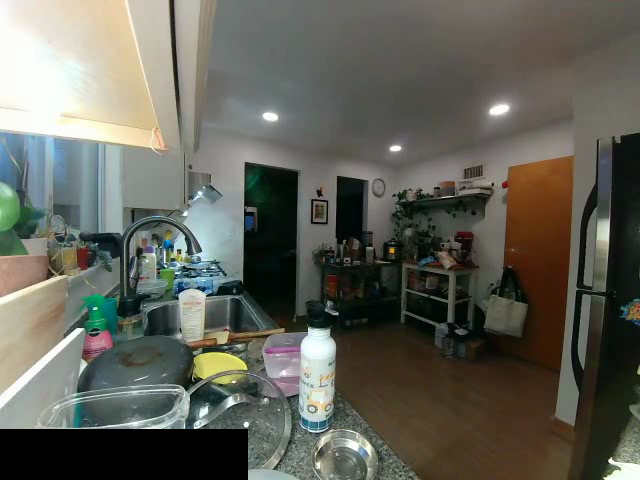}}
 \subfigure[Home-5]{
     \centering
     \includegraphics[width=0.19\textwidth]{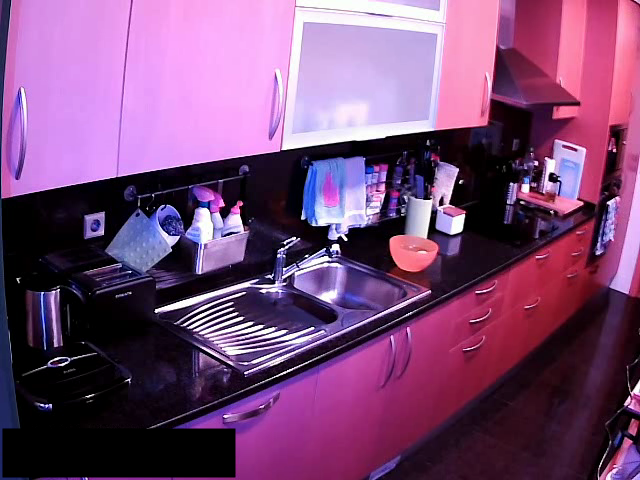}}
\end{center}
\caption{Kitchen environments used in real-world home deployments: (Home-1) compact galley layout captured from overhead diagonal perspective, (Home-2) medium-sized kitchen with distinct functional zones, (Home-3) narrow galley configuration from elevated angle, (Home-4) spacious kitchen with multiple countertops from horizontal perspective, and (Home-5) linear wall-mounted kitchen with integrated appliances from horizontal viewpoint.}
\Description[Five real-world home kitchen environments with varied layouts and camera perspectives]{Five photographs of residential kitchen spaces used for real-world deployment study. Home-1 shows overhead diagonal perspective of compact galley kitchen with light wood cabinets along left wall, white tile floor, gas stove on right side, and white refrigerator at back with doorway beyond. Home-2 displays horizontal view of medium-sized kitchen with dark gray countertops, sink area on left with dishes and cleaning supplies, white walls with recessed lighting, shelving units against back wall creating distinct functional zones, and open floor plan visible on right. Home-3 captures elevated angle view of narrow galley configuration with distinctive bright pink/magenta upper and lower cabinets, black countertops, stainless steel sink with integrated draining board in center, and pink color scheme throughout. Home-4 presents horizontal perspective of spacious kitchen with white upper and lower cabinets, white refrigerator on left, sink and counter area in center, white tile floor, doorway to adjacent room visible in background, and white electric stove on right providing multiple work surfaces. Home-5 shows horizontal viewpoint of linear wall-mounted kitchen with light wood cabinets, stainless steel sink in foreground with dishes and drying rack, white refrigerator in middle distance, gas stove beyond, and integrated appliances along single wall configuration with white walls throughout.}
\label{fig:camera-positioning-realworld}
\end{figure}

We extended our evaluation to examine \name's performance in actual home environments, where users engage in daily routines without experimental constraints. We deployed \name in 5 homes for 7 days, with participants collecting 6-8 sessions each based on their kitchen usage frequency. Participants used their kitchens normally—cooking, cleaning, snacking, socializing—without task restrictions or researcher guidance, capturing the complexity of real environments including activity interruptions and varying lighting conditions (See Figure \ref{fig:camera-positioning-realworld}).
The deployment generated approximately 11 hours of sensor data across diverse home layouts and activity patterns. For accuracy evaluation, we sampled 250 five-second segments per participant using temporal stratified random sampling: 60 from early sessions (1-2), 60 from middle sessions (3-5), and 130 from late sessions (6-8), enabling analysis of learning progression and mature system performance. We annotated ground truth using our established methodology (\S \ref{sec:datacollection}), mapping labels to discovered activities. Based on controlled evaluation results, we used the Balanced granularity setting ($\lambda=0.3$) and \textit{Ambient(Advanced)+Wearable(IMU)} sensor configuration for all real-world deployment analyses.

\begin{figure}[ht]
\centering
\hfill
\begin{minipage}[b]{0.48\textwidth}
    \centering
    \includegraphics[width=\textwidth]{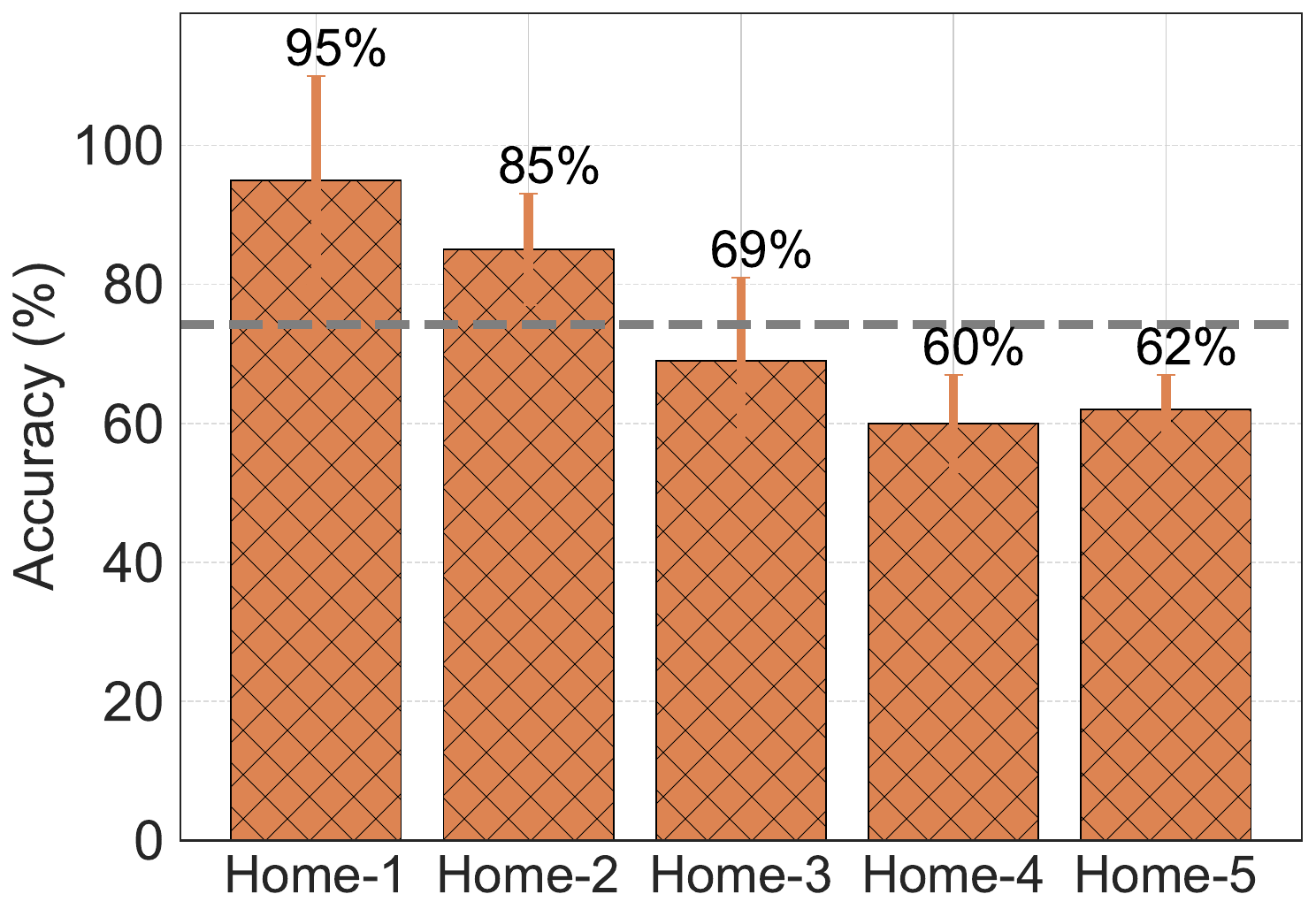}
    \caption{Overall recognition accuracy in real-world home deployments using leave-one-session-out cross-validation with Balanced granularity setting and \textit{Ambient(Advanced)+Wearable(IMU)} configuration. Each bar represents average performance when training on all other available sessions (including future ones) and testing on one held-out session. The horizontal dashed line indicates average accuracy (74.2\%) across all participants.}
    \Description[Bar chart showing overall recognition accuracy across five home deployments]{Bar chart displaying accuracy percentages (y-axis $0$-$100\%$) for five homes (x-axis) using leave-one-session-out cross-validation. All bars have orange crosshatch pattern with dark gray error bars. Home-1 achieves highest accuracy at $95\%$ with error bar extending from approximately $74\%$ to near $100\%$. Home-2 shows $85\%$ accuracy with error bar spanning roughly $74\%$ to $93\%$. Home-3 displays $69\%$ accuracy with error bar from approximately $55\%$ to $80\%$. Home-4 shows $60\%$ accuracy with error bar extending from roughly $48\%$ to $67\%$. Home-5 achieves $62\%$ accuracy with error bar spanning approximately $54\%$ to $68\%$. Horizontal dashed gray line at $74\%$ indicates average performance across all participants. Performance varies significantly across homes, with Home-1 and Home-2 exceeding average by substantial margins while Home-3, Home-4, and Home-5 fall below average.}
    \label{fig:deployment-overall}
\end{minipage}
\hfill
\begin{minipage}[b]{0.48\textwidth}
    \centering
    \includegraphics[width=\textwidth]{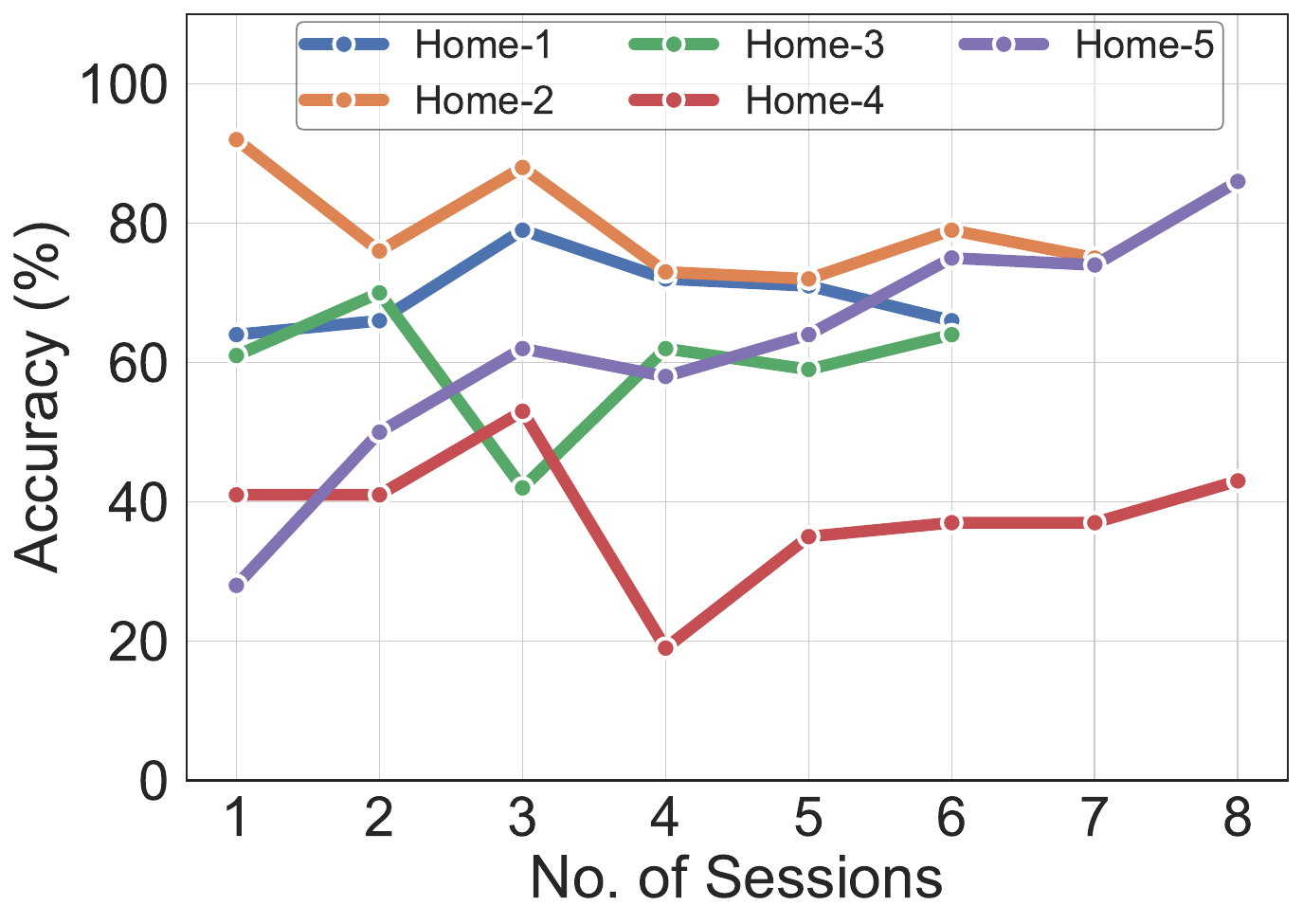}
    \caption{Incremental training trajectories showing forward-looking evaluation in chronological session order, where the system trains on sessions 1 through N and tests on all remaining future sessions (N+1 onwards). \textit{Note:} Performance patterns differ from Figure \ref{fig:deployment-overall} due to temporal dependencies, varying amounts of training/test data, and chronological activity pattern evolution—reflecting realistic deployment conditions where future data is unavailable.}
    \Description[Multi-line chart showing incremental learning trajectories across home deployments]{Line chart displaying accuracy percentages (y-axis $0$-$100\%$) across eight training sessions (x-axis $1$-$8$) for five homes with distinct line styles and colors. Home-1 (blue circles with solid line) starts at $64\%$ Session $1$, rises to $66\%$ Session $2$, peaks at $79\%$ Session $3$, then shows gradual decline through $72\%$ Sessions $4$-$5$ and $70\%$ Sessions $6$-$7$. Home-2 (orange squares with dashed line) begins highest at $92\%$ Session $1$, decreases to $77\%$ Session $2$, recovers to $88\%$ Session $3$, maintains $72$-$79\%$ through Sessions $4$-$7$. Home-3 (green triangles with dash-dot line) starts at $61\%$ Session $1$, increases to $69\%$ Session $2$, drops sharply to $43\%$ Session $3$, fluctuates between $58$-$64\%$ Sessions $4$-$6$, ending at $66\%$ Session $7$. Home-4 (red diamonds with dotted line) begins at $41\%$ Session $1$, remains stable around $41$-$53\%$ through Session $3$, shows dramatic drop to $18\%$ Session $4$, gradually recovers to $35$-$38\%$ Sessions $5$-$7$, reaching $43\%$ Session $8$. Home-5 (purple inverted triangles with dash-dot line) demonstrates strongest improvement starting at $28\%$ Session $1$, progressing through $50\%$ Session $2$, $62\%$ Session $3$, $59\%$ Session $4$, plateauing around $65$-$75\%$ Sessions $5$-$7$, achieving highest final accuracy at $86\%$ Session $8$. Variable trajectory patterns reflect different learning dynamics and activity pattern consistency across homes.}
    \label{fig:deployment-incremental}
\end{minipage}
\hfill
\end{figure}

\subsubsection{Overall Performance}

Figure \ref{fig:deployment-overall} presents the overall recognition accuracy across homes using leave-one-session-out cross-validation. \name achieves an average accuracy of 74.2\%±14.8\% across the five homes, with performance ranging from 60\% to 95\%. This represents a moderate decrease compared to the controlled evaluation results (78.8\%±8.9\% for the same sensor configuration and granularity setting), reflecting the additional complexity of unstructured home environments. Performance for individual homes varied significantly, with Home-1 achieving the highest accuracy at 95\%, followed by Home-2 at 85\%. Home-3, Home-4, and Home-5 achieved lower performance at 69\%, 63\%, and 60\% respectively. This variability appears partially related to both kitchen layout constraints and camera positioning challenges (See Figure \ref{fig:camera-positioning-realworld}). Home-4 and Home-5 utilized more lateral camera placements due to their linear kitchen configurations, which limited the VLM's ability to capture comprehensive activity context compared to the elevated diagonal perspectives used in Home-1 and Home-2. Additionally, Home-4's consistently dim lighting conditions throughout most sessions further degraded VLM performance during key moment analysis. These real-world constraints, which are difficult to replicate in controlled settings, highlight the importance of strategic camera positioning and adequate lighting for effective VLM-based activity discovery.

\subsubsection{Discovered Activity Patterns}
\label{subsubsec:activities-real-world}

Across all 5 homes, \name identified 4 common kitchen activities: ``stovetop cooking'', ``refrigerator interaction'', ``sink area tasks'', and ``countertop food preparation''. These represent behaviors that occur typically across these kitchen environments.
Beyond these commonly occurring behaviors, we discovered activities specific to usage in each home. Home-4 had ``coffee preparation'' routines with patterns for coffee station tasks and coffee machine operation. Home-5 showed ``rice cooker usage'' with separate patterns for preparation and serving phases, along with ``dishwasher interaction'' patterns. Home-3 was the only home with ``waste disposal'' detected as a standalone recurring activity. Home-2 demonstrated granular activity detection, with ``refrigerator door interactions'' separate from ``general refrigerator access''. This dual-layer discovery—common activities plus individual specific patterns—shows that users can expect baseline functionality while the system adapts to their specific routines.

\subsubsection{Incremental Training Analysis}

Figure \ref{fig:deployment-incremental} shows learning trajectories using forward-looking evaluation in chronological session order, where accuracy represents the system's performance on all future sessions after training exclusively on sessions 1 through N. The performance patterns may differ from Figure \ref{fig:deployment-overall} due to several temporal factors: (1) the amount of training data increases with each session while test data decreases, and (2) activity patterns evolve differently, making early sessions good/poor predictors of later behavior based on what kind of activities is performed in later sessions. While computationally expensive (and cost-prohibitive) combinatorial approaches (training on all possible session combinations) could provide alternative insights, this chronological evaluation reflects actual deployment scenarios. 
The results demonstrate varied adaptation across participants. Home-5 showed substantial improvement (28\% to 86\% accuracy), indicating successful learning of evolving activity patterns. Home-2 maintained consistently high performance (92\% to 72\%), reflecting stable routines that enable reliable prediction. Home-1 achieved stable moderate performance (64-79\% range). Home-3 and Home-4 exhibited variable patterns, suggesting the system requires additional training or user assistance when participants have irregular routines. These trajectories indicate that \name adapts effectively to individual activity patterns, with performance stability correlating to routine consistency—regular kitchen usage enables more stable recognition, while variable patterns present ongoing adaptation challenges.

\section{DISCUSSION}
\label{sec:discussion}

\subsection{User Story: From Initial Adoption to Customization}

\begin{figure}[t]
    \centering
    \includegraphics[width=0.6\textwidth]{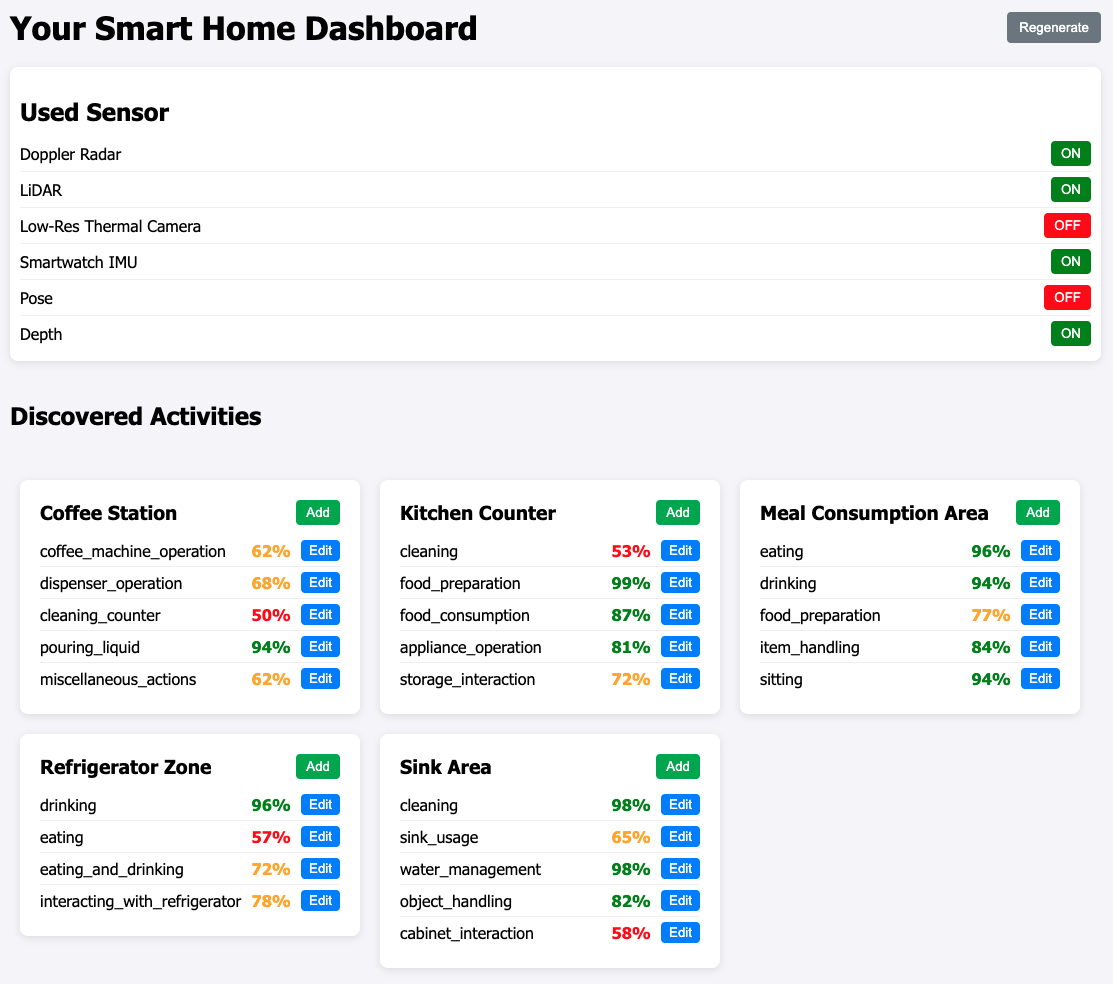}
    \caption{Interface for customizing the sensors to be used and activity labels. The percentage(\%) values show how well a deployed sensing configuration can predict activities when compared with discovered activity labels.}
    \Description[Smart home dashboard interface showing sensor configuration and discovered activities]{Web interface titled "Your Smart Home Dashboard" with Regenerate button in top right. Two main sections: Top section "Used Sensor" lists six sensors with ON/OFF toggle buttons: Doppler Radar (ON), LiDAR (ON), Low-Res Thermal Camera (OFF), Smartwatch IMU (ON), Pose (OFF), and Depth (ON). Bottom section "Discovered Activities" displays five white cards for different functional zones. Coffee Station card shows activities: coffee\_machine\_operation ($62\%$), dispenser\_operation ($68\%$), cleaning\_counter ($50\%$), pouring\_liquid ($94\%$), miscellaneous\_actions ($62\%$), each with blue Edit button and green Add button at top. Kitchen Counter card lists: cleaning ($53\%$), food\_preparation ($99\%$), food\_consumption ($87\%$), appliance\_operation ($81\%$), storage\_interaction ($72\%$). Meal Consumption Area card shows: eating ($96\%$), drinking ($94\%$), food\_preparation ($77\%$), item\_handling ($84\%$), sitting ($94\%$). Refrigerator Zone card displays: drinking ($96\%$), eating ($57\%$), eating\_and\_drinking ($72\%$), interacting\_with\_refrigerator ($78\%$). Sink Area card lists: cleaning ($98\%$), sink\_usage ($65\%$), water\_management ($98\%$), object\_handling ($82\%$), cabinet\_interaction ($58\%$). Percentages shown in color coding: green for high ($90\%$+), yellow/orange for medium ($60$-$89\%$), red for low (below $60\%$).}
    \label{fig:interface}
\end{figure}

\modified{\name is designed to incrementally adapt to users and their environments while preserving privacy.
First, by continuing data collection and analysis, we can surface representative examples of detectable activities for each environment, allowing users to form realistic expectations before deciding to deploy our system. The set of activities that we can typically detect in each location, \eg a kitchen, can be told to the user to manage their expectations (see \S \ref{subsubsec:activities-real-world}).
Upon installation, the system begins by identifying activities that are reliably detectable in the user's home using only privacy-preserving sensors.
As demonstrated in our deployment study, even with a few days of data, \name can detect 4–5 coarse activity categories using ambient sensors alone, and up to 8–9 fine-grained categories with additional modalities like IMU, depth, and pose.}

\modified{After this initial discovery phase, users can interact with our prototype customization interface (\figref{fig:interface}), which allows them to merge similar activity labels (\eg ``washing dishes'', ``rinsing hands'') into broader categories like ``sink-related tasks'', or define entirely new labels they want \name to detect.
While this interface has not yet been formally evaluated, it is technically feasible by reusing components from our label clustering and refinement pipeline.
Our contribution is not the interface itself, but the underlying paradigm shift, \ie \name surfaces candidate activity labels that are grounded in what the sensors can actually detect in each environment, which can then be reconciled with each user's preferences and goals.
Investigating how users make sense of, reconfigure, or reject such sensor-driven candidates represents a promising direction for future HCI work.}

\modified{Notably, \name does not require users to install all sensors we tested.
The framework is designed to be modality-agnostic, which allows flexibility to any combination of available sensors without algorithmic changes.
Users can select sensor combinations that balance privacy requirements, budget constraints, and recognition needs. The three-tier configuration framework tested in our evaluation can inform such cost considerations.}

\subsection{Applications in Healthcare and Assistive Technologies}

As we described in the Background (\S \ref{sec:bg}), \name is particularly promising for healthcare applications, where continuous monitoring can support individuals with special needs while respecting privacy.
For people with dementia and their caregivers, the system can notify household members when critical actions are forgotten (\eg leaving the refrigerator open) or summarize high-level activities for clinicians without invasive video monitoring~\cite{cipriani_daily_2020,autonomousprojectwebsite}.
Procedural task assistance is another example we are applying OrganicHAR to, for instance, to support wound care procedure for post-operative skin cancer patients~\cite{Vaccarello2024,DBLP:conf/uist/ArakawaPPLG25}.
Unlike existing approaches constrained by manually-predefined step sequences, \name could make procedural tracking more flexible by automatically identifying meaningful steps based on sensor capabilities rather than forcing predetermined sequences.
The hierarchical nature of our discovered activities (from coarse to fine-grained) enables adaptive assistance based on user needs, from general awareness of functional zone usage for those requiring minimal support, to detailed step-by-step guidance for those needing more comprehensive assistance.

\subsection{Potential Privacy Concerns and Mitigation Strategies}

\modified{While \name uses privacy-preserving sensors during deployment, using VLMs during training can raise privacy concerns about transmitting domestic visual information to third-party cloud services. While \name processes only 9-11\% of video data during key moments rather than continuous monitoring, users may worry about data misuse, unintended inferences about their private lives, and potential access by service providers or authorities \cite{zhang_evaluation_2025,ju_watch_2025,yun_what_2025}. Several techniques can mitigate these concerns without compromising labeling effectiveness. 
For example, since our framework focuses on actions, locations, and object interactions rather than personal identification, automatic face blurring, background anonymization, and removal of personally identifiable information can be applied. Additionally, user agency can be enhanced by providing transparent previews of what video segments would be transmitted to cloud services and enabling users to approve or decline each training session before any data is sent. 
Similarly, researchers have improved handling of sensitive content through targeted fine-tuning of VLMs~\cite{samson_little_2025}.
We will also explore emerging approaches such as federated learning and edge-based VLMs to minimize cloud dependencies while maintaining semantic understanding capabilities.}

\subsection{Limitations and Future Work}

\subsubsection{Towards In-the-Wild Deployment:}
\modified{Our real-world deployment revealed practical challenges: camera positioning constraints in linear kitchen layouts limited VLM effectiveness, varying lighting conditions affected recognition consistency, and participants' evolving routines created temporal adaptation challenges requiring ongoing refinement. Future work explores ways to improve OrganicHAR by considering these barriers, which will offer practical deployment guideline.}

\subsubsection{Beyond Discrete Activity Labels:}
Given challenges maintaining consistent activity labels across heterogeneous users and environments, we can explore alternatives to discrete classification. One approach involves direct translations between sensor streams and natural language descriptions. Rather than forcing classification between ``washing dishes'' or ``washing hands,'' the system could generate descriptions like ``using the sink with small movements, likely washing hands.'' This direction becomes increasingly feasible as small language models become more capable and deployable on edge devices~\cite{popov_overview_nodate}. While these compact models cannot perform complex reasoning, they could strategically interpret sensor patterns and generate contextual descriptions locally.

\subsubsection{Improving Fine-Grained Activity Recognition:}
\modified{Our evaluation reveals insights about current VLM capabilities and their downstream effects on HAR. While Conservative and Balanced settings achieve strong performance (72-89\% accuracy), the Relaxed setting (F1 scores of 48-72\%) explores current boundaries of fine-grained activity recognition. Since VLM labeling quality influences training data for privacy-preserving HAR models, this performance partially reflects VLM capabilities in distinguishing subtle activity differences. Analysis across three state-of-the-art VLMs—GPT-4.1, Gemini 2.5, and Claude Sonnet 4 (See~\appref{sec:vlm-analysis})—shows consistent patterns, representing general characteristics of current VLM technology rather than framework-specific limitations. This cascading relationship positions \name's sensor-first architecture to benefit from advances in VLM capabilities, particularly improved spatiotemporal reasoning and fine-grained visual understanding.}

\subsubsection{Hybrid Approaches: Combining Supervised and Discovery Methods}

\modified{Our clustering approach faces inherent tradeoffs: users must wait for sufficient data collection before meaningful clusters emerge (the "cold start" problem), and initial recognition performance remains lower until adequate training data accumulates. These limitations suggest promising hybrid approaches where supervised and discovery methods play complementary roles. Our real-world deployment revealed that certain activities occur consistently across diverse home environments, while others remain highly environment-specific. This observation suggests that supervised models could provide robust baseline recognition for common activities that generalize across homes, while unsupervised discovery captures the personalized routines that make each home unique. Future work should explore domain adaptation techniques that identify environment-invariant features for transferable activities while preserving the flexibility to adapt to environment-specific variations, potentially combining supervised representation learning with our sensor-first discovery paradigm.}

\section{CONCLUSION}
\label{sec:conclusion}

This work introduces \capname, a framework for at-home human activity recognition (HAR) that removes the reliance on pre-defined labels and continuous video processing by leveraging local, privacy-preserving sensors to identify key moments for targeted video bootstrapping and activity discovery.
This approach minimizes computational overhead and cloud dependency while maintaining robust recognition accuracy across varying levels of sensor granularity, as demonstrated in our evaluations across diverse kitchen environments.
By integrating multimodal sensor data with VLMs during an efficient training phase, our hierarchical bootstrapping technique successfully translates rich semantic information into discrete, sensor-compatible activity labels. 
Ultimately, our results validate the feasibility of scalable, privacy-aware HAR systems that can adapt to different sensing configurations, paving the way for more efficient and user-centric smart environments.

\section*{ACKNOWLEDGEMENTS}

This work was partially supported by NSF Awards SaTC-1801472 and CSR-1526237, the Translational Fellowship from CMU's Center for Machine Learning and Health (CMLH), and the Presidential Fellowship from CMU's CyLab Security and Privacy Institute. We gratefully acknowledge the generous gift from Trane Technologies supporting smart buildings research at Carnegie Mellon University. 
We extend our sincere thanks to the Longitudinal Prize for Dementia Foundation and the Lived Experience Advisory Panel (LEAP) for their valuable feedback on the overall project direction. We are deeply grateful to our Autonomous team for their assistance with data collection and their constructive feedback throughout the multiple iterations of the hardware development. 
We would also like to thank Ben Weinshel, Anu Sitaraman, Haozhe Zhou, Vimal Mollyn, and Shreya Bali for their insightful comments on the paper and invaluable input on the system design throughout this research. Finally, we express our appreciation to the anonymous reviewers for their thoughtful and constructive feedback that significantly improved this work.

\bibliographystyle{ACM-Reference-Format}
\bibliography{reference}

\newpage
\appendix
\section{APPENDIX}

\subsection{Featurization Pipelines for Various Sensing Modalities}
\label{sec:appendix-featurization}

This appendix provides detailed technical specifications for the featurization pipelines used for specific sensing modalities. The following detailed descriptions outline the specific features extracted from each sensor, and the rationale behind our design choices, providing the implementation details necessary for replicating our multimodal sensing approach.
\begin{itemize}
\item \textbf{Doppler Radar:} We extract temporal features (velocity statistics, directional changes), spatial features (distance from sensor, movement extent), complexity features (velocity entropy, motion transitions), and motion type (stationary, slow/fast movement \etc) distributions. While Doppler provides precise and high fidelity velocity sensing, it only captures motion in the radial direction and struggles with distinguishing movement at the same distance or complex spatial relationships.

\item \textbf{2D LiDAR:} Our approach first establishes a static boundary model in the horizontal 2D plane, then extracts features from deviations that represent dynamic objects. Features include centroid positions, velocities, point distributions, boundary interactions, and temporal patterns. LiDAR provides precise spatial mapping but with decreasing granularity at greater distances, and it cannot capture fine-grained movements or height information.

\item \textbf{Low-Resolution Thermal Camera:} From the 10×10 thermal array, we compute spatial features (temperature statistics, gradients) and signature-based features that identify thermal patterns from humans, appliances, and ambient sources. While thermal sensing excels at detecting heat-generating activities and object interactions, it is limited to coarse human movements or thermal changes.

\item \textbf{IMU (Wearables):} We leverage the pretrained SAMoSA \cite{mollyn_ubicomp2022_samosa} model to extract 128-dimensional motion features from 2.88-second windows with a stride of 0.21 seconds. Wearable IMUs capture detailed hand and arm movements but are limited to the wearer's perspective and require consistent device wearing.

\item \textbf{Pose:} From the raw 2D pose data over 25 body keypoints (in 640x480 frame) at 6-8Hz, we extract biomechanical features that characterize body positioning and movement, including joint velocities and accelerations, inter-joint configurations, working zones (based on torso and hand movements), and movement complexity. Despite providing rich body movement data, pose estimation is constrained by occlusions and the limited field of view (less than 60 degrees), sometimes missing key interactions outside the camera's perspective.

\item \textbf{Depth:} We extract motion patterns and statistical features from the 10×10 depth array, establishing boundary maps to identify deviations representing people and objects. While effective for detecting presence and basic movements, depth sensing is limited by its narrow field of view and cannot capture fine-grained interactions.

\end{itemize}

\subsection{Prompts for LLMs and VLMs}
\label{sec:appendix-prompts}

This appendix provides the detailed prompts used in the \name framework for activity discovery. 
The pipeline employs five main prompts for Vision Language Models (VLMs) and Large Language Models (LLMs)  that work in sequence to translate video clips into consistent activity labels at appropriate granularity levels.
The exact prompts can be found in the Supplemental File as a text file.
Here, we describe the key design for each prompt to address specific challenges in the pipeline. 

\begin{enumerate}
    \item The first step leverages a VLM to generate rich semantic descriptions from video frames captured during key moments. This prompt instructs the VLM to analyze short video clips focusing on actions, objects, locations, and activity structure. It specifically directs the model to focus only on clearly observable elements, provide confidence scores, and exclude low-confidence observations.
    \item Since VLM outputs contain diverse location references (\eg ``at sink,'' ``by counter''), our second step employs an LLM to consolidate these into consistent functional zones. This prompt directs the LLM to cluster locations based on supported activities, ensuring that areas like ``sink area,'' ``counter area,'' and ``coffee machine area'' are consistently identified. This spatial context is crucial because activities have different meanings in different locations.
    \item After establishing functional zones, we use a two-step process to transform the unstructured VLM descriptions into discrete activity labels. First, we employ the clustering prompt to create initial activity clusters for each functional zone. This prompt directs the LLM to preserve task-specific distinctions while minimizing the overall number of clusters.
    \item Next, we match each VLM description to the appropriate cluster using the matching prompt. This prompt directs the LLM to evaluate matches based on action alignment (60\%), object consistency (25\%), and location match (15\%), ensuring that descriptions are consistently mapped to appropriate activity labels.
    \item Finally, to enable granularity control through the relaxation parameter $\lambda$, we analyze each activity's semantic properties across multiple dimensions. This prompt directs the LLM to break down each activity in terms of action type, objects involved, sub-location, purpose, related activities, contrasting activities, adjacent objects, and access patterns. These dimensions form the basis for computing pairwise similarity scores between activities, which are then used with the relaxation parameter $\lambda$ to determine which activities should be merged at different granularity settings.
\end{enumerate}

\subsection{\modified{VLM Variability and Configuration Analysis}}
\label{sec:vlm-analysis}

\modified{We conducted additional analyses examining how different VLM models and frame rate configurations impact label generation performance. These evaluations provide insights into design choices and system adaptability across different VLM capabilities.}

\subsubsection{\modified{Performance of Label Discovery pipeline with different VLMs}}

\modified{We evaluated \name using three state-of-the-art vision language models as of July 2025: OpenAI's GPT-4.1, Google's Gemini 2.5 Flash, and Anthropic's Claude Sonnet 4. These models represent different architectures and training approaches currently available. For consistency, all models used identical prompts with minor formatting adaptations for model-specific requirements.}
\begin{figure}[ht]
    \centering
    \hfill
    \begin{minipage}[t]{0.5\textwidth}
        \centering
        \renewcommand{\arraystretch}{1.2}
        \setlength{\tabcolsep}{3pt}
        \begin{tabular}{|c|c|c|c|c|}
        \hline
        \multicolumn{5}{|c|}{\textbf{\begin{tabular}[c]{@{}c@{}}VLM Model\\Comparison\end{tabular}}} \\
        \hline
        \textbf{Granularity} & \textbf{Model} & \textbf{\begin{tabular}[c]{@{}c@{}}Acc\\(\%)\end{tabular}} & \textbf{\begin{tabular}[c]{@{}c@{}}F1\\(\%)\end{tabular}} & \textbf{\begin{tabular}[c]{@{}c@{}}Detection\\Rate (\%)\end{tabular}} \\
        \hline
        \multirow{3}{*}{\begin{tabular}[c]{@{}c@{}}\textbf{Conservative}\\ \textbf{($\lambda=0.4$)}\end{tabular}} 
        & GPT-4.1 & 84.9 & 72.5 & 92.0 \\
        \cline{2-5}
        & Gemini 2.5 & 80.9 & 69.7 & 50.2 \\
        \cline{2-5}
        & Claude Sonnet 4 & 59.3 & 50.8 & 89.2 \\
        \hline
        \multirow{3}{*}{\begin{tabular}[c]{@{}c@{}}\textbf{Balanced}\\ \textbf{($\lambda=0.3$)}\end{tabular}} 
        & GPT-4.1 & 83.9 & 69.4 & 92.0 \\
        \cline{2-5}
        & Gemini 2.5 & 79.0 & 67.2 & 50.2 \\
        \cline{2-5}
        & Claude Sonnet 4 & 60.4 & 55.4 & 89.2 \\
        \hline
        \multirow{3}{*}{\begin{tabular}[c]{@{}c@{}}\textbf{Relaxed}\\ \textbf{($\lambda=0.2$)}\end{tabular}} 
        & GPT-4.1 & 72.0 & 55.4 & 92.0 \\
        \cline{2-5}
        & Gemini 2.5 & 69.0 & 54.0 & 50.2 \\
        \cline{2-5}
        & Claude Sonnet 4 & 48.7 & 35.6 & 89.2 \\
        \hline
        \end{tabular}
        \vspace{0.5em}
        \captionsetup{width=\textwidth}
        \caption{\modified{Label discovery performance across VLM models and semantic granularity settings. Detection rate represents the \% of key moments that generate useful activity descriptions.}}
        \Description{\modified{Label discovery performance across VLM models and semantic granularity settings. Detection rate represents the \% of key moments that generate useful activity descriptions.}}
        \label{tab:vlm-model-comparison}
    \end{minipage}
    \hfill
    \begin{minipage}[t]{0.45\textwidth}
        \centering
        \renewcommand{\arraystretch}{1.2}
        \setlength{\tabcolsep}{4pt}
        \begin{tabular}{|c|c|c|c|c|}
        \hline
        \multicolumn{5}{|c|}{\textbf{\begin{tabular}[c]{@{}c@{}}Frame Rate\\Impact Analysis\end{tabular}}} \\
        \hline
        \textbf{\begin{tabular}[c]{@{}c@{}}Frame\\Rate\end{tabular}} & \textbf{Granularity} & \textbf{\begin{tabular}[c]{@{}c@{}}Acc\\(\%)\end{tabular}} & \textbf{\begin{tabular}[c]{@{}c@{}}F1\\(\%)\end{tabular}} & \textbf{\begin{tabular}[c]{@{}c@{}}Detection\\Rate (\%)\end{tabular}} \\
        \hline
        \multirow{3}{*}{\textbf{1 FPS}} 
        & Conservative & 84.9 & 72.5 & 92.0 \\
        \cline{2-5}
        & Balanced & 83.9 & 69.4 & 92.0 \\
        \cline{2-5}
        & Relaxed & 72.0 & 55.4 & 92.0 \\
        \hline
        \multirow{3}{*}{\textbf{3 FPS}} 
        & Conservative & 82.9 & 70.5 & 91.3 \\
        \cline{2-5}
        & Balanced & 80.6 & 62.5 & 91.3 \\
        \cline{2-5}
        & Relaxed & 69.0 & 52.4 & 91.3 \\
        \hline
        \multirow{3}{*}{\textbf{5 FPS}} 
        & Conservative & 82.4 & 67.5 & 90.6 \\
        \cline{2-5}
        & Balanced & 79.2 & 65.1 & 90.6 \\
        \cline{2-5}
        & Relaxed & 66.7 & 53.1 & 90.6 \\
        \hline
        \end{tabular}
        \vspace{0.5em}
        \captionsetup{width=\textwidth}
        \caption{\modified{Impact of frame rate on label discovery performance using GPT-4.1. Detection rates remain consistent across frame rate settings.}}
        \Description{\modified{Impact of frame rate on label discovery performance using GPT-4.1. Detection rates remain consistent across frame rate settings.}}
        \label{tab:frame-rate-analysis}
    \end{minipage}
    \hfill
\end{figure}

\modified{\tabref{tab:vlm-model-comparison} shows notable variations across models. GPT-4.1 and Gemini 2.5 Flash demonstrate similar accuracy levels in Conservative and Balanced settings (80-85\%), while Claude Sonnet 4 shows lower accuracy but reasonable performance in the Balanced setting (60.4\% accuracy, 55.4\% F1). However, accuracy and F1 scores alone do not capture whether VLMs can consistently provide meaningful activity descriptions. To measure this consistency, we examine the detection rate—the percentage of video segments for which the VLM generates actual location and action information rather than empty responses. GPT-4.1 maintains a high detection rate (92.0\%) across all settings, while Gemini 2.5 Flash achieves only 50.2\% detection rate despite competitive accuracy when successful, and Claude Sonnet 4 maintains a high detection rate (89.2\%) but with lower overall accuracy. For our use case, GPT-4.1 performs the best in terms of both accuracy and reliability, making it the most suitable choice for consistent activity discovery across diverse scenarios.}

\subsubsection{\modified{Frame Rate Impact Analysis}}

\modified{Based on GPT-4.1's combination of high accuracy and detection rate, we selected it for frame rate analysis. Using GPT-4.1, we evaluated performance across three frame rates (1, 3, and 5 FPS) using identical 5-second video clips from our dataset. \tabref{tab:frame-rate-analysis} shows an interesting pattern: increasing frame rate from 1 to 5 FPS results in modest performance decreases rather than improvements. Accuracy drops from 84.9\% to 82.4\% in the Conservative setting, with similar trends observed across all granularity levels. This finding suggests that kitchen activities may operate at coarse timescales where meaningful changes occur over multi-second intervals. Additional frames often capture intermediate postures, repetitive motions, or transitional states that may not provide additional semantic information useful for activity classification. The 1 FPS configuration appears to capture key semantic moments while avoiding potential temporal noise. Frame rate selection also has practical implications for deployment. Processing at 5 FPS requires five times the input context in the VLM API calls compared to 1 FPS, with corresponding increases in latency, bandwidth usage, and cloud processing costs. In conclusion, our analysis indicates that this increased computational overhead does not translate to improved recognition performance under current VLM capabilities.}

\subsubsection{\modified{Implications for System Design}}

\modified{These evaluations provide several insights for OrganicHAR deployments. The detection rate metric reveals that model selection involves trade-offs between accuracy when successful and reliability of processing, where consistent processing across all key moments may be more important than marginal accuracy improvements on successfully processed segments. The frame rate analysis supports our original configuration choices (1 FPS, 640×480 resolution) as a reasonable balance between performance and computational efficiency. The consistent performance patterns across multiple VLMs indicate that fine-grained activity recognition limitations (F1 scores of 48-72\% in Relaxed settings) reflect current VLM capabilities rather than weaknesses in our approach, as similar performance degradation occurs across different model architectures when distinguishing semantically similar activities. OrganicHAR's modular architecture enables adaptation to different VLMs without modifying sensor processing components, allowing the framework to incorporate future advances in VLM temporal reasoning and fine-grained visual understanding while maintaining efficiency for current deployments.}

\subsection{Activity Confusion Analysis (Other Sensor Configuration)}
\label{sec:appendix-activity-confusion}

\begin{figure}[ht]
\begin{center}
 \subfigure[Conservative ($\lambda=0.4$)]{
    \centering
    \includegraphics[width=0.29\textwidth]{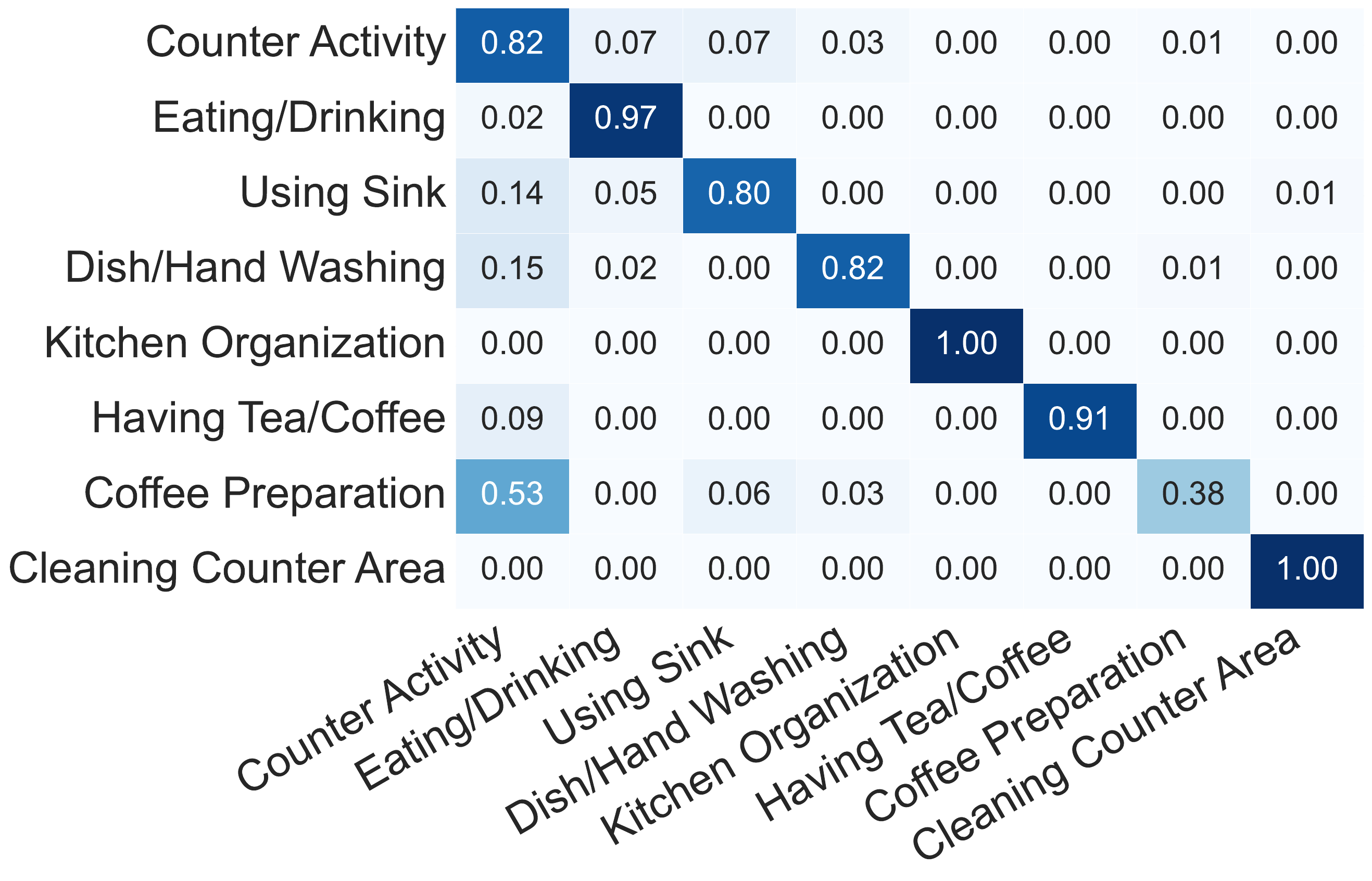}}
 \subfigure[Balanced ($\lambda=0.3$)]{
     \centering
     \includegraphics[width=0.29\textwidth]{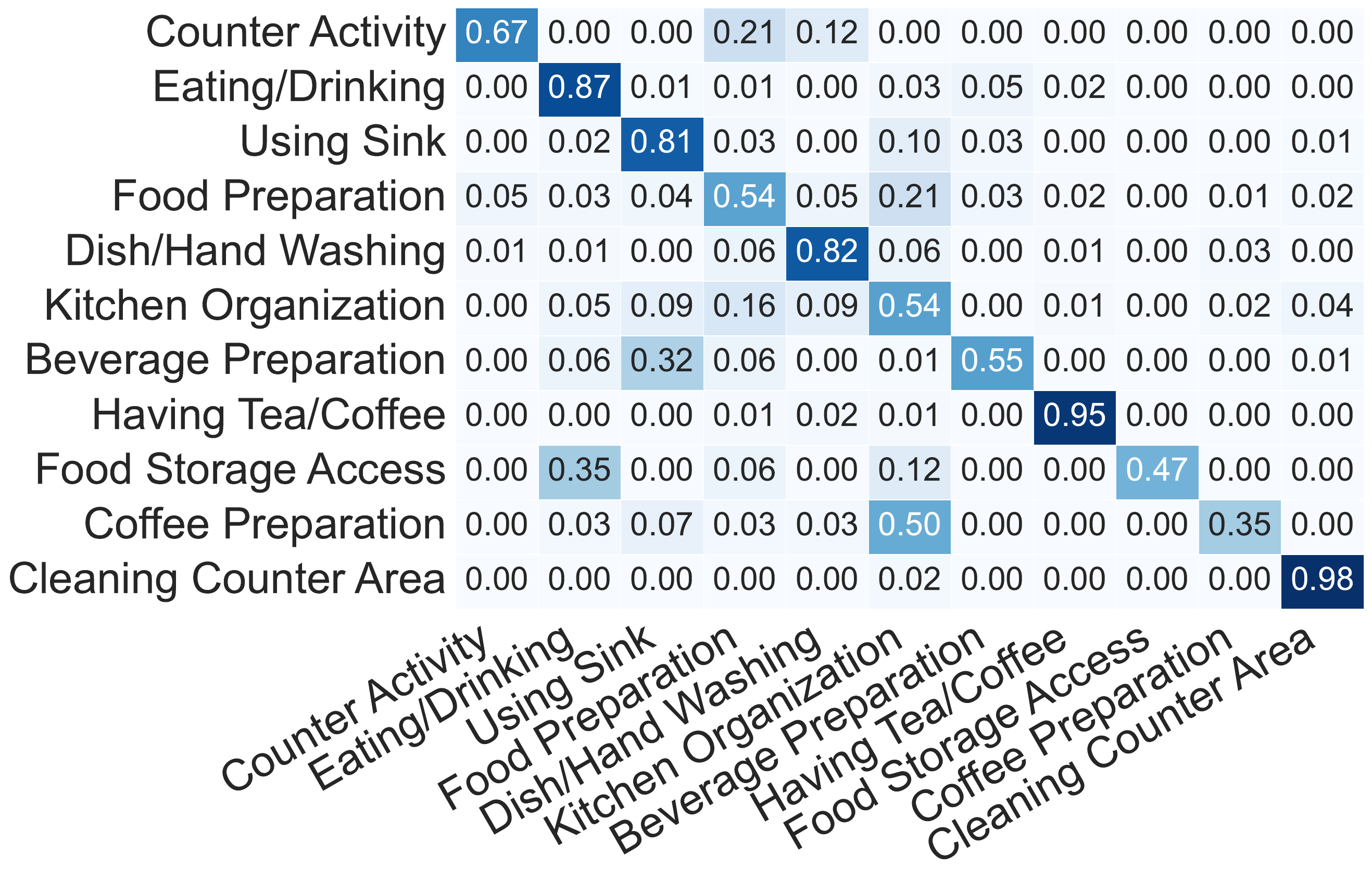}}
 \subfigure[Relaxed ($\lambda=0.2$)]{
     \centering
     \includegraphics[width=0.35\textwidth]{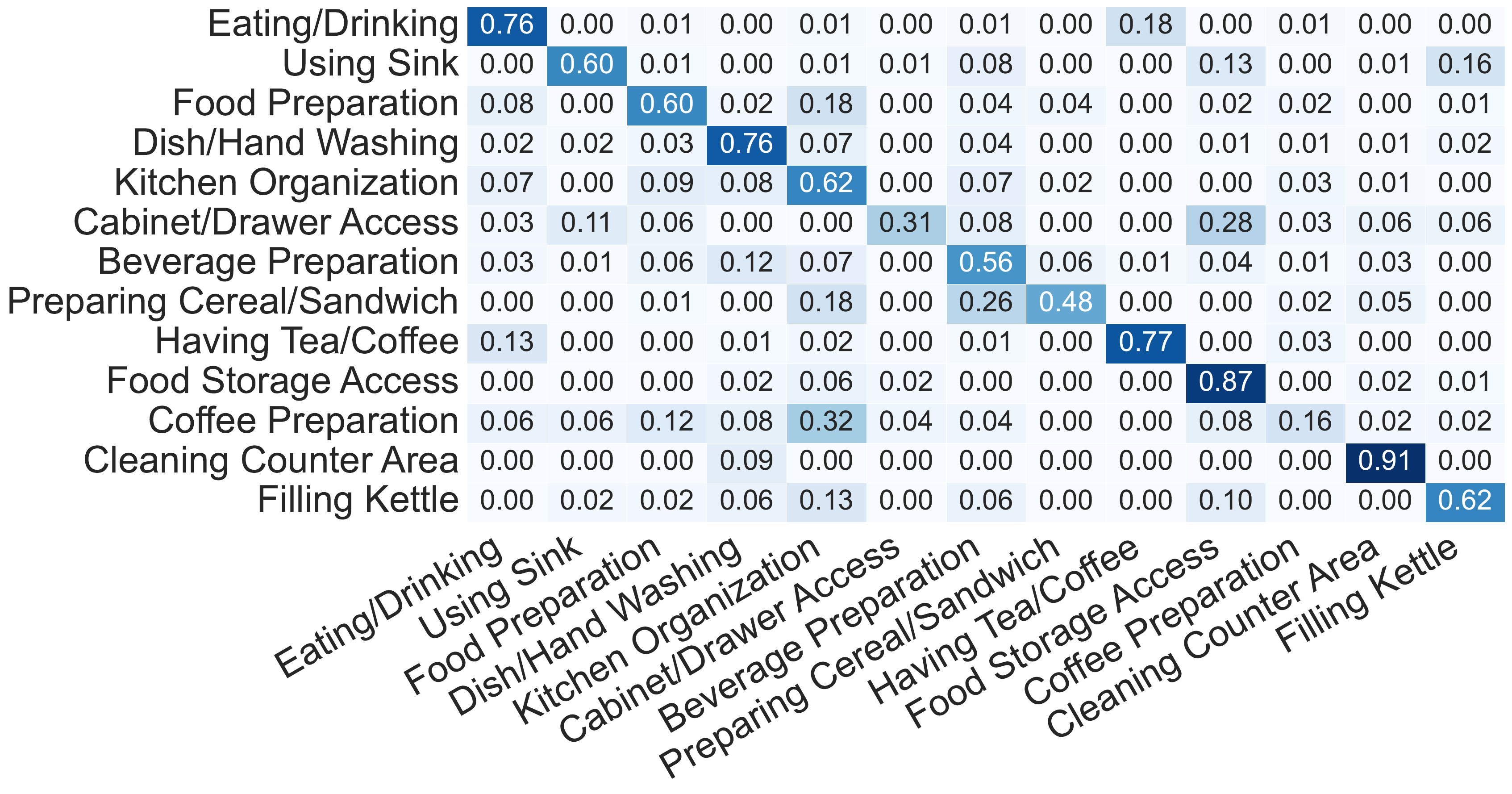}}
\end{center}
\caption{Confusion matrices showing activity recognition performance for \textit{Ambient(Basic)+Wearable(IMU)} configuration across three granularity settings.}
\Description[Three confusion matrices for \textit{Ambient(Basic)+Wearable(IMU)} configuration across granularity settings]{Three confusion matrices with blue color gradients showing activity recognition performance. Conservative matrix ($\lambda=0.4$, $8 \times 8$) shows strong diagonal performance with Eating/Drinking ($0.97$), Kitchen Organization ($1.00$), Having Tea/Coffee ($0.91$), and Cleaning Counter Area ($1.00$) achieving excellent recognition, while Coffee Preparation shows lower accuracy ($0.53$) with confusion to Counter Activity ($0.53$). Balanced matrix ($\lambda=0.3$, $13 \times 13$) with finer-grained activities displays diagonal values ranging $0.31$-$0.91$, with Cleaning Counter Area ($0.91$), Food Storage Access ($0.87$), Having Tea/Coffee ($0.77$), and Dish/Hand Washing ($0.76$) performing well, while Cabinet/Drawer Access ($0.31$) and Preparing Cereal/Sandwich ($0.48$) show substantial confusion. Relaxed matrix ($\lambda=0.2$, $11 \times 11$) demonstrates highest accuracy for Cleaning Counter Area ($0.98$), Having Tea/Coffee ($0.95$), Eating/Drinking ($0.87$), and Using Sink ($0.81$), but lowest for Food Storage Access ($0.47$, confused with Eating/Drinking $0.35$), Kitchen Organization ($0.54$), Food Preparation ($0.54$), Beverage Preparation ($0.55$, confused with Using Sink $0.32$), and Coffee Preparation ($0.35$, confused with Kitchen Organization $0.50$).}
\label{fig:confusion-matrices-imu}
\end{figure}

\begin{figure}[ht]
\begin{center}
 \subfigure[Conservative ($\lambda=0.4$)]{
    \centering
    \includegraphics[width=0.29\textwidth]{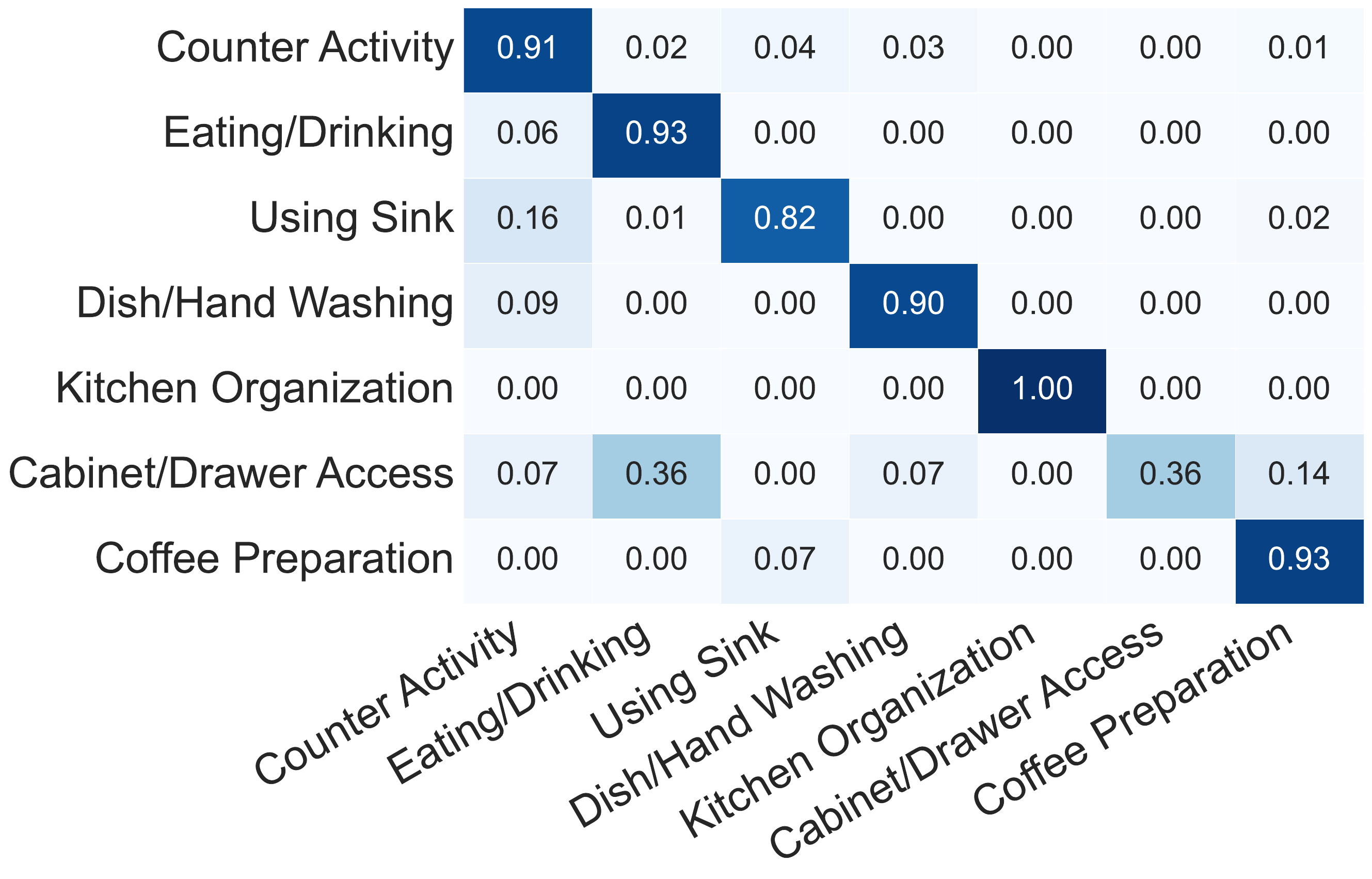}}
 \subfigure[Balanced ($\lambda=0.3$)]{
     \centering
     \includegraphics[width=0.29\textwidth]{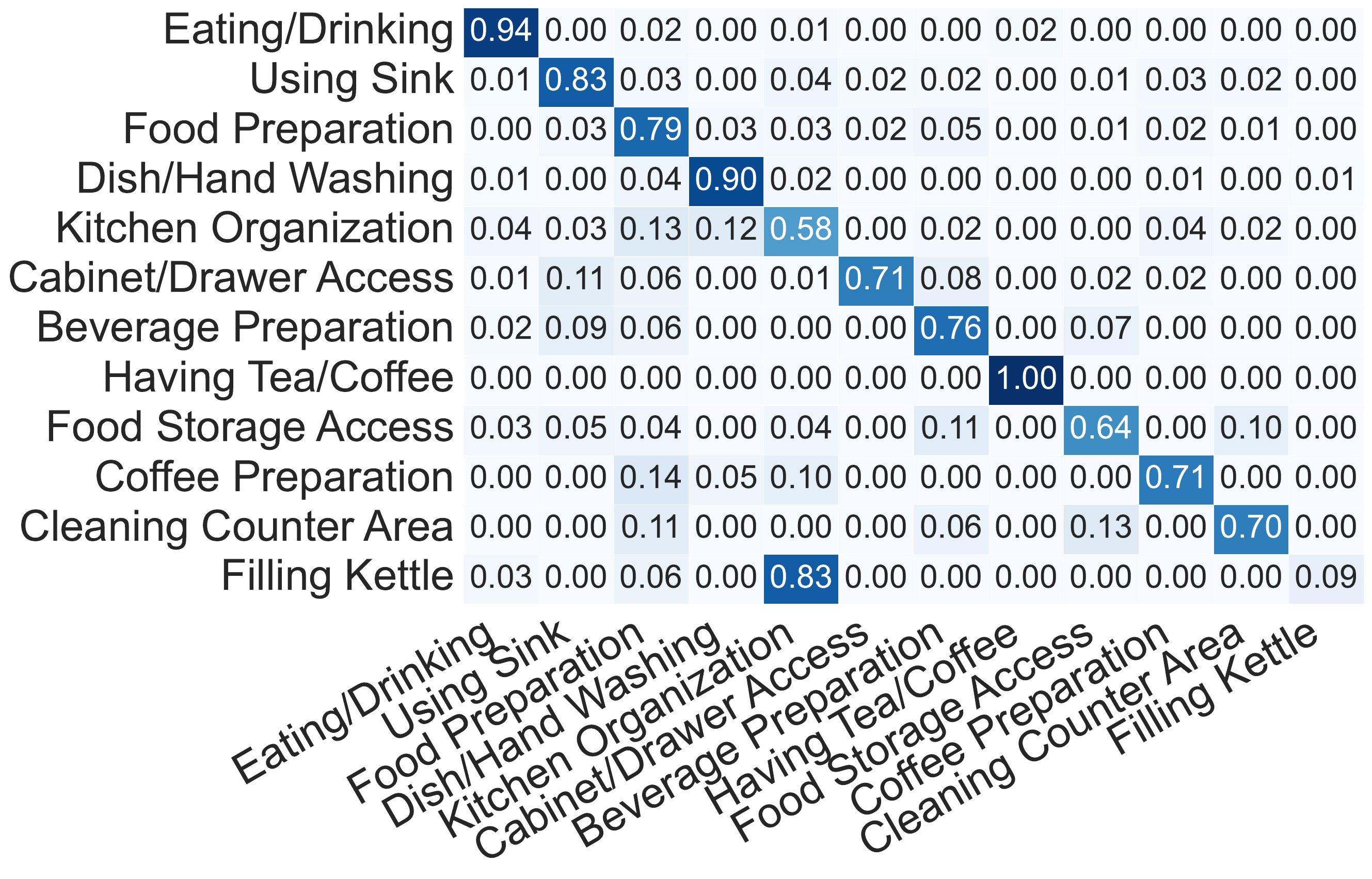}}
 \subfigure[Relaxed ($\lambda=0.2$)]{
     \centering
     \includegraphics[width=0.35\textwidth]{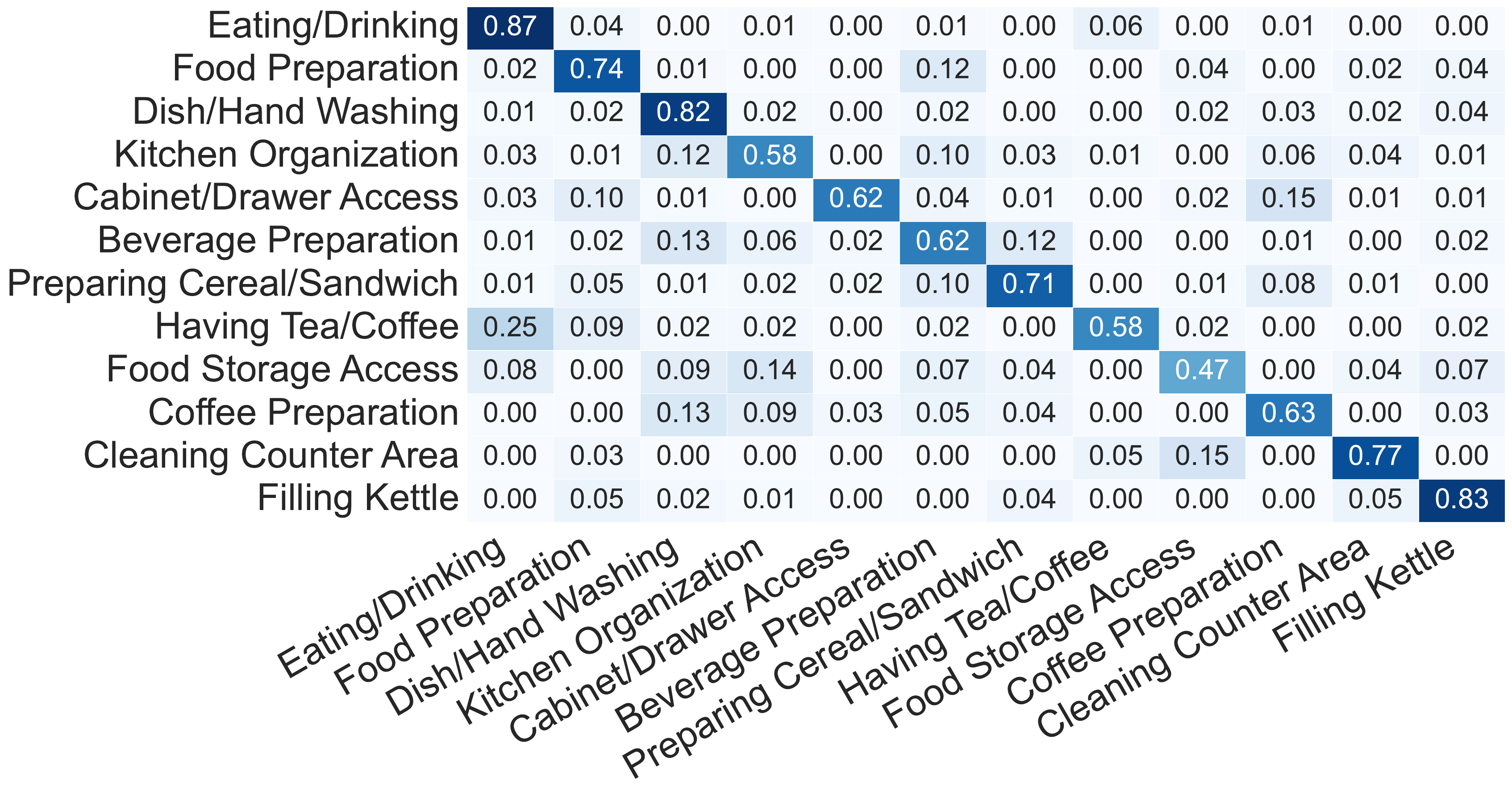}}
\end{center}
\caption{Confusion matrices showing activity recognition performance for \textit{Ambient(Advanced)+Wearable(IMU)} configuration across three granularity settings.}
\Description[Three confusion matrices for \textit{Ambient(Advanced)+Wearable(IMU)} configuration across granularity settings]{Three confusion matrices with blue color gradients for advanced sensor configuration. Conservative matrix ($\lambda=0.4$, $7 \times 7$) shows excellent performance with Kitchen Organization ($1.00$), Eating/Drinking ($0.93$), Coffee Preparation ($0.93$), Dish/Hand Washing ($0.90$), Counter Activity ($0.91$), and Using Sink ($0.82$), but Cabinet/Drawer Access struggles ($0.36$, confused with Eating/Drinking $0.36$). Balanced matrix ($\lambda=0.3$, $12 \times 12$) demonstrates diagonal values ranging $0.47$-$0.87$, with strong performers including Eating/Drinking ($0.87$), Dish/Hand Washing ($0.82$), Filling Kettle ($0.83$), and Cleaning Counter Area ($0.77$), while lower performers include Food Storage Access ($0.47$), Having Tea/Coffee ($0.58$), and Kitchen Organization ($0.58$). Relaxed matrix ($\lambda=0.2$, $12 \times 12$) achieves highest accuracies for Having Tea/Coffee ($1.00$), Eating/Drinking ($0.94$), and Dish/Hand Washing ($0.90$), with most activities maintaining $0.70$+ diagonal values except Kitchen Organization ($0.58$) and Food Storage Access ($0.64$).}
\label{fig:confusion-matrices-advanced}
\end{figure}

We present the confusion matrices for the other configurations in our evaluation.
First, adding wearable IMU sensing to basic ambient sensors reveals complementary strengths and persistent challenges (\figref{fig:confusion-matrices-imu}).
Activities with distinctive motion patterns like ``Kitchen Organization'' and ``Cleaning Counter Area'' show excellent recognition across granularity settings.
However, activities sharing similar motion signatures in different locations remain challenging to differentiate.
In Conservative settings, ``Coffee Preparation'' frequently confuses with broader ``Counter Activity'' categories.
The Balanced setting reveals confusion between ``Beverage Preparation'' and ``Using Sink,'' while the Relaxed setting shows increased confusion between semantically related activities like ``Cabinet/Drawer Access'' and ``Food Storage Access.''
These patterns demonstrate that wearable sensing effectively complements ambient sensors for activities with unique motion signatures but cannot fully resolve spatial ambiguities for activities with similar hand movements performed in different contexts.

Secondly, the most advanced configuration combining ambient sensors with wearable IMU, pose estimation, and depth sensing shows notable improvements in activity disambiguation (\figref{fig:confusion-matrices-advanced}).
With pose and depth information, activities that were previously confused due to spatial ambiguity show clearer separation.
Certain activities like ``Having Tea/Coffee'' and ``Eating/Drinking'' achieve particularly strong recognition rates across settings.
However, even with this rich sensing configuration, some challenging distinctions remain—particularly for activities with subtle differences like ``Filling Kettle'' (which often confuses with general ``Kitchen Organization'') and ``Food Storage Access'' (which sometimes confuses with ``Cleaning Counter Area'').
The advanced configuration particularly excels at disambiguating fine-grained activities that involve distinctive postures or spatial relationships, confirming that spatial awareness through pose and depth information provides a crucial complement to motion and environmental sensing. 
This highlights how different sensor modalities each contribute unique capabilities to the overall recognition system.

\end{document}